\documentclass{ws-rv9x6-nocrop}
\usepackage{subfigure}     
\usepackage{ws-rv-van}     
\usepackage{bm,graphicx}

\usepackage[normalem]{ulem}
\usepackage{enumitem}

\makeindex

\usepackage[usenames,dvipsnames]{color}

\newcounter{comm} 

\newcommand{\mylabel}[1]{\red{\hbox{\small\;\;[#1]}}\label{#1}} 

\renewcommand{\mylabel}[1]{\label{#1}}

\newcommand{\myfoot}[1]{\hspace{1.0pt}\footnote{#1}}

\begin{document}

\chapter[Flexoelectricity]{First-principles theory of flexoelectricity} 

\author{Massimiliano Stengel}
\address{ICREA - Instituci\'o Catalana de Recerca i Estudis Avan\c{c}ats, \\ 08010 Barcelona, Spain, and \\
Institut de Ci\`encia de Materials de Barcelona (ICMAB-CSIC), \\ Campus UAB, 08193 Bellaterra, Spain}

\author[M. Stengel and D. Vanderbilt]{David Vanderbilt}
\address{Department of Physics and Astronomy, Rutgers University, \\ Piscataway, New Jersey 08854-8019, USA}

\begin{abstract}
In this Chapter we provide an overview of the current
first-principles perspective on flexoelectric effects in crystalline
solids.
We base our theoretical formalism on the long-wave expansion
of the electrical response of a crystal to an acoustic phonon perturbation.
In particular, we recover the known expression for the piezoelectric
tensor from the response at first order in wavevector ${\bf q}$, and
then obtain the flexoelectric tensor by extending the formalism to second
order in $\bf q$.
We put special emphasis on the issue of surface effects, which
we first analyze heuristically, and then treat more carefully by
presenting a general theory of the microscopic response to an
arbitrary inhomogeneous strain.
We demonstrate our approach by presenting a full calculation
of the flexoelectric response of a SrTiO$_3$ film, where we point out
an unusually strong dependence of the bending-induced open-circuit voltage
on the choice of surface termination.
Finally, we briefly discuss some remaining open issues concerning the
methodology and some promising areas for future research.

\end{abstract}
\body

\makeatletter
\@addtoreset{footnote}{section}
\makeatother

\section{Introduction}\label{intro}

First-principles electronic structure calculations have played an
increasingly important role in our understanding of the properties
of materials and nanostructures in recent decades.  The phrase
``first principles'' is generally used in the condensed-matter
community to convey the notion that the calculations
are free of adjustable parameters, taking as input only some list
of atoms, their atomic numbers, and some initial guesses at their
coordinates in the unit cell.
One then solves the Schr\"odinger equation for the electrons in some
approximation, computes the relaxed atomic coordinates, and
calculates the desired properties of the crystal.  In the
condensed-matter community this is typically
done in the framework of density-functional theory
(DFT),\cite{jones-rmp89}
as shall be assumed below, but Hartree-Fock or other quantum-chemical
methods can also be used.

While the accuracy and efficiency of DFT methods have improved over
the years, of equal importance has been the increasing range of
quantities that can be computed.  In the context of dielectric
properties, the implementation of linear-response theory
for phonon and electric-field perturbations
in the 1980s and 1990s opened up the calculation of phonon frequencies,
dynamical charges, and both electronic and lattice contributions to
the dielectric constant.\cite{Baroni/deGironcoli/DalCorso:2001}
While there was initially some doubt
about whether the piezoelectric response was a bulk property at
all, a seminal paper of Martin laid this question to rest,\cite{Martin}
and the computation of the piezoelectric tensor is now a standard
feature of most DFT codes as well.  Strangely, although many of the above
properties can be computed as derivatives of the electric
polarization $\bf P$, a proper definition of the polarization
$\bf P$ itself proved more difficult; the physics was clarified, and
practical methods for computing it, were developed only in the
mid-1990s with the appearance of the ``modern theory of
polarization.''\cite{resta:92,King-Smith/Vanderbilt:1993,ferro:2007}
Related methods for computing the orbital magnetization of
ferromagnets and the properties of crystals in finite electric
fields have been developed since the 2000's.\cite{resta-jpcm2010}

The flexoelectric tensor has been among the few physical properties
to have resisted a proper first-principles formulation even until
today.  The theory of flexoelectricity was pioneered in the 1980s by
Tagantsev.\cite{Tagantsev,tagantsev-pt91}  However,
because it encodes a response to a strain gradient,
rather than just a strain, and because a strain gradient is
inconsistent with ordinary cell-periodic boundary conditions,
methods based on Bloch's theorem cannot be straightforwardly
applied in the first-principles context.  A serious attack on this
problem did not begin until 2010, when Hong and collaborators
presented the results of calculations on supercell configurations
containing strain gradients.\cite{hong-jpcm10}  Subsequent papers
of Resta\cite{Resta-10} and Hong and Vanderbilt\cite{hong-11}
clarified aspects of the electronic contribution to the flexoelectric
response.

More recently, Stengel\cite{artlin} and Hong and
Vanderbilt\cite{Hong-13} tackled the problem in a systematic way,
and working from slightly different perspectives, arrived consistently
at a nearly
complete framework for defining, and eventually
computing, the flexoelectric tensors fully from first-principles.
Some components
of the flexoelectric tensor that can be expressed only in terms
of bulk current responses, as opposed to charge responses,
still require care in their interpretation and await the development
of efficient methods for calculating them.  However, we can expect
these difficulties to be cleared up soon, so we can look forward
to a new era in which first-principles calculations of flexoelectric
responses can flourish and contribute to a fast-evolving experimental
field.
The purpose of this chapter is to outline the physical principles
underlying these advances in the understanding and computation
of flexoelectric responses, and to summarize a few of the preliminary
results that have been presented in the literature to date.

\section{Theory and methods}
\label{sec:theory}

\subsection{Strain, strain gradients, and responses}

We begin by establishing our notation.
In continuum mechanics, a deformation can be expressed as a three-dimensional (3D)
vector field, $u_\alpha ({\bf r})$, describing the displacement of a material
point from its reference position at ${\bf r}$ to its current location
${\bf r}'$,\myfoot{As in this Chapter we shall deal exclusively with linear
   flexoelectricity, we shall assume a regime of small deformations henceforth.}
$$
r_\alpha'({\bf r}) = r_\alpha + u_\alpha ({\bf r}),
$$
The \emph{deformation gradient} is defined as
the gradient of $u_\alpha$ taken in the reference configuration,
\begin{equation}
\tilde{\varepsilon}_{\alpha \beta}({\bf r}) = u_{\alpha,\beta}({\bf r}) = \frac{\partial u_\alpha({\bf r})}{\partial r_\beta}.
\mylabel{defgrad}
\end{equation}
$\tilde{\varepsilon}_{\alpha \beta}({\bf r})$ is often indicated in the literature as
``\emph{unsymmetrized} strain tensor'', as it generally contains a proper strain plus a rotation.
By symmetrizing its indices one can remove the rotational component, 
thus obtaining the \emph{symmetrized} strain tensor
$$
\varepsilon_{\alpha \beta} = \frac{1}{2} \left( u_{\alpha,\beta} + u_{\beta,\alpha} \right).
$$
This $\varepsilon_{\alpha \beta}$ is a convenient measure of local strain,
as it only depends on \emph{relative} displacements of two adjacent material points,
and not on their absolute translation or rotation with respect to some reference
configuration.

In this work we shall be primarily concerned with the effects of a
spatially inhomogeneous strain. The third-rank \emph{strain gradient} tensor
can be defined in two different ways, both important for the derivations that follow.
The first (\emph{type-I}) form consists in the gradient of the
\emph{unsymmetrized} strain,
\begin{equation}
\eta_{\alpha, \beta \gamma}({\bf r}) =
\frac{ \partial \tilde{\varepsilon}_{\alpha \beta}({\bf r})}{\partial r_\gamma} =
\frac{\partial^2 u_\alpha({\bf r})}{\partial r_\beta \partial r_\gamma}.
\mylabel{eta1}
\end{equation}
Note that $\eta_{\alpha, \beta \gamma}$, manifestly invariant upon $\beta \leftrightarrow \gamma$ exchange,
corresponds to the $\nu_{\alpha \beta \gamma }$ tensor of
Ref.~\citeonline{hong-11}, and to the symbol $\partial \epsilon_{ \alpha \beta} / \partial r_\gamma$
of Ref.~\citeonline{Tagantsev}.
Alternatively, the strain gradient tensor can be defined (\emph{type-II}) as the gradient of
the \emph{symmetric} strain, $\varepsilon_{\alpha \beta}$,
$$
\varepsilon_{\alpha \beta, \gamma}({\bf r}) = \frac{\partial \varepsilon_{\alpha \beta} ({\bf r})}{\partial r_\gamma},
$$
invariant upon $\alpha \leftrightarrow \beta$ exchange.
It is straightforward to verify that the two tensors contain exactly the
same number of independent entries, and that a one-to-one relationship can
be established to express the former as a function of the latter and vice versa,
\begin{equation}
\eta_{\alpha, \beta \gamma} = \varepsilon_{\alpha  \beta, \gamma} +
\varepsilon_{ \gamma \alpha, \beta} - \varepsilon_{ \beta \gamma,\alpha}.
\mylabel{eta2}
\end{equation}

The piezoelectric and flexoelectric tensors describe, respectively,
the macroscopic polarization response to a uniform strain and to a
strain gradient.  In type-I form, these are
\begin{eqnarray}
e_{\alpha \beta \gamma} &=& \frac{d P_\alpha}{d \varepsilon_{\beta \gamma}},
\mylabel{edef} \\
\mu^{\rm I}_{\alpha \beta, \gamma \lambda} &=& \frac{d P_\alpha}{d \eta_{\beta, \gamma \lambda}}.
\mylabel{muidef}
\end{eqnarray}
While the type-I form is more convenient to derive and calculate, the
type-II representation is often preferred in applications. The
type-II flexoelectric tensor is defined as
\begin{equation}
\mu^{\rm II}_{\alpha \lambda, \beta \gamma} = \frac{\partial P_\alpha}{\partial \varepsilon_{\beta \gamma, \lambda}}.
\mylabel{muiidef}
\end{equation}
Note that $\mu^{\rm I}$ and $\mu^{\rm II}$ are both symmetric
under the last two indices, and are related to each other via
Eq.~(\ref{eta2}) according to
\begin{eqnarray}
\mu^{\rm II}_{\alpha \lambda, \beta \gamma} &=& \mu^{\rm I}_{\alpha \beta, \gamma \lambda} +
\mu^{\rm I}_{\alpha \gamma, \lambda \beta} - \mu^{\rm I}_{\alpha \lambda, \beta \gamma},
\mylabel{muii-i} \\
\mu^{\rm I}_{\alpha  \beta, \gamma \lambda} &=& \frac{1}{2}\,\left(
   \mu^{\rm II}_{\alpha \lambda, \beta \gamma} +
   \mu^{\rm II}_{\alpha \gamma,  \beta \lambda}  \right) .
\mylabel{mui-ii}
\end{eqnarray}

\subsection{Long-wave approach}
\label{sec:longwave}

A macroscopic strain gradient breaks the translational
symmetry of the crystal lattice. For this reason, the response to
such a perturbation cannot be straightforwardly represented in periodic
boundary conditions. This makes the theoretical study of flexoelectricity
more challenging than other forms of electromechanical couplings such as
piezoelectricity.
To circumvent this difficulty, we shall base our analysis on
the study of long-wavelength acoustic phonons.
These perturbations, while generally incommensurate with the crystal lattice,
can be conveniently described
in terms of functions that are lattice-periodic, and therefore are formally and
computationally very advantageous.\cite{Baroni/deGironcoli/DalCorso:2001}

Consider a crystal lattice spanned by the real-space translation vectors
${\bf R}_l$ and by the basis vectors ${\bm \tau}_\kappa$, in such a way that
${\bf R}_{l\kappa} = {\bf R}_l + {\bm \tau}_\kappa$ indicates the location
of the atom of sublattice $\kappa$ and cell $l$.
In full generality, the atomic displacements along the Cartesian direction
$\alpha$ associated with a phonon eigenmode of wavevector ${\bf q}$ can be written as
\begin{equation}
u_{\kappa \alpha}(l,t)  = u^{\bf q}_{\kappa \alpha}  \, e^{i {\bf q} \cdot {\bf R}_{l\kappa} - i \omega t},
\mylabel{acoustic}
\end{equation}
where $u^{\bf q}_{\kappa \alpha}$ (independent of either $l$ or $t$) is an
eigenvector of the dynamical matrix at ${\bf q}$, and $\omega$ is the frequency.

A convenient description of arbitrary mechanical deformations can be established by
choosing an acoustic phonon branch, and by performing a long-wave (small ${\bf q}$)
expansion of its eigenvector in the vicinity of the $\Gamma$ point.
Provided that the long-range electrostatic fields are adequately screened
(see Sec.~\ref{sec:e-mac} for a discussion), the aforementioned
expansion can be written as
\begin{equation}
u^{\bf q}_{\kappa \alpha} = U_\alpha \left( \delta_{\alpha \beta} + i q_\gamma \Gamma^\kappa_{\alpha \beta \gamma}
 - q_\gamma q_\lambda N^\kappa_{\alpha \beta \gamma \lambda} + \ldots \right),
\mylabel{expanq}
\end{equation}
where ${\bf U}$ is a Cartesian vector, $\delta_{\alpha \beta}$ is the Kronecker delta,
and $\Gamma^\kappa_{\alpha \beta \gamma}$ and $N^\kappa_{\alpha \beta \gamma \lambda}$
are third- and fourth-order tensors, respectively.
(The dots stand for higher-order terms, which are irrelevant in the context
of the phenomena described here.)
At order zero in ${\bf q}$ the phonon eigenmode is a rigid translation of
the whole lattice along ${\bf U}$ (note the
absence of a sublattice index), while the first- and second-order terms
describe the internal-strain response of the lattice to a uniform strain
or to a macroscopic strain gradient, respectively.

Of course, to obtain the relevant electromechanical coupling coefficients,
the sole knowledge of the lattice distortions is not sufficient -- one needs
to establish a link between atomic displacements and macroscopic polarization.
While in a simplified point-charge model such a link is straightforward, in
the case of a more realistic quantum-mechanical description of a solid things are
significantly more involved, as one needs to understand how the electronic
wavefunctions, and not only the nuclei, respond to a macroscopic deformation.
If the deformation is sufficiently slow, which is
the case of the phenomena described in this chapter,
the electronic cloud responds adiabatically to atomic motion by generating
a microscopic current density (i.e., the quantum-mechanical probability
current). For example, if we displace by hand one atomic sublattice as
\begin{equation}
u_{\kappa \beta}(l,t) = \lambda(t) e^{i {\bf q} \cdot {\bf R}_{l\kappa}}
\mylabel{monochromatic}
\end{equation}
the microscopic current density that is linearly induced by such a
perturbation can be written as\myfoot{Recall that, in classical 
  electrostatics, the density of bound currents ${\bf J}$ and the 
  microscopic polarization ${\bf P}$ are related by 
  ${\bf J} = \partial {\bf P} / \partial t$.}
\begin{equation}
{\bf J}({\bf r},t) = \dot{\lambda}(t) \, {\bf P}_{\kappa \beta}^{\bf q}({\bf r}) \, e^{i {\bf q} \cdot {\bf r}}.
\end{equation}
The function ${\bf P}_{\kappa \beta}^{\bf q}({\bf r})$ is the microscopic polarization
response; its cell average,
\begin{equation}
\overline{\bf P}_{\kappa \beta}^{\bf q} = \frac{1}{\Omega} \int_{\rm cell} d^3 r {\bf P}_{\kappa \beta}^{\bf q}({\bf r}),
\mylabel{Pbar}
\end{equation}
where $\Omega$ is the cell volume, describes the contribution of atomic motion
to the macroscopic polarization, which is the quantity we are ultimately interested in.

To go from here to the electromechanical tensors we need one more step,
i.e., the small-${\bf q}$ expansion of $\overline{\bf P}_{\kappa \beta}^{\bf q}$.
Again expanding in powers of $\bf q$ and keeping terms up to second
order,\myfoot{Note the difference in sign convention between Eq.~(\ref{expanq})
  and Eq.~(\ref{pq}). In the former case, the choice of the sign was uniquely determined by
  the interpretation of $\bm{\Gamma}$ and ${\bf N}$ as internal-strain response tensors.
  In the latter case, the adopted convention allows one to identify the ${\bf P}^{(n)}$
  tensors with the real-space moments of the current-density response~\cite{artlin,Hong-13}.}
\begin{equation}
\overline{\bf P}_{\kappa \beta}^{\bf q} = \overline{\bf P}^{(0)}_{\kappa \beta}
-iq_\gamma \overline{\bf P}^{(1,\gamma)}_{\kappa \beta} -
\frac{q_\gamma q_\lambda}{2} \overline{\bf P}^{(2,\gamma \lambda)}_{\kappa \beta} + \ldots.
\mylabel{pq}
\end{equation}
The zero-th order term is the macroscopic polarization response to a macroscopic translation
of the sublattice $\kappa$ along the direction $\beta$. This corresponds precisely
to the definition of the Born dynamical charge tensor $Z^*$,
\begin{equation}
\overline{P}^{(0)}_{\alpha \, \kappa \beta} = \frac{Z^*_{\kappa,\alpha \beta}}{\Omega}.
\end{equation}
The remaining ${\bf P}$-tensors can be regarded as higher-order counterparts of
the Born charges. (Physically they are directly related to the moments of the
current density induced by the displacement of an isolated atom.~\cite{artlin,Hong-13})

Multiplying the lattice-polarization coupling tensors with the
phonon eigendisplacements, we can collect terms order-by-order in $\bf q$.
The zero-order term (rigid translation) vanishes due to the acoustic sum rule.
At first order in ${\bf q}$, we obtain the explicit
expression for the piezoelectric tensor\cite{artgr} 
\begin{eqnarray}
e_{\alpha \beta \gamma} &=& - \sum_\kappa
\overline{P}_{\alpha, \kappa \beta}^{(1,\gamma)} + \frac{Z^*_{\kappa, \alpha \rho}}{\Omega}
   \Gamma^\kappa_{\rho \beta \lambda},
\end{eqnarray}
where the first and second terms are the electronic (frozen-ion) and
lattice-mediated terms respectively.\myfoot{
  The unsymmetrized strain is $\tilde{\varepsilon}_{\beta \gamma}({\bf r}) =
  i U_\beta q_\gamma e^{i {\bf q \cdot r}}$; this can be replaced with the
  symmetrized strain tensor after observing that both terms on the right-hand side
  are invariant with respect to $\beta \gamma$ exchange.}
Collecting the terms at second order in $\bf q$ gives the flexoelectric
response, which is again a sum
\begin{equation}
\mu^{\rm I}_{\alpha \beta, \gamma \lambda} = \bar{\mu}^{\rm I}_{\alpha \beta, \gamma \lambda} +
\mu^{\rm I,mix}_{\alpha  \beta, \gamma \lambda} +
\mu^{\rm I,latt}_{\alpha \beta, \gamma \lambda},
\mylabel{mui}
\end{equation}
of electronic and lattice terms
\begin{eqnarray}
\bar{\mu}^{\rm I}_{\alpha \beta, \gamma \lambda} &=&
  \frac{1}{2} \sum_\kappa \overline{P}_{\alpha, \kappa \beta}
       ^{(2,\gamma \lambda)} , \mylabel{mui-el} \\
\mu^{\rm I,mix}_{\alpha \beta, \gamma \lambda} &=&
  -\frac{1}{2} \left( \Gamma^\kappa_{\rho \beta \gamma} \overline{P}_{\alpha, \kappa \rho}^{(1,\lambda)} +
    \Gamma^\kappa_{\rho \beta \lambda} \overline{P}_{\alpha, \kappa \rho}^{(1,\gamma)} \right),\\
\mu^{\rm I,latt}_{\alpha \beta, \gamma \lambda} &=& \frac{Z^*_{\kappa, \alpha \rho}}{\Omega}   N^\kappa_{\rho \beta \gamma \lambda},
\end{eqnarray}
where the bar symbol on the first term indicates a purely electronic response and
`mix' and `latt' refer to ``mixed'' and ``lattice-mediated'' contributions,
respectively.
While the piezoelectric and flexoelectric responses have been developed
in parallel until now, we will henceforth concentrate on the
latter, referring the reader to Refs.~[\citeonline{artlin,Hong-13}] for
the detailed treatment of the piezoelectric response.
The corresponding type-II flexoelectric responses are
\begin{equation}
\mu^{\rm II}_{\alpha \lambda, \beta \gamma} = \bar{\mu}^{\rm II}_{\alpha \lambda, \beta \gamma} +
\mu^{\rm II,mix}_{\alpha \lambda, \beta \gamma} +
\mu^{\rm II,latt}_{\alpha \lambda, \beta \gamma},
\mylabel{muii}
\end{equation}
where
\begin{eqnarray}
\bar{\mu}^{\rm II}_{\alpha \lambda, \beta \gamma} &=&  \frac{1}{2} \sum_\kappa \left(
\overline{P}_{\alpha, \kappa \beta}^{(2,\gamma \lambda)} + \overline{P}_{\alpha,
\kappa \gamma }^{(2,\lambda \beta)} - \overline{P}_{\alpha, \kappa \lambda}^{(2,\beta \gamma )}
\right), \mylabel{II-EL} \\
\mu^{\rm II,mix}_{\alpha \lambda, \beta \gamma} &=& -\Gamma^\kappa_{\rho \beta \gamma}
\overline{P}_{\alpha, \kappa \rho}^{(1,\lambda)}, \\
\mu^{\rm II,latt}_{\alpha \lambda, \beta \gamma} &=&  \frac{Z^*_{\kappa, \alpha \rho}}{\Omega}
\left( N^\kappa_{\rho \beta, \lambda \gamma}
      +N^\kappa_{\rho \gamma, \lambda \beta}
      -N^\kappa_{\rho \lambda, \beta \gamma} \right).
\mylabel{II-LM}
\end{eqnarray}
For later convenience we rewrite Eq.~(\ref{II-LM}) as
\begin{equation}
\mu^{\rm II,latt}_{\alpha \lambda, \beta \gamma} =
\frac{Z^*_{\kappa, \alpha \rho}}{\Omega}   L^\kappa_{\rho \lambda, \beta \gamma }
\end{equation}
where $L^\kappa_{\rho \lambda, \beta \gamma }$ (the type-II counterpart of the
type-I internal-strain tensor ${\bf N}$) is the quantity in parentheses
on the right-hand side of Eq.~(\ref{II-LM}).

To summarize, according to Eq.~(\ref{acoustic}) a long-wavelength sound
wave is comprised of a 
lattice-periodic distortion pattern
$u_{\kappa\alpha}^{\bf q}$ modulated by a time- and space-dependent
complex phase factor. At zero order in $\bf q$ the deformation can be
described as purely ``elastic,'' but at higher orders
(i.e., when moving away from the zone center),
internal relaxations of the basis atoms in the primitive cell occur,
as described by the tensors $\bm{\Gamma}$ and ${\bf N}$ (or $\bf L$) at
first and second orders in $\bf q$, respectively.
These are related to how the crystal locally
responds to a macroscopic strain (first order, ``piezo'') or
strain gradient (second order, ``flexo'').
The reader is referred to
Refs.~[\citeonline{artlin,Hong-13}] for the derivation of explicit expressions
for these tensors, but we shall highlight the main conceptual issues
associated with them in Sec.~\ref{sec:latt-resp}.
Each consecutive order in Eq.~(\ref{expanq}) gives rise
to a corresponding term in the expressions for the flexoelectric
tensor in Eqs.~(\ref{mui}) and (\ref{muii}).

Regarding the purely electronic term, $\bar{\bm{\mu}}$,
which is associated with the purely elastic part (order-zero in $\bf q$,
also referred to as ``frozen ion deformation''), we defer its detailed
discussion to Section~\ref{sec:electronic}.
It can be shown that the ``mixed'' $\mu^{\rm II,mix}_{\alpha \lambda, \beta \gamma}$
term involving $\Gamma^\kappa_{\rho \beta \gamma}$ is active only in
crystals that are characterized by Raman-active phonons,\cite{Hong-13}
which is not the case for simple systems such as cubic
rocksalt or perovskite crystals. (Again, we refer the reader to
Refs.~[\citeonline{artlin,Hong-13}] for the explicit discussion of this term.)
By contrast, the $\mu^{\rm II,latt}_{\alpha \lambda, \beta \gamma}$ term
is present in any insulator with IR-active phonons;
as this term is very important in practical applications of the
flexoelectric effect, we shall discuss it shortly in Sec.~\ref{sec:latt-resp}.
First, however, we shall briefly comment on an important issue that
is relevant to the above discussion, concerning
the treatment of
the macroscopic electric fields in the long-wave phonon analysis.

\subsection{Macroscopic electric fields}
\label{sec:e-mac}

Depending on their polarity, long-wave phonons in a crystalline
insulator generally produce macroscopic electric fields.
These are due to the charge perturbation
that is generated by the lattice distortion, and have a nonanalytic behavior
in the vicinity of the $\Gamma$ point.
For example, for a monochromatic perturbation such as that of
Eq.~(\ref{monochromatic}), at the lowest order in ${\bf q}$
the macroscopic electric field
tends to a direction-dependent constant,
\begin{equation}
\overline{\bf E}^{{\bf q} \rightarrow 0}_{\kappa \beta} \sim -\frac{{\bf q}}{\epsilon_0 \Omega}
\frac{ ({\bf q} \cdot {\bf Z}^* )_{\kappa \beta} } { {\bf q} \cdot \bar{\bm{\epsilon}}_{\rm r} \cdot {\bf q} },
\mylabel{e-mac0}
\end{equation}
where $\overline{\bf E}^{\bf q}_{\kappa \beta}$ is defined in
analogy with Eq.~(\ref{Pbar}) and $\bar{\bm{\epsilon}}_{\rm r}$
is the purely electronic relative permittivity tensor.  The main physical consequence of
this
is the well-known frequency splitting between longitudinal optical (LO) and transverse
optical (TO) phonons in polar crystals.
In particular, due to the contribution of Eq.~(\ref{e-mac0}) to the dynamical matrix,
the LO dispersion curves behave nonanalytically already at zero order in ${\bf q}$;
that is, the eigenvalue and eigenvector associated with an LO branch generally
depends on the direction along which one approaches $\Gamma$.
Such a nonanaliticity propagates directly to the electronic and lattice
response functions described in the previous Section, and needs to be
adequately treated in order to be able to apply the Taylor expansions
described in Eq.~(\ref{expanq}) and~(\ref{pq}).\myfoot{
  The response to an acoustic phonon in a nonpiezoelectric insulator is nonanalytic
  only at second order in ${\bf q}$, so the situation appears here, at first sight, less serious
  than in the case of optical phonons. Recall, however, that the flexoelectric
  tensor is precisely an $\mathcal{O}(q^2)$ property, and therefore it is directly
  affected by such issues.}

There are several ways to approach this problem.  For example, the
theory of Ref.~\citeonline{Hong-13} was developed for purely transverse
and longitudinal phonons separately, leading to flexoelectric
coefficients defined at fixed $\bf E$ and $\bf D$ (electric displacement
field) respectively.  Here, we take the approach of removing the
macroscopic ${\bf E}$-fields\myfoot{It is desirable to remove
  the macroscopic fields not only for practical reasons, i.e., to
  make the aforementioned Taylor expansions possible, but also because
  electromechanical tensors are traditionally defined in short-circuit
  electrical boundary conditions.}
in a physically meaningful way by assuming, following Martin,~\cite{Martin}
that a very low density of free carriers is present in the insulating crystal,
and that these are
allowed to redistribute adiabatically in response to a phonon perturbation.
In particular, within the Thomas-Fermi approximation, we write the
free-carrier density as
\begin{equation}
\rho^{\rm free} ({\bf r}) = - \epsilon_0  k_{\rm TF}^2 V({\bf r}),
\mylabel{rhofree}
\end{equation}
where $V({\bf r})$ is as usual the electrostatic potential, and
we suppose that the Fermi wavevector $k_{\rm TF}$ is much smaller
than any reciprocal lattice
vector of the crystal.
In such a regime, the ground-state charge density and wavefunctions
are essentially unaffected by the additional screening
provided by the free-electron gas.
Conversely, in the long-wave limit, the presence of the free carriers
drastically alters the electrostatics; for example, the field of
Eq.~(\ref{e-mac0}) becomes~\cite{artlin}
\begin{equation}
\overline{\bf E}^{{\bf q} \rightarrow 0}_{\kappa \beta} \sim -\frac{{\bf q}}{\epsilon_0 \Omega}
\frac{ ({\bf q} \cdot {\bf Z}^* )_{\kappa \beta} } { k_{\rm TF}^2 + {\bf q} \cdot \bar{\bm{\epsilon}}_{\rm r} \cdot {\bf q} }.
\mylabel{e-macscr}
\end{equation}
Such a modification has the following effects:
\begin{itemize}
\item The macroscopic electric fields, and hence all the response properties of the crystal,
      become \emph{analytic} functions of ${\bf q}$.
\item The macroscopic electric fields vanish at zero and first order in
      ${\bf q}$, and also at second order in $\bf q$ provided that
      we are considering an acoustic phonon branch.
\item Both the piezoelectric and flexoelectric tensors calculated in the presence
      of the free carrier gas are independent of $k_{\rm TF}$, and
      therefore can be unambiguously interpreted as the short-circuit
      versions of the corresponding electromechanical response functions.
\end{itemize}
In the first-principles calculations, this is done in practice by simply
suppressing the ${\bf G}=0$ contribution to the electrostatic energy
when computing the self-consistent linear response; this has the same
effect as introducing a low-density electron gas as described above.

Based on the above discussion, it would be tempting to conclude that the
flexoelectric tensor, like the piezoelectric tensor, is well defined
under short-circuit electrical boundary conditions.
In writing down Eq.~(\ref{rhofree}), however, we assumed a particular
type of carriers, namely electrons (not holes), and moreover that the
band edge for those carriers, the conduction-band minimum (CBM),
tracks with the macroscopic electrostatic potential of the crystal.
In general, however, the CBM energy may shift relative to the local
macroscopic potential as a result of a strain gradient,
via the so-called deformation-potential effect.  Thus, we
can obtain a different flexoelectric tensor
depending on what band feature (CBM or other) we choose as the
energy reference.
We shall come back to this point in Sec.~\ref{sec:spherical}.

For a given energy reference, the bulk flexoelectric tensor $\bm{\mu}$ is well defined
in short-circuit (fixed ${\bf E}$) boundary conditions.
If fixed ${\bf D}$ boundary conditions are imposed along a specific
direction $\hat{\bf q}$, the induced electric field (defined as the
tilt of the corresponding reference potential)
can be then easily calculated as\myfoot{Strictly speaking,
  this is the contribution from bulk effects; one cannot exclude
  surface contributions to the internal field, as we shall see
  in the later sections.}
\begin{equation}
{\bf \Delta E}^{\rm bulk} = -\frac{{\bf q}}{\epsilon_0}
  \frac{ q_\alpha \, \mu^{\rm II}_{\alpha \lambda, \beta \gamma} \, \varepsilon_{\beta \gamma, \lambda} }
       { {\bf q} \cdot \bm{\epsilon}_{\rm r} \cdot {\bf q} }.
\mylabel{fixedd}
\end{equation}
Note that Eq.~(\ref{fixedd}) cannot be written in tensorial form, except
for the simplest case of crystals with cubic symmetry, where the denominator
reduces to a direction-independent constant.

\subsection{Lattice response}
\label{sec:latt-resp}

To gain some insight into the nature of the lattice-mediated flexoelectric
effect it is necessary to understand, in broad terms, the physics behind
the internal-strain response (as described by the tensors ${\bf N}$
or ${\bf L}$) to a strain gradient deformation.
To that end, suppose that we perform a computational experiment where
we statically freeze in a lattice distortion that corresponds to
an acoustic\myfoot{We assume that the long-range Coulomb fields
  have been removed; see Sec.~\ref{sec:e-mac} for details.}
phonon truncated to first
order in ${\bf q}$, i.e., to the uniform-strain level,
\begin{equation}
u^l_{\kappa \alpha} = \left( \delta_{\alpha \beta} + i q_\gamma \Gamma^\kappa_{\alpha \beta \gamma} \right)
                   U_\beta e^{i {\bf q \cdot R}_{l\kappa} }.
\end{equation}
 (In the simplest crystal structures,
where the $\bm{\Gamma}$ tensor identically vanishes, this corresponds to a purely elastic wave.)
As we have perturbed the crystal from its equilibrium configuration, each atom in the lattice
(identified, as usual, by a cell index $l$ and a basis index $\kappa$) will experience a
restoring force $f^l_{\kappa \alpha}$.
If the amplitude of the deformation is small (linear-response regime), such forces
can be described, as usual, by a lattice-periodic (i.e., $l$-independent) function
that is modulated by a complex phase with the same wavevector ${\bf q}$ as the perturbation.
For small ${\bf q}$, it can be shown that the magnitude of the induced forces scales
as $\mathcal{O}(q^2)$ (first-order terms cannot be present, as we have assumed that
uniform-strain effects are already included), and can be written as
\begin{equation}
f^l_{\kappa \alpha} \sim -q_\gamma q_\lambda U_\beta T^\kappa_{\alpha \beta, \gamma \lambda} e^{i {\bf q \cdot R_{l\kappa}}}.
\mylabel{qqUT}
\end{equation}
Here $T^\kappa_{\alpha \beta, \gamma \lambda}$ is, by construction, the type-I flexoelectric
force-response tensor. (The detailed derivation can be found in
Ref.~[\citeonline{artlin,Hong-13}].)

Now one would be tempted, in close analogy to the piezoelectric case, to
define the internal-strain response tensor ${\bf N}$ by means of the
following linear system of equations,
\begin{equation}
\Phi^{(0)}_{\kappa \alpha \kappa' \rho} N^{\kappa'}_{\rho \beta,\gamma \lambda} \stackrel{?}{=} T^\kappa_{\alpha \beta, \gamma \lambda},
\mylabel{eqstat}
\end{equation}
where $\Phi^{(0)}_{\kappa \alpha \kappa' \rho}$ is the zone-center
force-constant matrix.\myfoot{$\Phi^{(0)}$ is the ${\bf q} \rightarrow 0$ 
  limit of the matrix $\Phi^{\bf q}_{\kappa \alpha \kappa' \rho}$, which is
  essentially a dynamical matrix with the mass prefactors set to unity.
  As for other quantities, $\Phi^{(0)}$ is defined at
  vanishing macroscopic electric field, i.e., closed-circuit boundary
  conditions, appropriate for computing transverse optical phonon
  frequencies.}
Unfortunately, the above system is generally not solvable: the sublattice- ($\kappa$-) sum of the
${\bf T}$-tensor does not vanish, and the $\Phi^{(0)}$ matrix is singular. (It is
always characterized by three null eigenvalues, corresponding to rigid translations
of the crystal as a whole.)
As negative as it sounds, this is nonetheless an important result: it tells us that the internal-strain
response to a \emph{static} strain-gradient deformation is \emph{generally} ill-defined.
(We shall see later on that there are notable exceptions to this statement, though.)

To understand what went wrong, let's start all over again, but instead of considering
a static (frozen-in) deformation, take a dynamical one, i.e., a phonon mode.
By performing a long-wave expansion of the equations of motion one obtains~\cite{artlin,Hong-13},
for the second-order eigendisplacements,
\begin{equation}
\Phi^{(0)}_{\kappa \alpha \kappa' \rho} N^{\kappa'}_{\rho \beta,\gamma \lambda} = T^\kappa_{\alpha \beta, \gamma \lambda}
 - \frac{m_\kappa}{M} \sum_{\kappa'} T^{\kappa'}_{\alpha \beta, \gamma \lambda},
\mylabel{eqdyn}
\end{equation}
where $m_\kappa$ are atomic masses and $M=\sum_\kappa m_k$.
Eq.~(\ref{eqdyn}) is in all respect analogous to Eq.~(\ref{eqstat}), except for
the additional term that appears on the right-hand side (rhs) of the latter.
It is trivial to check that the sublattice sum of the rhs now correctly
vanishes, providing us with well-defined values (modulo a rigid translation)
for the ${\bf N}$-tensor components.
This confirms our earlier suspicions that, unlike piezoelectricity, flexoelectricity
is a genuinely \emph{dynamical} effect: only in a sound wave are the internal
strains well defined, and these internal strains depend explicitly on atomic masses.
In retrospect, this conclusion is not entirely surprising. A uniform strain can
always be generated and sustained by applying an appropriate distribution of
external loads to the surface of the sample.
This is not the case for a strain gradient: in general, a uniform force field
applied to each material point of the sample is necessary to generate a given
component of $\varepsilon_{\beta \gamma,\lambda}$.
Such a uniform force can be, e.g., generated by a gravitational field~\cite{Hong-13}
or, as in the above example of the sound wave, by the acceleration of each material
point during its periodic oscillation.~\cite{artlin}
In either case, the result directly depends on the atomic masses.

To gain further insight into the physical nature of the mass-dependent
term in Eq.~(\ref{eqdyn}), it is useful to write the same equation in
type-II form,
\begin{equation}
\Phi^{(0)}_{\kappa \alpha \kappa' \rho} L^{\kappa'}_{\rho \lambda, \beta \gamma} =  C^{\kappa}_{\alpha \lambda, \beta \gamma} -
\frac{m_\kappa}{M} \Omega \mathcal{C}_{\alpha \lambda, \beta \gamma}.
\mylabel{massdep}
\end{equation}
Here, 
$C^{\kappa}$ is the type-II flexoelectric force-response tensor,
linked to ${\bf T}$ via the usual permutation of indices,
\begin{equation}
C^{\kappa}_{\alpha \lambda, \beta \gamma} =
  T^{\kappa}_{\alpha \beta, \gamma \lambda}
+ T^{\kappa}_{\alpha \gamma, \lambda \beta}
- T^{\kappa}_{\alpha \lambda, \beta \gamma}
\mylabel{CTTT}
\end{equation}
and $\mathcal{C}_{\alpha \lambda, \beta \gamma}$
is the macroscopic elastic tensor.
To write Eq.~(\ref{massdep}) we have made use of the result
\begin{equation}
\sum_\kappa C^{\kappa}_{\alpha \lambda, \beta \gamma} = \Omega \mathcal{C}_{\alpha \lambda, \beta \gamma},
\mylabel{sumrule}
\end{equation}
which directly relates flexoelectricity to elasticity.~\cite{artlin}
To justify such a sum rule recall that, in
the context of linear elasticity, the stress tensor $\sigma_{\alpha \beta}$
(which we allow to be inhomogeneous in space) is directly related to the elastic and
strain tensors via
\begin{equation}
\sigma_{\alpha \beta} ({\bf r}) = \mathcal{C}_{\alpha \beta \gamma \lambda} \, \varepsilon_{\gamma \lambda} ({\bf r}).
\end{equation}
Recall also that the divergence of the stress tensor integrated over a finite region of
space yields the net force acting on the corresponding volume element of the
material,
\begin{equation}
f_\alpha = \int_\Omega d^3 r \, \nabla_\beta \sigma_{\alpha \beta} ({\bf r}),
\end{equation}
By assuming that the crystal is homogeneous (i.e., that the elastic tensor is
a constant), and by assuming that the deformation varies slowly over the volume of a
primitive cell, we have
\begin{equation}
\sum_\kappa f^\kappa_\alpha ({\bf r}) = \Omega \mathcal{C}_{\alpha \beta \gamma \lambda} \varepsilon_{\gamma \lambda, \beta} ({\bf r}).
\mylabel{totfor}
\end{equation}
Assuming that the force on individual atoms is exclusively produced by
strain-gradient effects (which is justified, as the relaxations due to the
local strain are already included), we can replace $f^\kappa_\alpha$ with the
definition of the flexoelectric force-response tensor, and easily recover
Eq.~(\ref{sumrule}).
Thus, in a hand-waving way,
one can say that the type-II flexoelectric force-response tensor is
a ``sublattice-resolved'' version of the macroscopic elastic coefficients.

The dynamical nature of the flexoelectric tensor is worrisome if we are to
use this theory to rationalize typical experiments -- these are typically
performed statically.
As we shall see in the following, this is not a real issue.
In a material is at static equilibrium there might be nonvanishing stress
fields due to the application of external loads; nevertheless, the force
acting on a material point must vanish everywhere in space.
This leads to the following condition on the \emph{strain-gradient} field,
\begin{equation}
\sum_{\beta \gamma \lambda} \mathcal{C}_{\alpha \lambda, \beta \gamma} \, \varepsilon_{\beta \gamma, \lambda} ({\bf r}) = 0.
\mylabel{static}
\end{equation}
This means that two or more strain-gradient components will
typically be present in any inhomogeneous strain field, in such a way
that their respective net forces mutually cancel.
By using Eq.~(\ref{static}) it is
straightforward to see that the mass dependence disappears from
the resulting polarization field (as obtained by multiplying
the flexoelectric tensor by the local strain-gradient tensor),
confirming the internal consistency of the theory.

The important message here is that, at the static level, we can
define a number of \emph{effective} flexoelectric coefficients;
each of them will correspond to a linearly independent set of
strain-gradient components that satisfies Eq.~(\ref{static}).
(An explicit example is provided in Sec.~\ref{sec:results}.)
It is easy to see that the number of such effective static coefficients
is always smaller than the number of independent components of the
$\bm{\mu}$-tensor.
This means that the latter contains, in fact, more information than
is actually needed to predict the outcome of a static measurement.
This also means that, in order to determine the full flexoelectric
tensor, one cannot rely on static experiments only; additional
dynamical data need to be combined with the static results.\cite{Pavlo}
The resulting values of the tensor components are
always inherently dynamic quantities, even if static
data are, in part, used to compute them.

\subsection{Electronic response}
\label{sec:electronic}

While the lattice-mediated response has a straightforward physical
interpretation (i.e., in terms of a polar distortion of the basis atoms that
is induced by the macroscopic strain gradient), the purely electronic
response (given by the tensor $\bar{\mu}^{\rm II}_{\alpha \lambda, \beta \gamma}$)
is far less intuitive, and therefore deserves a separate
discussion.
First, recall that $\bar{\mu}^{\rm II}_{\alpha \lambda, \beta \gamma}$
is defined in terms of the second-order ${\bf P}$-tensor,
$\overline{P}_{\alpha, \kappa \beta}^{(2,\gamma \lambda)}$.
To understand the physical meaning of the latter, consider the
microscopic current density $J_\alpha({\bf r})$ that is adiabatically
induced when displacing an isolated atom $(l,\kappa)$
with velocity $\dot{u}^l_{\kappa \beta}$ in the 
Cartesian direction $\beta$,~\cite{hong-11,artlin}
\begin{equation}
\mathcal{P}_{\alpha,\kappa \beta}({\bf r}) = \frac{\partial J_{\alpha}({\bf r} + {\bf R}_{l\kappa})}{\partial \dot{u}^l_{\kappa \beta}},
\end{equation}
Provided that the macroscopic electric fields have been appropriately screened,~\cite{artlin}
one can introduce~\cite{Hong-13} the moments of the vector field $\mathcal{P}_{\alpha,\kappa \beta}({\bf r})$
at an arbitrary order $n$,
\begin{eqnarray}
J^{(n,\gamma_1 \ldots \gamma_n)}_{\alpha, \kappa \beta} &=& \int d^3 r \, \mathcal{P}_{\alpha, \kappa \beta} ({\bf r}) r_{\gamma_1} \ldots r_{\gamma_n}.
\end{eqnarray}
Then, one can show~\cite{artlin} that the resulting ${\bf J}$-tensors coincide with the $\overline{P}$-tensors
of the same order apart from a trivial factor of volume,
\begin{eqnarray}
J^{(n,\gamma_1 \ldots \gamma_n)}_{\alpha, \kappa \beta} &=& \Omega \overline{P}^{(n,\gamma_1 \ldots \gamma_n)}_{\alpha,\kappa \beta}.
\end{eqnarray}
This result tells us that the ``frozen-ion'' (in the sense specified
in Ref.~\citeonline{hong-11}) contributions to the piezoelectric and
flexoelectric tensors are given in terms of the first and second
moments of the current-density response to atomic displacements,
respectively.

Direct calculation of the $\overline{P}$-tensors is technically challenging at the time of
writing -- the required current-density response functions are presently not available in the
existing implementations of DFPT.
To avoid this complication altogether, Resta~\cite{Resta-10} proposed to
determine the frozen-ion flexoelectric tensor via the sole knowledge of
the \emph{charge-density} response to an acoustic phonon, in close analogy
with Martin's classic treatment of the piezoelectric problem~\cite{Martin}.
In particular, for an elemental crystal (this result was later generalized to
arbitrary crystals by Hong and Vanderbilt~\cite{hong-11}) Resta demonstrated that
the longitudinal component of the response to a longitudinal strain gradient is given by
\begin{equation}
\mu_{\hat{\bf q}} = \frac{1}{6 \Omega} Q^{(3)}_{\hat{\bf q}}.
\mylabel{longq3}
\end{equation}
Here $\hat{\bf q}$ indicates the spatial direction of interest, and
$Q^{(3)}$ indicates the corresponding \emph{third} moment of the charge-density
response to atomic displacement (dynamical octupole).

To derive this result in the context of the formalism of
Sec.~\ref{sec:longwave}, it is useful to introduce the charge-density
response to the monochromatic lattice perturbation of Eq.~(\ref{monochromatic}),
\begin{equation}
\overline{\rho}_{\kappa \beta}^{\bf q} =  
-iq_\gamma \overline{\rho}^{(1,\gamma)}_{\kappa \beta} -
\frac{q_\gamma q_\lambda}{2} \overline{\rho}^{(2,\gamma \lambda)}_{\kappa \beta} +
i \frac{q_\gamma q_\lambda q_\delta}{6} \overline{\rho}^{(3,\gamma \lambda \delta)}_{\kappa \beta} + \ldots,
\mylabel{rhoq}
\end{equation}
where the overline symbol implies cell averaging as in Eq.~(\ref{Pbar}),
and we have pushed the expansion up to \emph{third} order in ${\bf
q}$. (The zero-order term vanishes because of the condition of
charge conservation.) The $\overline{\rho}$ tensors are trivially related
via
\begin{equation}
\overline{\rho}^{(n,\gamma_1 \ldots \gamma_n)}_{\kappa \beta} = \frac{1}{\Omega} Q^{(n,\gamma_1 \ldots \gamma_n)}_{\kappa \beta}.
\end{equation}
to the moments
\begin{equation}
Q^{(n,\gamma_1 \ldots \gamma_n)}_{\kappa \beta} = \int d^3 r \, f_{\kappa
\beta} ({\bf r}) r_{\gamma_1} \ldots r_{\gamma_n}
\mylabel{Qdef}
\end{equation}
of the charge-density response function $f_{\kappa \beta} ({\bf r})$,\myfoot{
  The function $f$ is, in all respects, analogous to that introduced
  by Martin in his seminal work on piezoelectricity~\cite{Martin}.}
defined as the change in charge density resulting from
a single ionic displacement $\kappa\beta$.
One can also show that the $\overline{\bf P}$-tensors and
$\overline{\rho}$-tensors are related by~\cite{artlin}
\begin{equation}
\overline{\rho}^{(n,\gamma_1 \ldots \gamma_N)}_{\kappa \beta} =
\sum_l \overline{P}^{(n-1,\gamma_1 \ldots [\gamma_l] \ldots \gamma_n)}_{\gamma_l,\kappa \beta} \qquad (n \geq 1),
\end{equation}
where the symbol $[\gamma_l]$ indicates the absence of the element $l$
in the list.
Then one immediately has, for $n=3$,
\begin{equation}
J^{(2, \gamma \lambda)}_{\alpha, \kappa \beta} +
J^{(2, \alpha \gamma)}_{\lambda, \kappa \beta} +
J^{(2, \lambda \alpha)}_{\gamma, \kappa \beta} =
Q^{(3, \alpha \gamma \lambda)}_{\kappa \beta}.
\mylabel{jjj}
\end{equation}
By applying Eq.~(\ref{mui-el}) and Eq.~(\ref{jjj}) to the case of
a longitudinal strain gradient oriented along $\hat{\bf q}$, one easily
recovers Eq.~(\ref{longq3}).

Unfortunately, it is not possible to invert Eq.~(\ref{jjj}) and extract all
components of the ${\bf J}^{(2)}$-tensor from the octupolar response tensor,
${\bf Q}^{(3)}$.
(The fact that ${\bf J}^{(2)}$ contains more information than ${\bf Q}^{(3)}$
can be already appreciated by counting the maximum number of independent entries in
either tensor: 54 in the former, 30 in the latter.~\cite{Hong-13})
Therefore, working only with the charge-density response at the bulk level
is not a viable route to achieving full information over the frozen-ion (electronic)
flexoelectric tensor, $\bar{\bm{\mu}}$.

Such a limitation can be circumvented, at least in cubic crystals,
by considering a more general class of deformations
that cannot be straightforwardly described as bulk acoustic phonons.
For example, as we shall see in the next Section, the open-circuit internal 
field that is linearly induced by bending a free-standing slab is a bulk property 
of the material. 
Since the electric field is uniquely determined by the induced charge density 
this gives us, in principle, access to the \emph{transverse} component of the 
electronic flexoelectric tensor without the need for calculating the polarization 
response.
A bending deformation can be conveniently simulated (although at the price
of a significantly higher computational cost) by adopting a slab geometry,
and by performing a long-wave analysis analogous to that described here to the
corresponding slab supercell.

A formal derivation clarifying whether such a procedure
does indeed yield the same transverse flexoelectric component as the
${\bf P}$-response theory is still missing, due to subtleties at
both the conceptual and technical levels. We shall refer to these
issues again in
the discussion following Eq.~(\ref{delhp}).
In the remainder of the Chapter we shall disregard such issues, and 
provisionally assume that this relationship holds, i.e., that
the bending-induced open-circuit (OC) ${\bf E}$-field and the corresponding 
component of the flexoelectric tensor are related by
\begin{equation}
\frac{\partial E^{\rm OC}_x}{\partial \varepsilon_{yy,x} } = -\frac{ \mu^{\rm II}_{xx,yy} } {\epsilon_0 \bar{\epsilon}_{\rm r} }
\mylabel{E-transverse}
\end{equation}
[for a beam bent as in Fig.~\ref{sketch}(b)], where $\bar{\epsilon}_{\rm r}$ is the
(isotropic) relative permittivity of the material.
We shall use Eq.~(\ref{E-transverse}) from now on, whenever necessary,
to resolve the aforementioned indeterminacy in the transverse components
of $\bar{\bm{\mu}}$.

Note that this issue does not apply to the simpler case of the piezoelectric
response. In fact, one can write that
\begin{equation}
J^{(1,\gamma)}_{\alpha \beta} + J^{(1,\alpha)}_{\gamma \beta} = Q^{(2,\alpha \gamma)}_\beta.
\end{equation}
(Recall that the basis sum of the ${\bf J}^{(1)}$ tensors essentially coincides with
the frozen-ion piezoelectric tensor, and that ${\bf Q}^{(2)}$ is the dynamical
quadrupole tensor.) The above equation can be readily inverted,
\begin{equation}
J^{(1,\gamma)}_{\alpha \beta} = \frac{1}{2} \left[ Q^{(2,\alpha \gamma)}_\beta + Q^{(2,\alpha \beta)}_\gamma
- Q^{(2, \beta \gamma)}_\alpha    \right],
\label{rich}
\end{equation}
which provides an alternative derivation of Martin's theory~\cite{Martin} of piezoelectricity.
For completeness, it is useful to mention that, at order zero, the relationship
between ${\bf J}$- and ${\bf Q}$-tensors is even more direct,
\begin{equation}
J^{(0)}_{\alpha, \kappa \beta} = Q^{(1,\alpha)}_{\kappa \beta} = Z^*_{\kappa,\alpha \beta},
\end{equation}
where ${\bf Z}^*_\kappa$ is the Born effective charge tensor associated
with the $\kappa$ sublattice.
Thus, both  ${\bf J}^{(n)}$ and ${\bf Q}^{(n+1)}$ can be regarded as higher-order
generalizations of the dynamical charge concept, although starting from $n\!=\!2$
(which is relevant for flexoelectricity) the former quantities generally carry more
information than the latter ones.

\subsubsection{Spherical term, pseudopotential dependence, and
the noninteracting spherical-atom paradox}
\label{sec:spherical}

\begin{figure}[!tb]
\begin{center}
\includegraphics[width=4in]{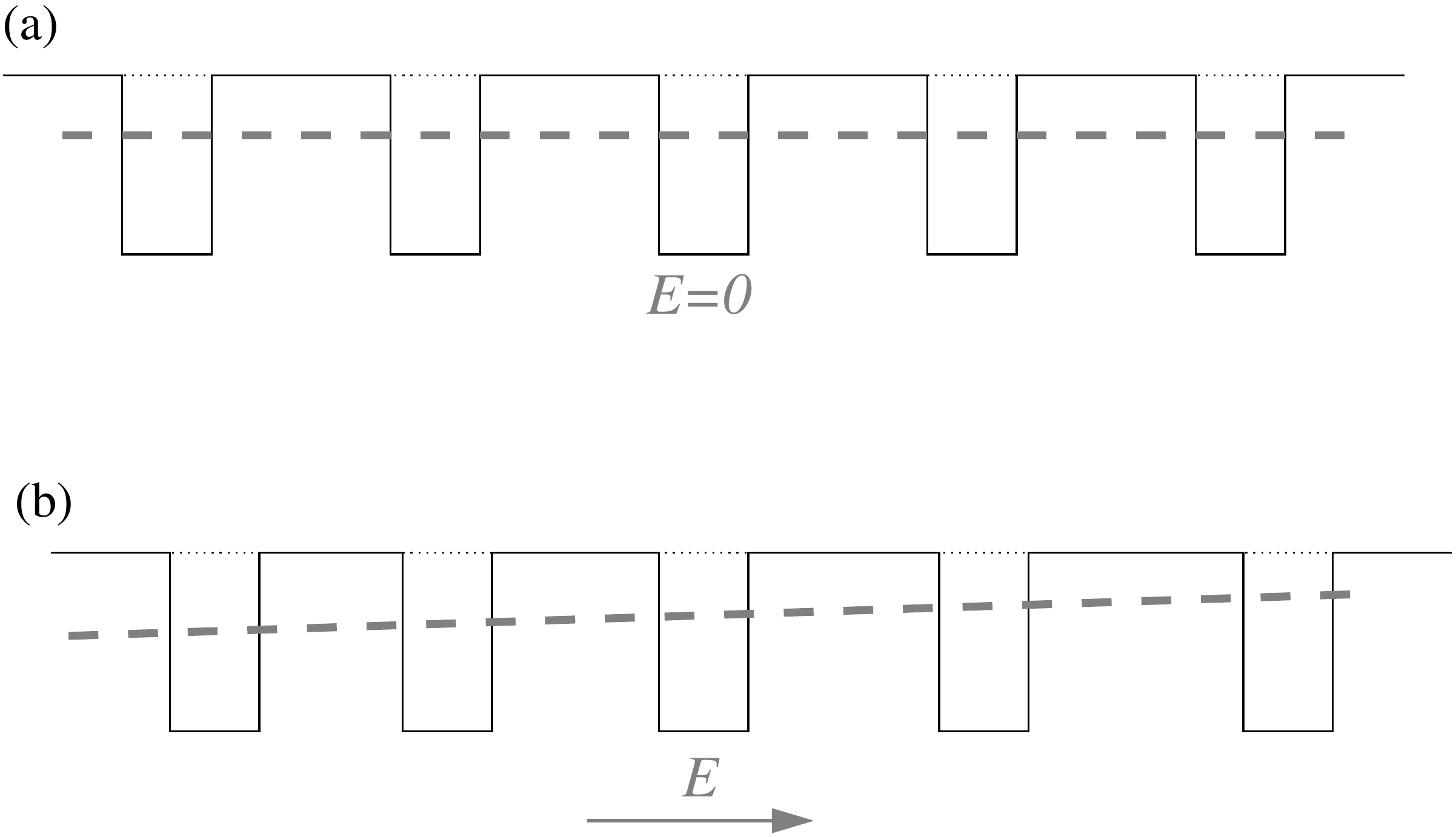}
\end{center}
\caption{Simplified sketch of the planar-averaged electron potential
energy, $-e \overline{V}(x)$ (black curves) for (a) an undistorted
crystal, and (b) a crystal with a uniform longitudinal strain
gradient. \label{figlong} The red dashed lines show the macroscopic
averages of the aforementioned functions.
}
\end{figure}

As an illustration of the above derivations, it is useful in this
context to work out a simple toy model that can be solved analytically;
this will be also useful to point out some unconventional aspects
of the flexoelectric response that have no counterpart
in earlier theories of electromechanical effects in solids.
We consider a rocksalt ionic crystal such as NaCl or MgO,
and suppose that a longitudinal strain gradient develops along the (100) direction.
Here we shall focus on electronic effects only, so that the atomic $x$ coordinates
undergo displacements that are a predetermined quadratic function of $x$ with
no further relaxations.
For the time being we shall also assume that the crystal is perfectly
ionic, i.e., that its electronic charge density can be approximated
by a superposition of spherical closed-shell ions whose shape is not
altered by changes in bond distances, etc.
With the above assumptions in mind, one can perform an average of the
electrostatic potential in the $yz$ planes, and express the
result as a one-dimensional function of $x$. The atomic planes
will appear as a periodic arrangement of potential wells (each
well corresponding to a single charge-neutral monolayer), whose
shape will reflect the radial distribution of electrons in the
constituent ionic species.
For the present purposes, the fine details of the potential
wells are irrelevant; the only important quantity will be
\begin{equation}
K = -e \int_{-\infty}^\infty \overline{V}_\textrm{ML}(x)\,dx \,,
\end{equation}
where $\overline{V}_\textrm{ML}(x)$ is the
$yz$-averaged electric (Hartree) potential $V_\textrm{ML}({\bf r})$
generated by one monolayer, and $-e\overline{V}_\textrm{ML}(x)$ is
the corresponding electron potential energy.
Thus, for purposes of illustration we can represent the potential-energy
wells as nonoverlapping rectangular dips of
area $|K|$, whose shape is fixed and independent of the surrounding
neighbors, as sketched in Fig.~\ref{figlong}(a).
As the wells are all identical and their separations are uniform in the
undistorted crystal, the macroscopic electron potential energy
obtained by convoluting the corresponding microscopic function with
an appropriate low-pass filter~\cite{Baldereschi-88}, shown as a dashed
line in Fig.~\ref{figlong}(a), is constant.
After freezing in the strain gradient deformation pattern, as shown
in Fig.~\ref{figlong}(b),
the interlayer distance increases linearly along the chosen axis, leading to
a constant slope in the macroscopic electrostatic potential and, hence, to
a uniform electric field throughout the bulk crystal.
This result points to a nonzero flexoelectric coefficient of purely
electronic origin, since we explicitly neglected possible internal strains.

There are several important questions that naturally arise at this stage.
The first obvious one is whether (and if yes, how) the outcome of Fig.~\ref{figlong}
can be rationalized in the context of the theory developed in this Chapter.
A second and less obvious issue arises in pseudopotential-based
first-principles calculations, where one may wonder whether and how the
results depend on choice of pseudopotential.
The third question concerns the physical nature of the electric
field that we describe in
Fig.~\ref{figlong}(b).  Does it, for example, produce a direct force on
charged particles such as electron and hole carriers and ionic cores?

To answer the first question, it suffices to suppose that the
electrostatic potential wells are generated by a regular lattice
of spherical charge distributions. To make things simple,
consider a monatomic lattice, as for a rare-gas
solid, that we construct by periodically repeating a spherical
charge distribution $\rho_0({\bf r})$.
We assume that the volume of the unit cell $\Omega({\bf r})$ depends smoothly on
space as a result of an inhomogeneous macroscopic deformation.
One can show (see Supplementary Note 1 of Ref.~\citeonline{artgr})
that the resulting macroscopically averaged (in three dimensions)
electric potential is given by
\begin{equation}
V({\bf r}) = -\frac{1}{6 \epsilon_0 \Omega({\bf r})}
 \int d^3 s \, s^2 \rho_0(s) = -\frac{1}{6 \epsilon_0 \Omega({\bf r})} O_{\rm L},
\end{equation}
where
$O_{\rm L}=4\pi\int ds \, s^4 \, \rho_0(s)$
is the isotropic quadrupole moment\myfoot{That is, the trace of the
   3$\times$3 second-moment tensor.}
of the static charge distribution $\rho_0({\bf r})$. Equivalently,
it is the longitudinal component $O_{\rm L}= \sum_\kappa Q^{(3,xxx)}_{\kappa x}$
of the dynamical octupole tensor defined in Eq.~(\ref{Qdef}), as
follows from straightforward algebra.

In the linear regime (small deformations) we have
\begin{equation}
\Omega({\bf r}) \simeq \Omega (1 + \det{[\varepsilon({\bf r})]}),
\end{equation}
which leads to the variation in the macroscopic electrostatic
potential induced by the deformation,
\begin{equation}
\Delta V({\bf r}) = \frac{1}{6 \epsilon_0 \Omega} \det{[\varepsilon({\bf r})]} O_{\rm L}
\mylabel{Vmacro}
\end{equation}
Assuming that the crystal has cubic symmetry
and making use of Eq.~(\ref{E-transverse}), this yields,
after some algebra,
two of the three independent components of the
flexoelectric tensor
\begin{equation}
\bar{\mu}^{\rm II}_{\alpha \alpha, \beta \beta} =
\frac{O_{\rm L}}{6\Omega}.
\mylabel{bulkfxe}
\end{equation}
(In the case of a biatomic ionic crystal one simply needs to replace
$O_{\rm L}$ with the sublattice sum of the dynamical octupoles of the
individual atoms.)
By using the relationship between ${\bf J}$-tensors and ${\bf Q}$-tensors discussed earlier in
this Section, it is not difficult to deduce that the third component,
$\bar{\mu}^{\rm II}_{\alpha \beta,\alpha \beta}$ with $\alpha \neq \beta$, must
be zero.
Summarizing the above, the three independent components (longitudinal, transverse and shear)
in a rigid-sphere crystal read as
\begin{equation}
\bar{\mu}^{\rm II}_{xx,xx} = \frac{O_{\rm L}}{6\Omega}, \qquad
\bar{\mu}^{\rm II}_{xx,yy} = \frac{O_{\rm L}}{6\Omega}, \qquad
\bar{\mu}^{\rm II}_{xy,xy} = 0.
\mylabel{mu-sphere}
\end{equation}
This demonstrates that the effect illustrated in Fig.~\ref{figlong} is
indeed a natural consequence of the theory developed in this Chapter.

We now turn to the second question, concerning the use of
pseudopotentials in first-principles calculations, as discussed
in Ref.~\citeonline{hong-11}.  One aspect of the pseudopotential
approximation is the replacement of the all-electron charge
density $\rho^{\rm AE}(r)$ by a pseudo charge density
$\rho^{\rm PS}(r)$ in the core region of the atom.  Since these
charge densities are essentially rigid and spherically symmetric,
the above considerations apply to them.  As a result, to compensate
for the use of the pseudopotential, one should add a ``rigid core
correction''
\begin{equation}
O_{{\rm L},\kappa}^{\rm RCC}=4\pi\int ds\;s^4\,
  \left[ \rho_\kappa^{\rm AE}(r) - \rho_\kappa^{\rm PS}(r) \right]
\mylabel{rcc}
\end{equation}
to the longitudinal dynamic octopole of each atom $\kappa$ to
recover the all-electron result.  This propagates into a change
$\Delta Q^{(3,xxx)}_{\kappa x} = 3 \Delta Q^{(3,xyy)}_{\kappa x} =
O_{{\rm L},\kappa}^{\rm RCC}$, and to a change of
$\bar{\mu}^{\rm II}_{xx,xx}$
and $\bar{\mu}^{\rm II}_{xx,yy}$
(but not $\bar{\mu}^{\rm II}_{xy,xy}$) by $\sum_\kappa
O_{{\rm L},\kappa}^{\rm RCC}/6\Omega$.\myfoot{Recall that
  we work in the framework of Eq.~(\ref{E-transverse}), i.e., we
  extract the flexoelectric tensor components from the induced electrostatic
  potential, rather than from the polarization response.}

This rigid-core correction is not small, and is not independent of
the details of pseudopotential construction. Therefore, two different
calculations of the bulk flexoelectric response cannot be directly
compared unless this correction has been applied in both cases.
Nevertheless, as long as the same pseudopotential is consistently
used in the calculation, predictions of physical, experimentally
measurable quantities should not be affected by this correction.
In particular, we shall see in Sec.~(\ref{sec:surface}) than
$O_{{\rm L},\kappa}^{\rm RCC}$ makes an equal and opposite contribution
to the \textit{surface} contribution.  Because of this cancellation,
the total (bulk and surface) flexovoltage response [see Eq.~(\ref{varphi1})] 
can be computed without the need for including this correction.

The third question, regarding the physical
nature of the resulting electric
field, requires taking a closer look at some earlier works on the theory
of \emph{absolute deformation potentials}.\cite{resta-prb91,Resta-DP}
(These can be regarded as the foundation
of the modern theory of flexoelectricity, even if they were aimed
at addressing a slightly different physical problem.) In a nutshell, if we
wish to draw a band diagram of a crystal subjected to a strain-gradient
deformation, knowledge of the macroscopic electrostatic field is not
sufficient. Indeed, the relative location of
the valence-band maximum or conduction-band minimum with
respect to the electrostatic reference is itself a function of the local strain
(via the so-called \emph{band-structure term}), which implies that each band
will ``see'' a different electric field.
This means that
one band edge
may be perfectly flat, and the corresponding carriers feel no force whatsoever,
even while the other band edge and/or the mean electrostatic potential
can be strongly tilted.\myfoot{The tilt of the
  mean electrostatic potential will also depend on choice of
  pseudopotentials when these are employed, but the tilt of the
  valence and conduction band edges will not.}
In fact, even a metal subjected to a strain gradient will generally
have a nonzero internal macroscopic electric field arising from a
gradient in the mean electrostatic potential, although no current will flow.
Thus, one should be careful not to interpret the macroscopic electric field
produced by the flexoelectric effect in a longitudinal acoustic wave as
a ``real'' physical field; it is just the tilt of some arbitrary reference
energy that may have little to do with the phenomenon of interest in a given
specific case.
Just as for the notion of a ``flexoelectric field,''
care must be used when speaking of ``short-circuit'' and ``open-circuit''
electrical boundary conditions, as these are ambiguous in
the nonperiodic strain-gradient world.

\begin{figure}[!t]
\begin{center}
\includegraphics[width=2in]{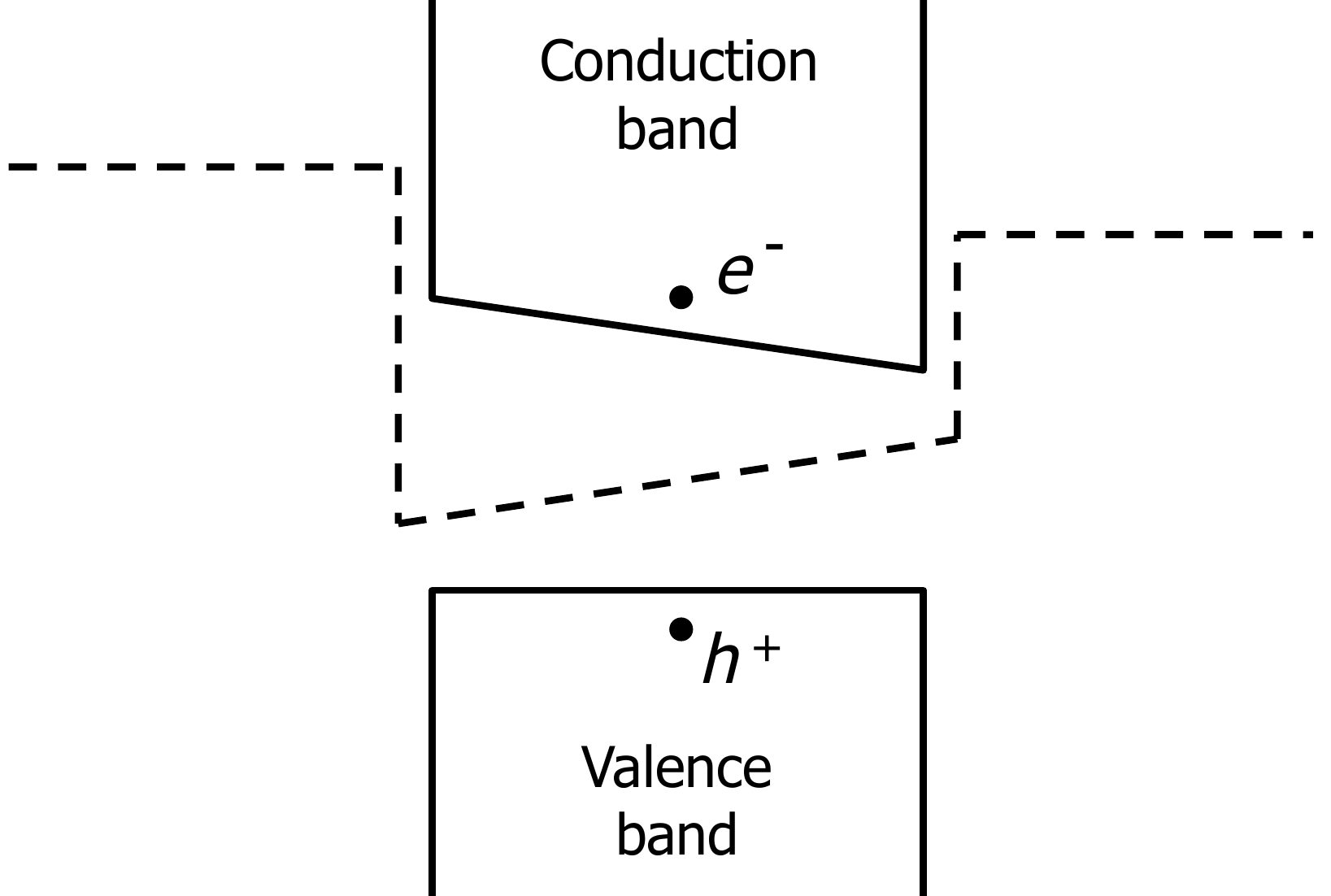}
\end{center}
\caption{Sketch of valence bands (VB) and conduction bands (CB) for a slab
with a strain gradient across its width.  Dashed line indicates
the macroscopic electron potential energy $V(x)=-e\phi(x)$.  Hole carriers
feel no force because the VB maximum is flat, while electron carriers
feel a force to the right, both of which are contrary to naive
expectations based on the electric field pointing to the right in the
interior of the slab.
\label{fig:bands} }
\end{figure}

In light of the above arguments, it is legitimate to wonder
whether the bulk flexoelectric effect is experimentally measurable at all.
In fact, there are good reasons to believe that the tilt of the mean electrostatic
potential does \emph{not} provide a realistic description of the response -- at least
no more realistic than other reference energies (e.g., the conduction band bottom, or
the valence band top, or the Fermi level).
First, as we have argued above, the present theory yields a
finite open-circuit ``flexoelectric field'' even in a metal, which is
physically inconsistent.
%
Second, if we go back to the example of the noninteracting spherical atoms,
there are apparent inconsistencies as well:
Since we have assumed that each potential well is independent of its environment,
its motion cannot, in principle, be detected by an electrode that is placed at the
far-away surface of the sample -- and yet, the bulk flexoelectric coefficients do not vanish.
We have, therefore, a sort of paradoxical situation, where the
presence of a macroscopic electric field inside the material is indisputable,
but at the same time there cannot be any open-circuit voltage, because of the
hypothesis of rigid potential wells (which excludes long-range effects).
To resolve these paradoxes, and place the present theory in
the right context regarding experimental measurements, it is
necessary to account for surface effects.
We shall see how to do this in the following Sec.~\ref{sec:surface}.

\subsection{Surface effects}
\label{sec:surface}

Knowing whether a given physical property is sensitive to the details of the
sample surfaces is a matter of central importance in condensed matter
theory.
In the majority of cases (e.g., piezoelectricity), surfaces typically start to matter
only at small length scales, where they are responsible for deviations in the measured
property from the corresponding bulk value. There are situations, however, where such
a sensitivity to the crystal termination persists up to the macroscopic scale;
flexoelectricity belongs to this category.
In the present Section we shall elaborate on this statement from a heuristic
point of view, which is anyway sufficient to illustrate the most relevant
physical ideas. A more formal discussion, based on a microscopic theory of
the response to deformations, will be presented in Sec.~\ref{sec:microscopic}
and Sec.~\ref{sec:surf_micro}.

In order to calculate the flexoelectric response of a finite object
such as a slab it is appropriate to consider, rather than the
induced macroscopic polarization, the open-circuit voltage
$\Delta V$ produced by the deformation.\myfoot{We indicate here
  by $\Delta V$ the total potential step that builds up, as a
  consequence of the mechanical deformation, between the two vacuum
  regions located at either side of the slab.}
We shall only focus, in the following,
on contributions that tend to a finite constant in the limit of
infinite thickness $t$, and introduce the \emph{flexovoltage}
coefficient,
\begin{equation}
\varphi_{x \lambda, \beta \gamma} = \lim_{t \rightarrow \infty} \frac{1}{t}
\frac{\partial \Delta V}{\partial \varepsilon_{\beta \gamma, \lambda}}.
\mylabel{varphi1}
\end{equation}
Recall that $\varepsilon_{\beta \gamma, \lambda} =
\partial \varepsilon_{\beta \gamma} / \partial r_\lambda$
is the gradient of the symmetric strain tensor
along the Cartesian direction $r_\lambda$, and $x$ indicates
the direction normal to the surface.\myfoot{In spite of its
  notation, $\varphi_{x \lambda, \beta \gamma}$ should not be thought as
  a tensor. First, the surface contribution depends on the specific details of
  the crystal termination, and is therefore not a simple function of the 
  surface plane orientation. Second, the bulk contribution is defined 
  in fixed-${\bf D}$ boundary conditions and therefore it has 
  a nonanalytic behavior [see Eq.~(\ref{fixedd})] in all materials 
  except those characterized by cubic crystal symmetry.}
For simplicity, here we shall also restrict our analysis to strain
gradients of the type $\varepsilon_{\alpha \alpha,x}$, i.e., a diagonal
(either longitudinal or transverse)
component of the symmetric strain
tensor that is linearly growing across the slab thickness.
(These are sufficient to describe the bending of a free-standing slab; a
more general analysis, including the shear component, is deferred to
Sec.~\ref{sec:surf_micro}.)
We shall write the flexovoltage coefficient as a sum of bulk
and surface-specific contributions,
\begin{equation}
\varphi_{x x, \alpha \alpha} = \varphi^{\rm bulk}_{x x, \alpha \alpha} + \varphi^{\rm surf}_{x x, \alpha \alpha},
\mylabel{varphi}
\end{equation}
whose explicit forms will be derived in the following paragraphs.

\subsubsection{Electronic surface response}
\label{sec:surf-electronic}
%

First let us consider only the purely electronic (frozen-ion) response.
Strain gradients of the type $\varepsilon_{\alpha \alpha,x}$
are governed by Eq.~(\ref{E-transverse}); in our present notation
this implies that the
(open-circuit) uniform electric field that builds up in
the interior of the slab as a consequence of the deformation is uniquely given
in terms of the bulk flexoelectric coefficient of the material and its macroscopic
dielectric constant by
\begin{equation}
\frac{ \partial {E}^{\rm slab}_{x}}{\partial \varepsilon_{\alpha \alpha,x} } \Big|_{\rm frozen-ion} =
- \frac{ \bar{\mu}^{\rm II}_{xx,\alpha \alpha} } {\epsilon_0 \bar{\epsilon}_{\rm r} }.
\end{equation}
Here $\epsilon_0$ and $\epsilon_{\rm r}$ are the vacuum and relative
permittivities, respectively, while $\mu^{\rm II}$ is the type-II flexoelectric
tensor; as before, we use the bar symbol to distinguish frozen-ion
quantities from fully relaxed ones.
Since the electric field is minus the derivative of the potential, the
bulk internal field contribution to the overall open-circuit voltage
is then proportional to $t$, leading to a finite contribution to the overall
flexovoltage coefficient that we identify with $\bar{\varphi}^{\rm bulk}$,
\begin{equation}
\bar{\varphi}^{\rm bulk}_{x x, \alpha \alpha}
= -\frac{ \partial {E}^{\rm slab}_{x}}
        {\partial \varepsilon_{\alpha \alpha,x} }
  \Big|_{\rm frozen-ion}
= \frac{ \bar{\mu}^{\rm II}_{xx,\alpha \alpha} }
       {\epsilon_0 \bar{\epsilon}_{\rm r} }.
\mylabel{mubii}
\end{equation}

\begin{figure}[!t]
\begin{center}
\includegraphics[width=4in]{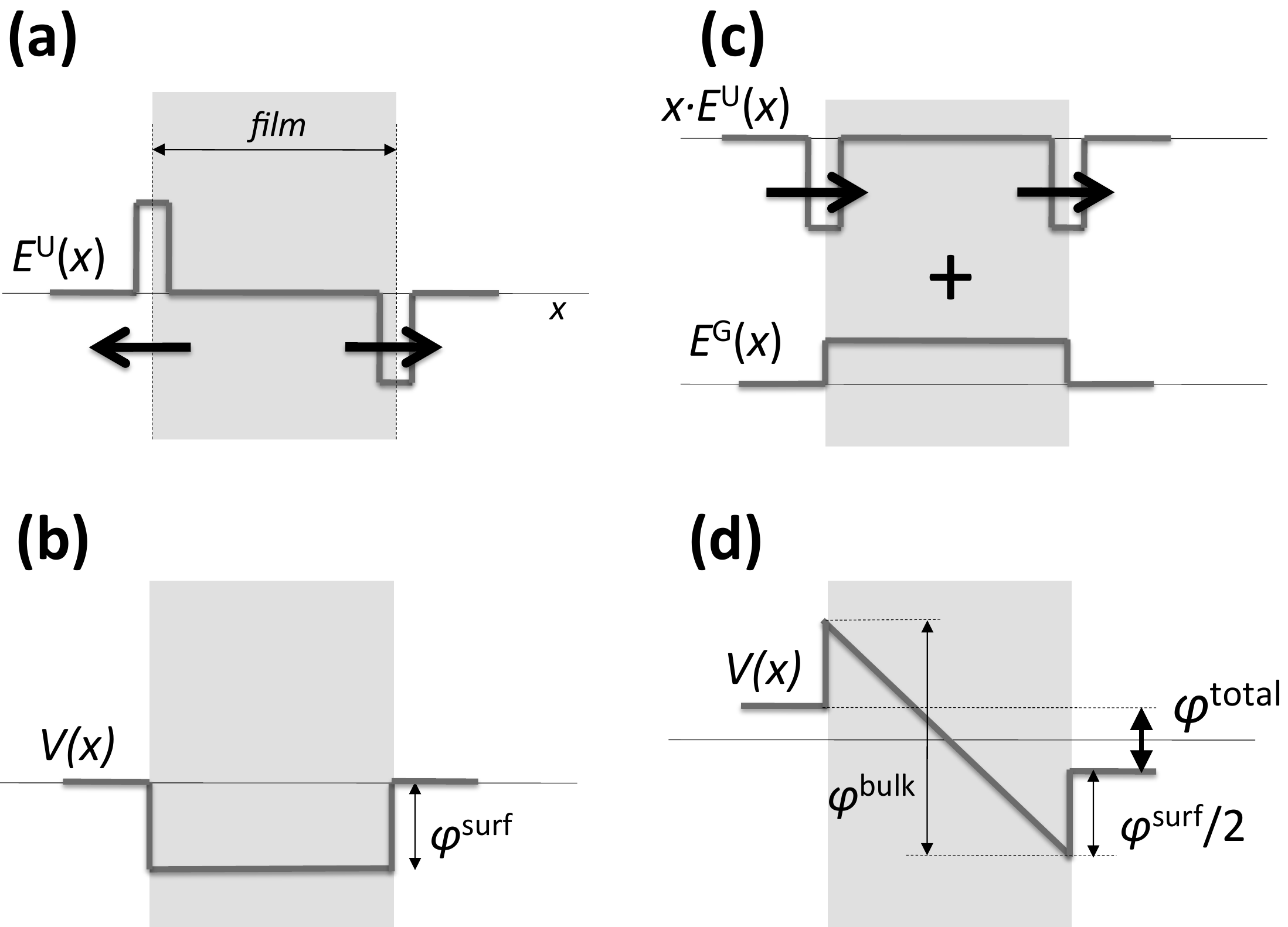}
\end{center}
\caption{ (a-b): Linear response of the electric field (a) and
the electric potential (b)
to a \emph{uniform} strain applied to
a film of thickness $t$.  The function $E^{\rm U}$ (where U indicates
``uniform'') can be regarded as representing
a surface piezoelectric response (surface dipole layer appearing
in response to strain).
(c-d): Linear response of the field (c) and the potential (d) in
the case of a strain \emph{gradient}.  The sketch shows the derivative
with respect to $(\varepsilon_{\alpha \alpha,x}t)$ for a strain
variation $\varepsilon_{\alpha \alpha}(x)=x\varepsilon_{\alpha \alpha,x}$
in a slab extending from $-t/2<x<t/2$.
The field response contains two contributions. The first is given
by $E^{\rm U}$, appropriately scaled by the linearly varying local
strain. Note that the
induced surface dipoles, schematically illustrated by arrows, now
point in the \emph{same} direction. The other is given by genuine
strain-gradient effects contained in $E^{\rm G}$, which essentially
reflects the bulk flexovoltage response. The resulting potential
response in (d) thus consists in a macroscopic internal field plus a
surface dipole contribution.}
\label{slabpotential}
\end{figure}

The surface contribution $\varphi^{\rm surf}$ in Eq.~(\ref{varphi})
originates from the fact that a surface can always be characterized
by a potential offset $\phi$ between the macroscopic potential
just inside and just outside the surface, and that this offset
is different for the two surfaces in the presence of a strain
gradient.
Consider first the case of a \emph{uniform} strain
$\varepsilon_{\alpha \alpha}$ applied to a slab of thickness $t$,
as shown in Fig.~\ref{slabpotential}(a-b).
The figure shows the derivative of the macroscopic electric
field (panel a) and electron potential energy (panel b) with respect
to the applied uniform strain $\varepsilon_{\alpha \alpha}$,
and $\varphi^\textrm{surf}$ is the corresponding derivative of
the potential offset $\phi$.
The variation of $\phi$ with strain can be regarded as resulting
from the fact that the surface,  by virtue of its lack of inversion
symmetry, is locally piezoelectric.\myfoot{
  In another language, we are basically describing a strain dependence
  of the surface work function, although technically the latter
  is referenced to the valence band maximum rather than the
  average potential in the subsurface region.}
For the slab as a whole, however, a uniform strain does not
produce a net voltage, since the induced potential offsets on
either side of the slab cancel each other, consistent with
the fact that the overall slab is nonpiezoelectric.

In the case of a \emph{strain-gradient} deformation, on the other hand,
the local strains at the opposite surfaces are opposite in sign, and do
not cancel out, as illustrated in Fig.~\ref{slabpotential}(c-d).
The slab is taken to extend over $-t/2<x<t/2$ with local strain
$\varepsilon_{\alpha \alpha}(x)=x\varepsilon_{\alpha \alpha,x}$,
reaching values of
$\varepsilon_{\alpha \alpha}=\pm(t/2) \varepsilon_{\alpha \alpha,x}$
at the two surfaces.  The figure shows the derivative of the
field (panel c) and potential (panel d) with respect
to $(\varepsilon_{\alpha \alpha,x}t)$.
This means that the induced potential offsets at the two opposite surfaces
have the same sign and add up in a flexoelectric experiment, leading
to a surface contribution of the form
\begin{equation}
\bar{\varphi}^{\rm surf}_{x x, \alpha \alpha} = \frac{\partial \phi}{ \partial \varepsilon_{\alpha \alpha}} \Big|_{\rm frozen-ion}.
\end{equation}
The total flexovoltage coefficient then reads as
\begin{equation}
\bar{\varphi}_{x x, \alpha \alpha} =
  \frac{ \bar{\mu}^{\rm II}_{xx,\alpha \alpha} } {\epsilon_0 \bar{\epsilon}_{\rm r} } + \bar{\varphi}^{\rm surf}_{x x, \alpha \alpha}.
\mylabel{flexotot}
\end{equation}

The above derivation allows us to solve the paradoxes that we
mentioned at the end of the previous Section.
First, recall that we encountered some difficulties in giving a physical
interpretation to the ``internal electric field'' that is induced by
a strain gradient, as such a field depends on the reference energy
(i.e., Bloch electrons in different eigenstates do not experience the
same electrical force).
This issue is easily solved by observing that the surface potential
offset $\phi$ suffers from the same ambiguity as the bulk
flexoelectric field;
we defined it relative to the macroscopically
averaged electrostatic potential under the surface, but we could have
used the valence or conduction band edge instead.
It is easy to see that the respective ambiguities contained in the surface and
bulk terms exactly cancel, yielding an overall flexovoltage coefficient that
is uniquely defined.
Next, we have observed that there is an apparent physical inconsistency
in the rigid-spherical-atom model, in that there should be no overall
voltage response to a strain gradient, and yet the bulk flexoelectric
coefficient does not vanish.
It is easy to see that, once the surface contribution is taken into account,
the total flexovoltage response of a slab made of noninteracting spherical
atom is zero as it should be. Indeed, when such a slab is subjected to
a uniform strain, its surface potential voltage response is
\begin{equation}
\bar{\varphi}^{\rm surf}_{xx,xx} = \bar{\varphi}^{\rm surf}_{xx,yy}
= -\frac{O_{\rm L}}{6 \epsilon_0 \Omega} \,,
\end{equation}
%
%
since a positive
longitudinal or transverse strain increases the spacing between the atomic
spheres and thereby reduces the surface potential offset.
But, using Eqs.~(\ref{bulkfxe}) and (\ref{mubii}) (and the fact
that $\bar{\epsilon}_{\rm r}=1$ for this model),
this is exactly
$-\bar{\varphi}^{\rm bulk}_{xx,\alpha \alpha}$, leading to the
claimed cancellation in Eq.~(\ref{flexotot}).
This cancellation also explains why the replacement of the
all-electron by the pseudo core charge density in the context
of pseudopotential calculations has no effect on the total
flexovoltage response, so that the rigid-core correction of
Eq.~(\ref{rcc}) can be neglected, as was claimed
in Sec.~\ref{sec:spherical}.

Note that the spherical atom model, in spite of its simplicity, is crucial
to understand how flexoelectricity works in real materials.
As we shall see in the results section, there generally tends to
be a large cancellation between surface and bulk contributions
to the flexoelectric effect.
This happens because, even in covalently-bonded materials, the electronic
charge distribution that is dragged along by each atom during its motion is largely
constituted by a spherical shell, with comparatively smaller aspherical components.
Spherical objects do not contribute to the overall flexovoltage
coefficient of a slab, hence the aforementioned cancellation.

This gives a measure of the importance of the surface contribution -- only when it
is correctly taken into account together with the bulk term we obtain a meaningful
physical quantity.
Therefore, asking whether the surface contribution is ``large or small''
compared to the bulk effect is a poorly formulated question;
the two must always go hand in hand.
Instead, a more physically meaningful question is ``How strong is the
dependence of the surface contribution on its atomic and electronic
structure?''

Based on these considerations, one can attempt to give an answer to a
long-standing question that has been somewhat controversial in recent
years: ``Is flexoelectricity a bulk property?''
As we said above, if by ``flexoelectricity'' we refer to the result of
a typical flexoelectric experiment (i.e., where the induced current upon bending
a short-circuited slab is measured), the answer is \emph{no}.
Conversely, if by the same name we call the current flowing through the bulk
of the material while well-defined internal electrical boundary conditions
are imposed, then the answer is \emph{yes}.
The problem is that the internal electrical boundary conditions depend
on the externally-applied ones in a way that is
surface-dependent, and unlike in the case of most known material properties,
such a dependence persists in the limit of a macroscopically thick sample.
All in all, in the present context we would rather stay away from the
traditional rigid classification into bulk properties and surface properties,
as flexoelectricity, strictly speaking, does not belong cleanly to either category.

\subsubsection{Lattice surface response}
\label{sec:surf-lattice}

We now discuss how the above conclusions need to be modified when full
ionic relaxations are incorporated -- these are, of course, of the utmost
importance for a realistic description of the flexoelectric effect.
Essentially, the above conclusion still hold, except for two
important details: (i) the frozen-ion quantities (flexoelectric coefficient,
dielectric constant, surface potential response) need to be replaced with
their relaxed-ion counterparts; (ii) an \emph{effective} deformation, given
by an appropriate linear combination of, e.g., a longitudinal and transverse
strain gradient, need to be considered in place of the individual tensor components.

To illustrate the implications of (i) and (ii) in a practical situation, it is
useful to work out the explicit formulas for the simplest case of an unsupported
slab subjected to bending.\myfoot{We shall exclusively focus,
  for the time being, on the \emph{plate}-bending regime, where any
  deformation (e.g., anticlastic bending) along the main bending axis
  is forbidden. More general situations will be considered in the later
  Sections.}
Linear elasticity dictates that a transverse strain gradient (corresponding to
a ``frozen-ion'' bending deformation) at static equilibrium must be
accompanied by a longitudinal strain gradient, which for most
materials will have opposite sign compared to the transverse one.
In fact, the top layers of the slab (``top'' here means furthest from the
bending center) are under tensile strain, and this typically induces
a longitudinal contraction of such layers, whose magnitude is related to the Poisson's
ratio of the material. Conversely, the bottom layers
are transversely compressed, and will therefore expand longitudinally by an
equal amount.
This means that, to calculate the static flexovoltage coefficient of a bent slab,
we need to consider the ``effective'' deformation
\begin{equation}
\varepsilon_{yy,x} = \varepsilon_{{\rm eff},x}; \quad \varepsilon_{xx,x} = -\nu \varepsilon_{{\rm eff},x},
\end{equation}
rather than the individual strain-gradient tensor components,
where
\begin{equation}
\nu = \frac{ \mathcal{C}_{xx,yy} } { \mathcal{C}_{xx,xx} }
\end{equation}
is uniquely given by the elastic constants of the bulk material.
Consequently, when the ions are relaxed, we shall be concerned with an effective
flexovoltage coefficient reflecting the aforementioned mechanical equilibrium
condition,
\begin{equation}
\varphi_{x x, {\rm eff}} =
  \frac{ \mu^{\rm II}_{xx,{\rm eff}} } {\epsilon_0 \epsilon_{\rm r} }
+ \frac{\partial \phi}{ \partial \varepsilon_{\rm eff}},
\mylabel{rel-ion}
\end{equation}
where
\begin{equation}
\varepsilon_{yy} = \varepsilon_{{\rm eff}};
\quad \varepsilon_{xx} = -\nu \varepsilon_{{\rm eff}}
\end{equation}
refers to an analogous linear combination of the \emph{uniform} strain
components.

The fact that, even at the level of the surface contribution, we have
an \emph{effective} response to a combined transverse and longitudinal
strain is fully consistent with the behavior of an unsupported slab
subjected to uniform in-plane tension.
In such a situation, the relaxation will affect not only the surface
atoms, but will also extend to the entire slab, leading to a contraction
in the third dimension proportional to the bulk coefficient $\nu$.
Thus, for a free film in a relaxed-ion context, it is only meaningful
to consider the response of the surface potential offset $\phi$
to $\varepsilon_{{\rm eff}}$, and not to the individual
$\varepsilon_{yy}$ or $\varepsilon_{xx}$ components;
the former is precisely the quantity that enters the total
flexovoltage coefficient in Eq.~(\ref{rel-ion}).

Of course, one generally needs to consider more realistic
mechanical boundary conditions than that of a free-standing
film. In such cases, some of the specifics of the above example
are no longer valid (e.g., the absence of surface loads).
Still, the points (i) and (ii) are applicable to the most general case.

\subsection{Electronic and lattice response revisited: Curvilinear coordinates}
\label{sec:microscopic}

In the early Sections of this Chapter we have described a fundamental theory
of the bulk flexoelectric effect, based on a first-principles quantum-mechanical
description of the insulating crystal.
Later, in Sec.~\ref{sec:surface} we have argued, based on heuristic
arguments, that there are important surface contributions to the flexoelectric
response of a finite sample, and that these need to be accounted for when discussing
experimental results.
Here, we shall put the derivations of Sec.~\ref{sec:surface} on
firmer theoretical grounds by developing an alternative approach.
In particular, we shall clarify how to
describe the microscopic charge and current responses to an arbitrary
inhomogeneous strain field in terms of cell-periodic response functions.
Such a formalism is necessary in order to treat, in full generality, the
response of a finite (and hence, spatially inhomogeneous) body to a
deformation.
As we shall see later, this will be useful not only for the
formal derivation of the surface contributions to the flexoelectric effect
in finite samples, but also for the practical calculation of the
transverse bulk components of the flexoelectric tensor. (Recall that
such components are presently difficult to access at the bulk level.)
Given its rather technical character, and the fact that the most relevant
physical results have already been presented in Section~\ref{sec:surface}, this
Section and the following can be skipped on a first reading.

\begin{figure}[!tb]
\begin{center}
\includegraphics[width=4in]{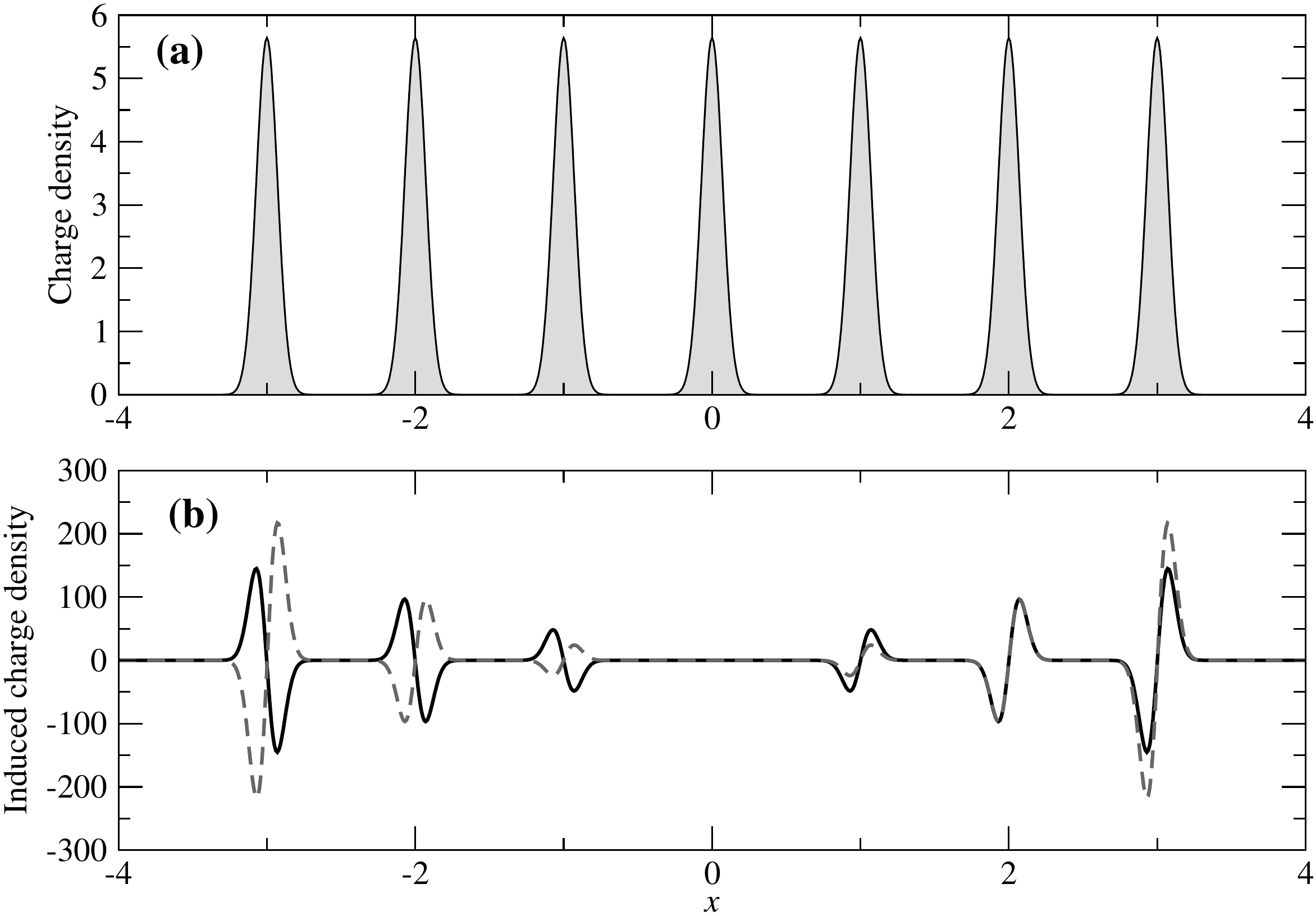}
\end{center}
\caption{(a) Unperturbed charge density of model 1D crystal composed
of Gaussian charge packets. (b) Change of charge density in linear
response to a uniform strain (solid line) or strain gradient
(dashed line). \label{figlatt1d} }
\end{figure}

\subsubsection{A simple one-dimensional example}

In order to establish a microscopic theory of deformations, the first
issue one needs to address concerns the proper representation of the
scalar and vector fields that describe the physical property of
interest (e.g., atomic positions, electronic charge density, etc.).
To appreciate the nature of the problem, it is useful to analyze the
charge-density response of a simple lattice to a macroscopic deformation.
Consider a one-dimensional chain of equally spaced atoms, which we
represent as a regular array of Gaussian charge distributions
as in Figure~\ref{figlatt1d}(a).  Its unperturbed charge density is
\begin{equation}
\rho(x) = \sum_n \rho_0 (x - R_n), \qquad \rho_0(x) =
   \frac{1}{\sigma \sqrt{\pi}} e^{-x^2/\sigma^2},
\end{equation}
where $R_n = na$ is an integer multiple of the lattice parameter $a$.
Now we apply a uniform expansion to the chain by displacing each atom by
\begin{equation}
u_n = \varepsilon R_n,
\end{equation}
and we look at how the charge density responds to such a perturbation .
In the linear limit (small lambda) we obtain the response function
$\partial \rho(x) / \partial \varepsilon$ that is plotted as the black curve in
Fig.~\ref{figlatt1d}(b).
The form of $\partial \rho(x) / \partial \varepsilon$ is manifestly
problematic: such a function grows linearly when moving away from the
origin, i.e., it is clearly nonperiodic,
which contrasts with the fact the system remains periodic after the
application of the perturbation.
Moreover, it introduces an undesirable dependence of the
result on the arbitrary location of the coordinate origin.
Such issues become even more severe when considering a strain-gradient
perturbation of the type
\begin{equation}
u_n = \frac{\eta}{2} R_n^2.
\end{equation}
The charge density response, plotted as the red curve in Fig.~\ref{figlatt1d}(b),
now grows \emph{quadratically} with the value of the unperturbed atomic
position, and extracting any relevant physical information from such a function
appears difficult.

The solution of the above problems comes from the realization that the
fixed laboratory frame is a poor choice of coordinate system if we wish
to represent the response to a macroscopic elastic deformation. In
such a frame, the boundary atoms in a large crystallite have to move
very far from their original location even if the applied strain is
small; if we naively take the difference in the charge density from the
original to the current state we obtain a result that has little physical
meaning, and most likely will strongly deviate from the linear regime that
we have in mind.
A viable alternative is to treat an elastic deformation as a deformation
of \emph{space}, rather than an atomic displacement pattern.
This implies applying a \emph{coordinate transformation} that exactly
reproduces the macroscopic elastic deformation.\myfoot{Recall that
  a deformation of a continuum is a 3D-3D mapping of each material point
  to its perturbed location, i.e., it has the exact same mathematical
  form as a coordinate transformation.}
From this viewpoint,
the atoms do not explicitly move from their original location,
although they do move with respect to the laboratory frame because the
coordinate system itself is changing.

\begin{figure}[!tb]
\begin{center}
\includegraphics[width=4in]{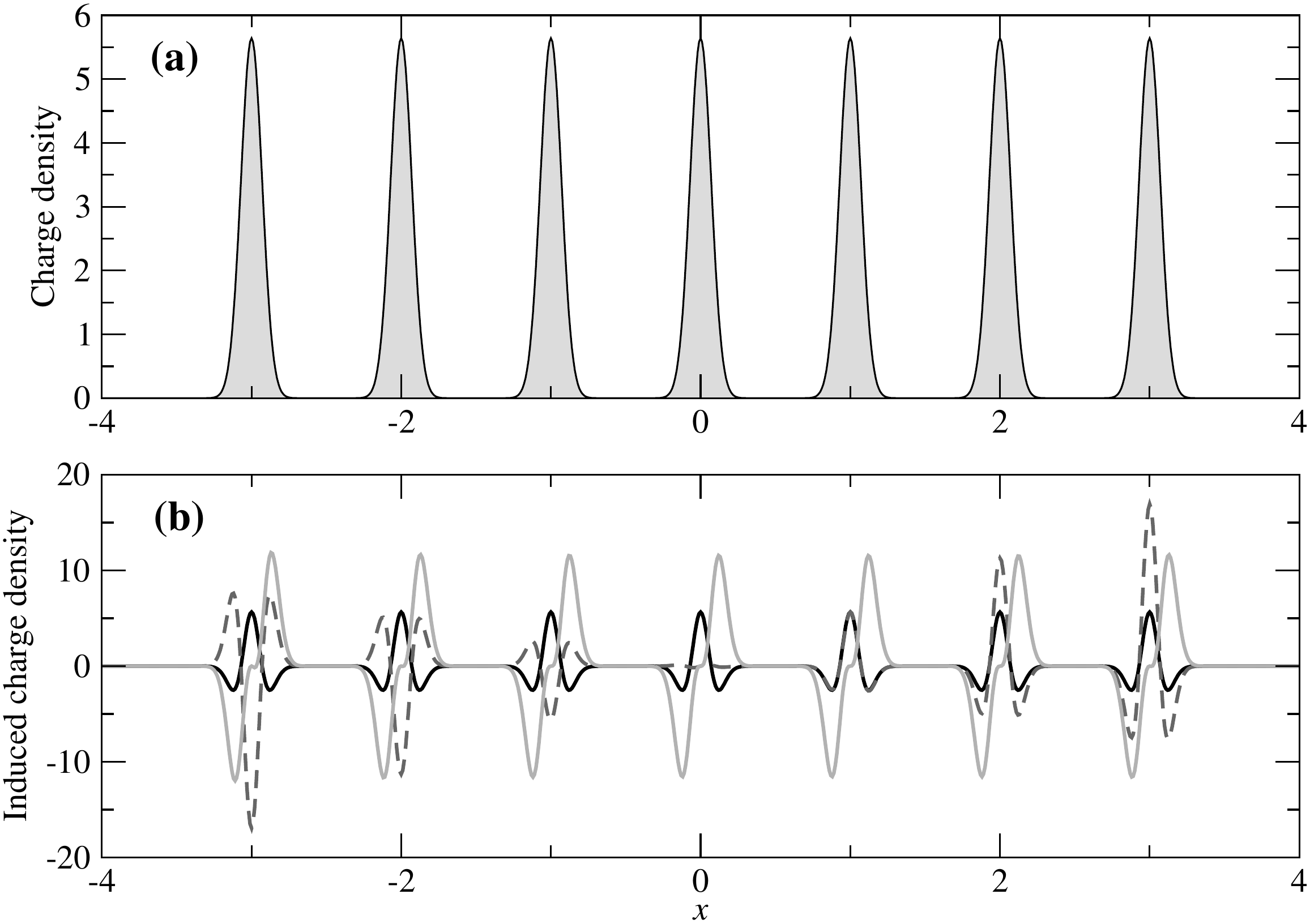}
\end{center}
\caption{(a) Same as in Fig.~\ref{figlatt1d}(a). (b) Change in charge
density, when expressed in transformed coordinates, for a uniform strain
(solid black line) or a strain gradient (dashed dark gray line).  The latter,
while not periodic, can be expressed as a sum of black and light gray
contributions (the latter was magnified by a factor of 50),
as explained in the text.}
\label{figlatt1d_met}
\end{figure}

To be explicit, consider a distortion ${\bf r}'={\bf r}+{\bf u}({\bf r})$
that maps point $\bf r$ in the original periodic crystal into point
${\bf r}'$ of the distorted crystal, and such that a nucleus at
${\bf R}_{l \kappa}$ would be carried to
${\bf R}'_{l \kappa} ={\bf R}_{l \kappa}+{\bf u}({\bf R}_{l \kappa})$
if one neglects the additional internal displacements arising
from the lattice effects described in Sec.~\ref{sec:latt-resp}.
If the initial charge
density $\rho_0({\bf r})$ were also carried along by this distortion,
the new charge density would be
\begin{equation}
\rho_\textrm{ref}( {\bf r}' ) = \rho_0( {\bf r}) \, {\rm det}^{-1}({\bf h})
\mylabel{trho-a}
\end{equation}
where the Jacobian factor involving
$h_{\alpha\beta}=\partial r'_\alpha/\partial r_\beta
= \delta_{\alpha \beta} + \partial u_\alpha/\partial r_\beta$
is needed to reflect the dilution or concentration of charge density.
In fact, the actual charge density $\rho({\bf r}')$ has to be computed
from the appropriate physical laws (e.g., first-principles DFT
calculations), so it will not be equal to $\rho_\textrm{ref}( {\bf r}' )$.
However, we may hope that the difference
$\rho({\bf r}')-\rho_\textrm{ref}( {\bf r}' )$ is small, and we
want to express this difference in terms of the \textit{original} spatial
variable $\bf r$.  This is conveniently done by defining
\begin{equation}
\hat{\rho}( {\bf r})
=
\rho( {\bf r}' ) \,\det({\bf h})
\mylabel{trho-b}
\end{equation}
so that our small quantity is
$\Delta \hat{\rho}({\bf r}) =\hat{\rho}({\bf r}) - \rho_0( {\bf r})$.
Note that $\hat{\rho}( {\bf r})$ describes the actual charge density
after the deformation, but transformed back to the original coordinate
system; the hat symbol is used henceforth to highlight
quantities that describe the transformed system from the
curvilinear-coordinate point of view.

In Fig.~\ref{figlatt1d_met} we again perform the same analysis as
in Fig.~\ref{figlatt1d}, illustrating how the use of coordinate
transformations effectively solves the problems that we pointed out
earlier.
Panel (a) shows the same charge density at rest.
As before, in this model we assume that the actual charge densities shift
rigidly with the nuclei.
In panel (b) we plot as the black curve
the induced density $\partial \hat{\rho}(x) / \partial \varepsilon$ for a uniform
strain. The
response is now periodic and much smaller in magnitude than before (note the scale
change). We shall denote this response function as $\rho^{\rm U}(x)$, where
`U' indicates a `uniform' strain.
The response to a strain gradient, shown as the dashed red curve, is still not periodic,
although it now
has a milder dependence on the spatial coordinate, growing only linearly rather
than quadratically with $x$.
Remarkably, however, we can write this response as a linear
combination of two \emph{cell-periodic} functions,
\begin{equation}
\frac{\partial \hat{\rho}(x)}{\partial \eta} = x \rho^{\rm U}(x) + \rho^{\rm G}(x),
\end{equation}
where $ \rho^{\rm U}(x)$ is the same as above (response to uniform strain), and
$\rho^{\rm G}(x)$ is a new quantity, reflecting the genuine strain-gradient effects
(shown as a thick light gray curve in the figure, where it has been magnified by a factor of
50 to better illustrate its functional form).
Since we are considering a uniform strain gradient above, we have
$\varepsilon(x) = x \eta$, so that we can write
\begin{equation}
\Delta \hat{\rho}(x) = \varepsilon(x) \rho^{\rm U}(x) +
\frac{d\varepsilon(x)}{d x} \rho^{\rm G}(x) + \ldots
\mylabel{drho1d}
\end{equation}
In other words, we have achieved a closed expression for the induced charge density
$\Delta \hat{\rho}(x)$ that depends only on proper measures of the local deformation,
with the only hypothesis that the local strain $\varepsilon(x)$ varies slowly on the
scale of the interatomic spacings.

Several questions naturally arise from the above discussion. First, how general is
such an analysis?
For our illustrative example above
we have used a trivially simple system, and a single
(longitudinal) strain (or strain gradient) type, so it is legitimate to wonder whether
the same procedure is applicable to a full first-principles simulation in 3D.
Second, what do we do with $ \Delta \hat{\rho}(x)$ once we have calculated it?
To make the discussion relevant for flexoelectricity it is necessary to trace a
direct link between $ \Delta \hat{\rho}(x)$ and measurable electrical quantities,
such as the macroscopic polarization in short circuit, or the induced voltage in
open circuit.
We shall address both questions in the remainder of this Section.

\subsubsection{General formalism in three dimensions}

Regarding the general applicability of the coordinate transformation
method, there are several conceivable ways to proceed. One could, for
example, directly incorporate the curvilinear-coordinates formalism
at the level of the Kohn-Sham equations (borrowing from the adaptive
coordinate scheme of Gygi~\cite{Gygi-93}) and, in a similar spirit as
in Ref.~\citeonline{Hamann-metric}, directly perform the perturbation
expansion with respect to the metric tensor and its gradients.
Alternatively -- and we shall follow this latter strategy throughout
this Chapter -- one can go back to the phonon analysis that we have
introduced in Sec.~\ref{sec:longwave}, this time focusing on the
microscopic charge-density response functions; the challenge here
lies in converting these to the curvilinear representation outlined
in this Section.
We thus consider a deformation
\begin{equation}
r'_\beta({\bf r}) = r_\beta+ U_\beta e^{i {\bf q \cdot r}}
\mylabel{sinusoid}
\end{equation}
which generates a simple frozen phonon
\begin{equation}
u^l_{\kappa \beta} = U_\beta e^{i {\bf q \cdot R}_{l\kappa} }.
\mylabel{frozenph}
\end{equation}
For the moment we neglect the internal displacements leading
to the lattice response of Sec.~\ref{sec:latt-resp},
so that Eq.~(\ref{frozenph}) is equivalent to
Eqs.~(\ref{acoustic}-\ref{expanq}) with the $\bf q$-dependent
terms neglected,
but they will be restored shortly in Sec.~\ref{sec:relax}.

In the linear limit, the charge density responds as
\begin{equation}
\rho({\bf r})= \rho_0({\bf r})
  + \rho^{\bf q}_\beta({\bf r})\,e^{i {\bf q \cdot r}}
\mylabel{rhoper}
\end{equation}
where the cell-periodic part $\rho^{\bf q}_\beta({\bf r})$
gets modulated by the same phase factor as in Eq.~(\ref{sinusoid}).
Inserting this in Eq.~(\ref{trho-b}) gives
\begin{equation}
\hat{\rho}({\bf r})=
\left( \rho_0({\bf r}')+U_\beta \rho_\beta^{\bf q}({\bf r}')
 e^{i {\bf q \cdot r}'} \right) \,
\left(1+iq_\gamma U_\gamma e^{i {\bf q \cdot r}'} \right)
\mylabel{intermed}
\end{equation}
where the last term in parentheses is the value of $\det({\bf h})$
resulting from Eq.~(\ref{sinusoid}).
We now expand the cell-periodic response function up to second order in ${\bf q}$,
\begin{equation}
\rho^{\bf q}_\beta({\bf r}) = \rho^{(0)}_\beta({\bf r}) - iq_\gamma \rho^{(1,\gamma)}_\beta({\bf r}) - \frac{q_\gamma q_\lambda}{2} \rho^{(2,\gamma \lambda)}_\beta({\bf r}).
\end{equation}
Since we are only collecting terms to first order in $\bf U$ in
Eq.~(\ref{intermed}), we can ignore the distinction between
$\bf r$ and $\bf r'$ in the cross terms,
but for the direct term we have
\begin{equation}
\rho_0({\bf r}')=\rho_0({\bf r}+{\bf U} e^{i {\bf q \cdot r}})
=\rho_0({\bf r})- U_\beta \rho_\beta^{(0)}({\bf r}) e^{i {\bf q \cdot r}}
\mylabel{details}
\end{equation}
where we have used that $\partial_\beta \rho_0({\bf r})=-\rho_\beta^{(0)}({\bf r})$.
Collecting all the terms linear in $\bf U$, we obtain
\begin{equation}
\Delta \hat{\rho}({\bf r}) =
U_\beta e^{i {\bf q \cdot r} } \left[ i q_\beta  \rho({\bf r})
 - iq_\gamma \rho^{(1,\gamma)}_\beta({\bf r})
 - \frac{q_\gamma q_\lambda}{2} \rho^{(2,\gamma \lambda)}_\beta({\bf r}) \right].
\end{equation}
The $\rho^{(0)}$ terms have now canceled, as expected from the
fact that the coordinate transformation has removed the translational
part from the response.

After observing that the unsymmetrized strain and strain gradient are related to partial derivatives of
the displacement field,
\begin{eqnarray}
\tilde{\varepsilon}_{\beta \gamma}({\bf r}) &=& i q_\gamma U_\beta e^{i {\bf q \cdot r} }, \\
\eta_{\beta, \gamma \lambda}({\bf r}) &=& - q_\gamma q_\lambda U_\beta e^{i {\bf q \cdot r} },
\end{eqnarray}
we can readily write
\begin{equation}
\Delta \hat{\rho}({\bf r}) =  \tilde{\varepsilon}_{\beta \gamma}({\bf r}) \left[ \delta_{\beta \gamma} \rho({\bf r})  - \rho^{(1,\gamma)}_\beta({\bf r})\right] + \frac{\eta_{\beta, \gamma \lambda}({\bf r})}{2} \rho^{(2,\gamma \lambda)}_\beta({\bf r}) .
\end{equation}
Finally, one can replace $\tilde{\varepsilon}_{\beta \gamma}$ with the symmetrized counterpart, $\varepsilon_{\beta \gamma}$ (the
quantity in the square brackets is invariant upon $\beta \gamma$ exchange~\cite{artlin}), and replace $\eta_{\beta, \gamma \lambda}$
with the type-II strain gradient tensor $\varepsilon_{\beta \gamma,\lambda}$. This leads to an expression that is in
all respects analogous to Eq.~(\ref{drho1d}),
\begin{equation}
\Delta \hat{\rho}({\bf r}) = \varepsilon_{\beta \gamma}({\bf r}) \rho^{\rm U}_{\beta \gamma}({\bf r}) +
\frac{\partial \varepsilon_{\beta \gamma}({\bf r})}{\partial r_\lambda} \rho^{\rm G}_{\beta \gamma,\lambda}({\bf r}),
\mylabel{delhrho}
\end{equation}
where the uniform (U) and gradient (G) terms are defined as follows,
\begin{eqnarray}
\rho^{\rm U}_{\beta \gamma}({\bf r}) &=& \delta_{\beta \gamma} \rho({\bf r})  - \rho^{(1,\gamma)}_\beta({\bf r}), \\
\rho^{\rm G}_{\beta \gamma,\lambda}({\bf r}) &=& \frac{1}{2} \left[ \rho^{(2,\gamma \lambda)}_\beta({\bf r}) +
 \rho^{(2, \lambda \beta)}_\gamma({\bf r}) - \rho^{(2,\beta \gamma )}_\lambda({\bf r}) \right].
\end{eqnarray}
This result formalizes and generalizes the arguments of the first part
of this Section: it shows that the microscopic charge density response
to an arbitrary inhomogeneous deformation can indeed be computed (and
rigorously expressed in terms of well-defined response quantities)
in a first-principles context, and for an arbitrary combination of the
relevant 3D deformation tensor components.

As a side note, one can show that the quantity $\rho^{\rm U}_{\beta
\gamma}({\bf r})$ essentially coincides (apart from a trivial scaling
factor) with the first-order charge-density as defined by Hamann, Wu,
Rabe and Vanderbilt~\cite{Hamann-metric} within their linear-response
theory of strain based on the metric tensor. It will be interesting in the
near future to draw even closer connections between the two formalisms,
which bear several similarities at the conceptual level.

\subsubsection{Microscopic polarization response}

In the above derivations we have focused on the charge-density response
of the system to an inhomogeneous deformation, but we could have worked
just as well with the microscopic polarization response instead.
This quantity is well-defined only for infinitesimal transformations,
otherwise it depends on the specific path followed by the system during its
evolution; this is not an issue here, since we are exclusively interested in the
linear-response regime.
The microscopic polarization ${\bf P}({\bf r})$ is related to the adiabatic
current-density response of the system to a time-dependent perturbation. Thus,
in order to construct an appropriate definition of this quantity in a generic
curvilinear frame, we need first to examine the transformation laws of the
current-density field ${\bf J}({\bf r})$.
To this end, let ${\bf r}' = {\bf r} + {\bf u}({\bf r},t)$ be a generic
time-dependent coordinate transformation, which we suppose to coincide,
as usual, with the displacement field associated with the mechanical
deformation of the sample.
(We suppose now that such a deformation happens slowly over a finite interval
of time.)
In a curvilinear framework the four-current, defined as $J^\mu = (\rho, j_1, j_2, j_3)$,
transforms as
\begin{equation}
\bar{J}^\mu = \frac{\partial \bar{x}^\mu} {\partial x^\alpha} \, J^\alpha \,
{\rm det}^{-1}\left[  \frac {\partial\bar{x}^\beta}{\partial x^\gamma} \right],
\end{equation}
where $x^\mu = (t, x_1, x_2, x_3)$ is the coordinate four-vector and the
barred (unbarred) symbols refer to the deformed (original) frame.
We work here in the nonrelativistic limit with $\bar{t}=t$ and $\bf r$
independent of $t$, so that
\begin{eqnarray}
\bar{\rho} &=& \rho \, {\rm det}^{-1} ({\bf h}), \\
\bar{J}_i &=& \left( \rho \frac{ \partial {u}_i }{\partial t} + h_{ij} J_j \right)
   \, {\rm det}^{-1} ({\bf h}).
\end{eqnarray}
(Latin indices refer to the three-dimensional Cartesian space.)
This leads to the definitions
\begin{eqnarray}
\hat{\rho} &=& \bar{\rho} \, \det ({\bf h}), \\
\hat{J}_i &=&   ({\bf h}^{-1})_{ij} \left[ \bar{J}_j
  - \bar{\rho} \frac{ \partial {u}_j }{\partial t} \right] \det ({\bf h}).
\mylabel{Jhat}
\end{eqnarray}
The above expression for the curvilinear-frame charge density $\hat{\rho}$
coincides with that postulated earlier in Eq.~(\ref{trho-b}), showing
that this definition is, in fact, dictated by the fundamental
transformation laws of a scalar density field.
Equation~(\ref{Jhat}), on the other hand, gives the desired expression for
the curvilinear-frame current density $\hat{J}_i$, which we will
use in the following to derive the microscopic polarization response to
an acoustic phonon perturbation.

As in the former case of the charge density response, we consider a
monochromatic acoustic phonon as in Eq.~(\ref{sinusoid}), again
without internal cell relaxations.
This time, however, we allow the amplitude of the displacement to depend on time,
\begin{equation}
{\bf r}' = {\bf r} + {\bf u}({\bf r},t), \qquad
    u_\beta ({\bf r},t) =  U_\beta (t) \, e^{i {\bf q \cdot r}}.
\end{equation}
(Henceforth we shall go back to using Greek indices for three-dimensional space
coordinates, as we did in the previous sections.)
In the adiabatic limit, one has
$${\bf J}({\bf r},t) = \dot{U}_\beta(t) \frac{ \partial {\bf P}({\bf r})}{\partial U_\beta},$$
and, after dropping all terms that are quadratic in either $U_\beta$ or $\dot{U}_\beta$ [this implies
setting $h_{ij}=\delta_{ij}$ in Eq.~(\ref{Jhat})],
$$\hat{\bf J}({\bf r},t) = \dot{U}_\beta(t) \frac{ \partial {\bf P}({\bf r})}{\partial U_\beta} - \dot{\bf U}(t) \rho_0({\bf r}).$$
Now, the microscopic polarization response can be written, as usual, as
a cell-periodic part times a phase,
${\bf P}({\bf r}) = U_\beta \, e^{i {\bf q \cdot r}} \,
 {\bf P}^{\bf q}_\beta({\bf r})$,
which immediately leads to
\begin{equation}
\hat{P}_\alpha({\bf r}) = U_\beta  e^{i {\bf q \cdot r}} \left[ {P}^{\bf q}_{\alpha,\beta}({\bf r}) -
  \delta_{\alpha \beta} \rho_0({\bf r}) \right].
\label{Phat}
\end{equation}
By following the same steps as for the case of the charge-density
response, we now proceed to expand ${P}^{\bf q}_{\alpha,\beta}({\bf r})$
in powers of ${\bf q}$,
\begin{equation}
{\bf P}^{\bf q}_\beta({\bf r}) = {\bf P}^{(0)}_\beta({\bf r}) - iq_\gamma {\bf P}^{(1,\gamma)}_\beta({\bf r}) - \frac{q_\gamma q_\lambda}{2} {\bf P}^{(2,\gamma \lambda)}_\beta({\bf r}).
\end{equation}
From translational invariance, it is then easy to show that the zeroth-order term
\begin{equation}
P^{(0)}_{\alpha,\beta}({\bf r}) = \delta_{\alpha \beta} \rho_0({\bf r}),
\end{equation}
exactly cancels with the
last term involving $\rho_0$ in Eq.~(\ref{Phat}).
Eventually, we arrive at a provisional result for the
linearly induced polarization currents in the curvilinear frame of the
deformed body in the form
\begin{equation}
\Delta \hat{P}_\alpha = \varepsilon_{\beta \gamma}({\bf r}) P^{\rm U}_{\alpha,\beta \gamma}({\bf r}) +
\frac{\partial \varepsilon_{\beta \gamma}({\bf r})}{\partial r_\lambda} P^{\rm G}_{\alpha \lambda,\beta \gamma}({\bf r}),
\mylabel{delhp}
\end{equation}
where the cell-periodic vector fields ${\bf P}^{\rm U,G}$ are
\begin{eqnarray}
P^{\rm U}_{\alpha,\beta \gamma}({\bf r}) &=&  - P^{(1,\gamma)}_{\alpha,\beta}({\bf r}), \\
P^{\rm G}_{\alpha \lambda, \beta \gamma}({\bf r}) &=& \frac{1}{2} \left[ P^{(2,\gamma \lambda)}_{\alpha,\beta}({\bf r}) +
 P^{(2, \lambda \beta)}_{\alpha,\gamma}({\bf r}) - P^{(2,\beta \gamma )}_{\alpha,\lambda}({\bf r}) \right].
\end{eqnarray}
To arrive at this equation, however, we have had to assume that the
currents generated by a global rotation are the same as those
obtained by rigidly rotating a classical charge density that is
equal to the true quantum-mechanical one.
This assumption,
which is implicit in Eq.~(\ref{E-transverse}),
was
used to conclude that $P^{(1,\beta)}_{\alpha,\gamma}({\bf r})=
P^{(1,\gamma)}_{\alpha,\beta}({\bf r})$, and hence to replace the
unsymmetrized ($\tilde{\varepsilon}$) with the symmetrized
($\varepsilon$) strain tensor (see Sec.~V.C of Ref.~\citeonline{artlin}).
While such an assumption
was indeed valid in the charge-density case, it is 
not obvious that it is
justifiable in the present case of the microscopic polarization
current.  Discussing this point in detail would take us far from
the scope of the present Chapter; nevertheless, the reader is warned
that there are still some unresolved formal issues in the theory
of the current-density response.

Regardless of such issues, one can show that the functions
$\Delta \hat{\rho}$ and $\Delta \hat{\bf P}$ enjoy the fundamental relationship
\begin{equation}
\hat{\nabla} \cdot \Delta \hat{\bf P} ({\bf r}) = - \Delta \hat{\rho}({\bf r}),
\end{equation}
where $\hat{\nabla} \cdot \hat{\bf A}$ indicates the divergence of the
vector field $\hat{\bf A}$ in the curvilinear frame.
(The hat is used to emphasize that the differentiation is with
respect to ${\bf r}$ rather than ${\bf r}'$.)
This reflects a well-known fact, which is important in the specific
context of flexoelectricity: the induced charge density can be readily
deduced from the polarization, but not the other way around. As a
matter of fact, taking the divergence annihilates the solenoidal part
of the $\Delta \hat{\bf P}$-field, which does contribute to the bulk
flexoelectric tensor.
(Recall the relationship between dynamical octupoles and second moments of
the current-density response: the additional information contained in the latter
can indeed be ascribed to divergenceless polarization currents that arise in
response to an atomic displacement.)

\subsubsection{Atomic relaxations}
\label{sec:relax}

We now return to the inclusion of the internal atomic relaxations,
and describe how they can be
conveniently incorporated in the aforementioned theory; the
practical implications regarding surface effects will be
discussed in Sec.~\ref{sec:surf_micro}.

The first important observation
is that, given a strain field $\varepsilon_{\beta \gamma}({\bf r},t)$ that
depends slowly on space and time, the internal-strain tensors that were introduced in
Section~\ref{sec:longwave} can readily be identified with the microscopic lattice response of the
crystal in the curvilinear frame of the deformed body,
\begin{equation}
u^l_{\kappa \alpha}(t) = \varepsilon_{\beta \gamma}({\bf R}_{l \kappa},t) \Gamma^\kappa_{\alpha \beta \gamma}
+ \frac{\partial \varepsilon_{\beta \gamma}({\bf R}_{l \kappa},t)}{\partial r_\lambda} L^\kappa_{\alpha \lambda, \beta \gamma} + \ldots
\end{equation}
In particular, for a macroscopic strain gradient, the above relationship reduces to
\begin{equation}
\frac{\partial u^l_{\kappa \alpha}}{\partial \varepsilon_{\beta \gamma,\lambda}} =
R_{l \kappa \lambda} \, \Gamma^\kappa_{\alpha \beta \gamma} + L^\kappa_{\alpha \lambda, \beta \gamma}.
\mylabel{atrel}
\end{equation}
Next, it is easy to show that Eq.~(\ref{delhp}) still holds,
provided that we replace the purely electronic response functions
with their relaxed-ion counterparts,
\begin{eqnarray}
P_{\alpha, \beta \gamma}^{\rm U}(x) \, & = & \, \bar{P}_{\alpha, \beta \gamma}^{\rm U}(x) +
  P_{\alpha, \kappa  \rho}^{(0)}(x) \Gamma^\kappa_{\rho \beta \gamma}, \label{p1rel} \\
P_{\alpha \lambda, \beta \gamma}^{\rm G}(x) \, & = & \, \bar{P}_{\alpha \lambda, \beta \gamma}^{\rm G}(x)
- P_{\alpha, \kappa \rho}^{(1,\lambda)}(x) \Gamma^\kappa_{\rho \beta \gamma}  +
 P_{\alpha, \kappa  \rho}^{(0)}(x) L^\kappa_{\rho \lambda, \beta \gamma}. \label{p2rel}
\end{eqnarray}
(We have used the bar symbol here to denote the purely electronic $P^{\rm U,G}$
response functions and thereby distinguish them from the fully relaxed quantities.)
This formally extends the microscopic linear-response theory discussed in this
Section to the relaxed-ion case.

\subsubsection{Electrostatics in a curved space}

Having established a convenient form for the microscopic charge and
polarization response functions, we still need to figure out how to use
them, e.g., how to calculate the voltage response $V({\bf r})$. This
requires some attention, given the fact that we are no longer working
in the Cartesian frame of the laboratory.
In particular, one needs to replace Gauss's law with its
``curvilinear'' generalization\cite{cloaks,genrel}
\begin{equation}
\hat{\nabla} \cdot (\hat{\bm{\epsilon}} \cdot \hat{\bf E}) = \hat{\rho}, 
\mylabel{maxwell}
\end{equation}
where the vacuum permittivity has been replaced with the tensor
\begin{equation}
\hat{\bm{\epsilon}} = \epsilon_0 \det {({\bf h})} {\bf g}^{-1},
\end{equation}
the hat on $\hat{\nabla}$ is again a reminder
that the gradient is in the curvilinear frame,
${\bf g} = {\bf h \cdot h}^{\rm T}$ is the metric of the deformation,
and
\begin{equation}
\hat{E}_\alpha ({\bf r}) = h_{\alpha \beta} E_\beta  ({\bf r}')
\end{equation}
is the transformed electric field.\myfoot{
  One can arrive at Eq.~(\ref{maxwell}) by observing that, in a curvilinear
  frame, Poisson's equation reads as
  $\sqrt{g}^{-1} \, \partial_\mu \left( \sqrt{g} g^{\mu \nu} \partial_\nu V \right)
      = -\rho / \epsilon_0$ where $g = \det {({\bf g})} = \det^2 {({\bf h})}$,
  i.e., the Laplacian must be replaced with the Laplace-Beltrami operator.
  Then, by defining
  $\hat{\rho} = \sqrt{g} \rho$, $\hat{\mathcal{E}}_\nu = -\partial_\nu V$,
  and $\hat{\epsilon}^{\mu \nu} = \epsilon_0 \sqrt{g} g^{\mu \nu}$,
  one immediately recovers Eq.~(\ref{maxwell}).}
As $\hat{\nabla} V({\bf r}) = - \hat{\bf E} ({\bf r})$, the
above strategy allows one to compute the induced electrostatic potential
from the induced polarization (or charge density).
\myfoot{It is interesting to note the close connection between the
  curvilinear-frame electrical quantities ($\hat{\bf E}$ and $\hat{\bf P}$)
  described here and the \emph{reduced} electrical variables (respectively
  $\bar{\varepsilon}_\alpha$ and $p_\alpha$) of Ref.~\citeonline{fixedd},
  where a linear mapping was implicitly used to connect lattice and
  Cartesian coordinates.}

In the linear limit of small deformations, one has
\begin{equation}
\epsilon_0 \hat{\nabla} \cdot ( \Delta \hat{\bf E} + \Delta {\bf E}^{\rm met}  ) =  \Delta \hat{\rho}({\bf r})
\mylabel{ehat}
\end{equation}
where $\Delta \hat{\bf E}({\bf r}) = -\hat{\nabla} [\Delta V({\bf r})]$ is
minus the (curvilinear) gradient of the induced electrostatic potential, $\Delta V$,
and $\Delta {\bf E}^{\rm met}$, coming from
the linearization of $\hat{\bm{\epsilon}}$, reads as
\begin{equation}
\Delta {E}^{\rm met}_\alpha ({\bf r}) = \varepsilon_{\beta \gamma}({\bf r})
\left[ \delta_{\beta \gamma} {E}_\alpha ({\bf r}) - \delta_{\alpha \beta} {E}_\gamma ({\bf r})
- \delta_{\gamma \alpha} {E}_\beta ({\bf r})
\right],
\mylabel{emet}
\end{equation}
where ${E}_\gamma ({\bf r})$ is the electric field in the
unperturbed system.
The choice of notation $\Delta {\bf E}^{\rm met}$ is meant
to suggest a ``metric contribution to the curvilinear-frame electric
field,'' but this is somewhat problematic as it is not an irrotational
field.  Alternatively, $\epsilon_0\Delta {\bf E}^{\rm met}$
could be regarded as a ``metric contribution'' to the polarization, since
one can rewrite Eq.~(\ref{ehat}) in terms of $\Delta {\bf P}$ as
\begin{equation}
\epsilon_0 \hat{\nabla} \cdot \Delta \hat{\bf E} = - \hat{\nabla} \cdot
  \left( \Delta {\bf P} + \epsilon_0 \Delta {\bf E}^{\rm met}  \right).
\mylabel{ehat2}
\end{equation}
However, this is not entirely satisfactory either, as
$\Delta {\bf E}^{\rm met}$ does not really originate from the
displacement of charged particles.
It is probably most appropriate to interpret
$\epsilon_0 \Delta {\bf E}^{\rm met}$
as a \emph{displacement current} arising from the effective change of
permittivity associated with the deformation of the reference frame.

In any case, Eq.~(\ref{ehat}) shows that the induced potential
$\Delta V({\bf r})$ contains, in addition to contributions from the
rearrangement of the electron cloud occurring during the deformation
(these are contained in $\Delta \hat{\rho}$), also a term that depends
on the local variation of the metric at fixed charge density. We shall
come back to this point in the discussion of surface contributions to
the flexoelectric effect in Section~\ref{sec:surf_micro}.
Note that, by construction, the microscopic electric field response
to an arbitrary deformation enjoys an analogous representation as the
charge density response,
\begin{equation}
\Delta \hat{{E}}_\alpha ({\bf r}) = \varepsilon_{\beta
\gamma}({\bf r}) {E}^{\rm U}_{\alpha,\beta \gamma}({\bf r}) +
\frac{\partial \varepsilon_{\beta \gamma}({\bf r})}{\partial r_\lambda}
{E}^{\rm G}_{\alpha \lambda,\beta \gamma}({\bf r}),
\mylabel{delhe}
\end{equation}
where both ${E}^{\rm U}$ and ${E}^{\rm G}$ are
lattice-periodic functions whose explicit expressions can be readily
derived from Eq.~(\ref{ehat}).

\subsubsection{Treatment of the macroscopic electric fields}

Eq.~(\ref{ehat}) specifies $\Delta \hat{\bf E}({\bf
r})$ modulo an ${\bf r}$-independent integration constant, $\Delta
\hat{\overline{\bf E}}$ whose value is fixed by the electrical
boundary conditions (EBC) of the problem.
The electronic response functions are typically defined (and calculated)
by assuming $\Delta \hat{ \overline{\bf E} }=0$,
i.e., short-circuit (SC) EBCs, but any other EBC choice can be recovered
if the charge-density (and/or polarization) response to a macroscopic
electric field is known.\myfoot{Note that
  $\Delta \hat{\overline{\bf{E}}}$ is first-order in the perturbation, and
  therefore its contribution to the electronic response functions is only due to the
  microscopic dielectric properties of the unperturbed system.}
For example, in the case of the polarization one can write
\begin{align}
\Delta \hat{P}_\alpha ({\bf r}) &=  \Delta \hat{P}_\alpha^{\rm SC} ({\bf r}) +
\Delta \hat{\overline{{E}}}_\beta P_\alpha^{{E}_\beta} ({\bf r}),
\label{efield}
\end{align}
where $P_\alpha^{{E}_\beta} ({\bf r}) = \partial P_\alpha({\bf r}) / \partial {E}_\beta$
is the microscopic $P$ response to an applied field along $\beta$.~\cite{puma}
The contribution of the macroscopic field can be readily incorporated
into the strain-gradient term (${\bf P}^{\rm G}$ in this case), as
the macroscopic electric field response in a nonpiezoelectric crystal
vanishes at the uniform-strain level.
Therefore Eqs.~(\ref{delhrho}), (\ref{delhp}), and (\ref{delhe})
remain valid in arbitrary EBC.

\subsection{Surface effects in curvilinear coordinates}
\label{sec:surf_micro}

Recall that, in order to quantify the flexoelectric response of a free-standing
slab, we have introduced the flexovoltage coefficient
$\varphi_{x \lambda, \beta \gamma}$.
Here we shall demonstrate how this quantity can be rigorously derived
in the context of the microscopic theory developed in the previous Section.
To express $\varphi_{x \lambda, \beta \gamma}$ in terms of well-defined
response functions of the system, we shall follow the strategy of
Sec.~\ref{sec:microscopic}, now specializing to the case of
a supercell geometry.
As before, we shall derive the microscopic response of the system (charge
density, polarization, and atomic displacements) to a strain-gradient
deformation via a long-wave analysis of its acoustic phonons.
With the help of a coordinate transformation to the curvilinear
frame of the perturbed body, one can express such microscopic
response functions in terms of ``proper'' measures of the local
deformation, i.e., in a translationally and rotationally invariant
form,
\begin{equation}
\Delta \hat{f} ({\bf r}) = \varepsilon_{\beta \gamma} ({\bf r}) f_{\beta \gamma}^{\rm U} ({\bf r}) +
\frac{ \partial \varepsilon_{\beta \gamma} ({\bf r})}{\partial r_\lambda} f_{\beta \gamma, \lambda}^{\rm G} ({\bf r}) + \ldots
\end{equation}
Here $\hat{f}$ can stand for the charge density ($\hat{\rho}$),
polarization ($\hat{\bf P}$) or electric field ($\hat{\bf E}$)
expressed in the curvilinear frame.
Note that the cell-periodic functions $f^{\rm U}$ and $f^{\rm G}$,
referring respectively to
uniform and gradient terms, are characterized by an oscillatory
behavior on the scale of an interatomic distance, due to the discreteness of
the atomic lattice. As it is customary in the space-resolved analysis of many
other physical properties (e.g., dielectric response), we shall assume in the
following (unless otherwise specified) that such oscillations have been filtered
out by means of an appropriate nanosmoothing~\cite{Baldereschi-88,Junquera-07} technique.

\subsubsection{Surface polarization and metric}

\label{sec:elec_surf}

\begin{figure}[!t]
\begin{center}
\begin{tabular}{c c c}
\raisebox{25pt}{\includegraphics[width=1.3in]{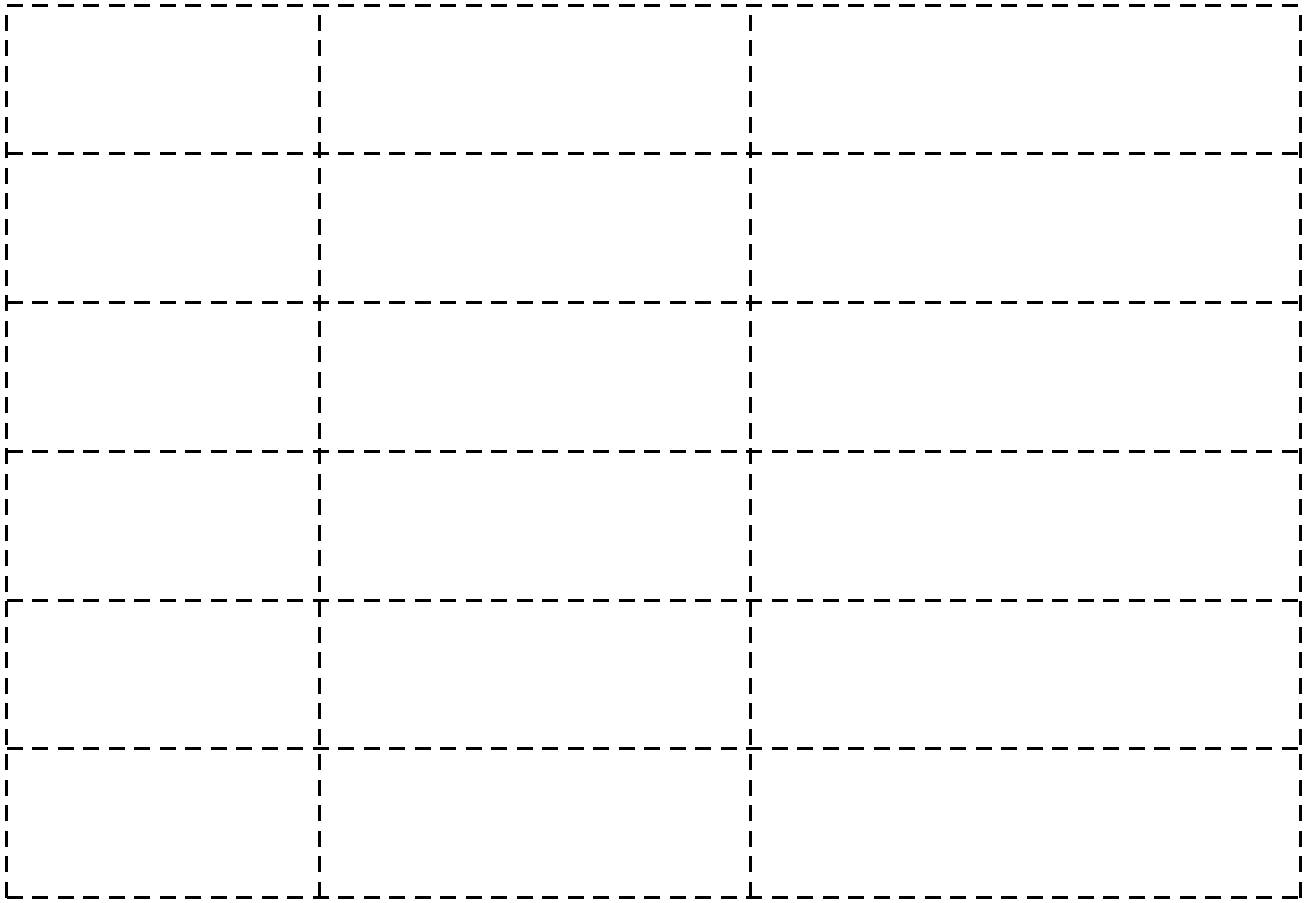}}  &
\includegraphics[width=1.1in]{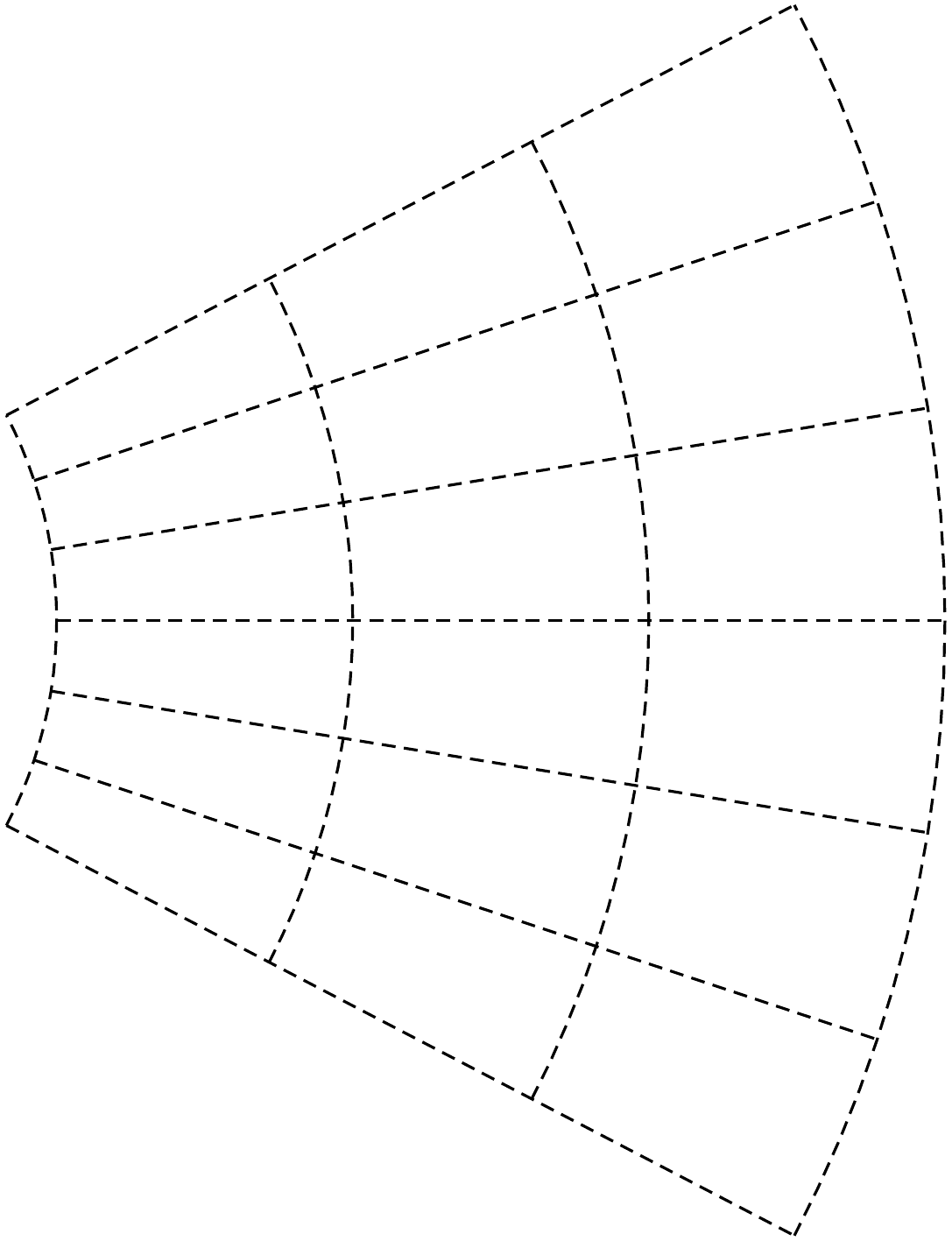} &
\raisebox{25pt}{\includegraphics[width=1.35in]{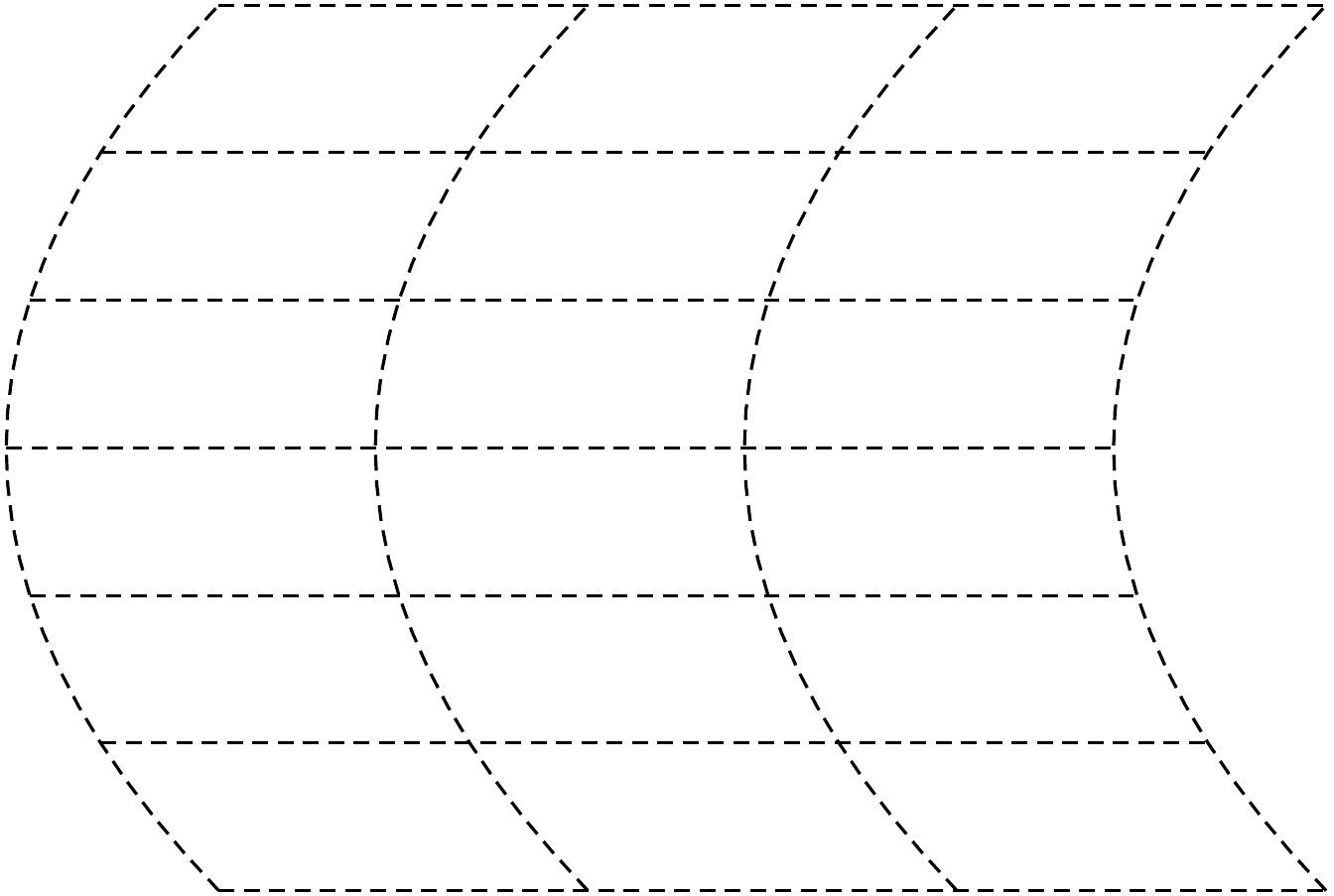}} \\
(a) & (b) & (c)
\end{tabular}
\end{center}
\caption{ \label{sketch}
Schematic representation of the three types of strain gradient described
in the text. (a) Longitudinal, $\varepsilon_{xx,x}$. (b) Transverse,
$\varepsilon_{yy,x}$. (c) Shear, $\varepsilon_{xy,y}$. The $x$ and $y$
axes correspond to the horizontal and vertical directions in the figure
respectively. (Adapted from Ref.~\citeonline{artcalc}.)
}
\end{figure}

To illustrate the above arguments in the present context, consider
a symmetrically terminated slab of a cubic material (we assume that the
surfaces are parallel to the $yz$ plane), and perturb it
with a strain-gradient deformation. We assume for the moment that
the ionic coordinates simply follow the deformation;
we shall lift this limitation in the next subsection.
In order to calculate the flexovoltage coefficient of the slab, we shall
first derive the electric field ${\bf E}({\bf r})$ induced
by the deformation under open-circuit electrical boundary conditions.
Then, by performing a line integral of ${\bf E}({\bf r})$
across the slab thickness, one can readily obtain the desired value of
$\varphi_{\alpha \lambda, \beta \gamma}$.
To calculate ${\bf E}({\bf r})$ we need, in turn, two basic
ingredients: the microscopic polarization response, $\Delta {\bf
P}({\bf r})$, and the ``metric'' contribution to the polarization,
$\Delta {\bf E}^{\rm met}$.
Regarding the former, after nanosmoothing ${\bf P}^{\rm U,G}$
are functions of $x$ only, and we can write
\begin{equation}
\frac{\partial \hat{P}_\alpha (\bf r)}{\partial \varepsilon_{\beta \gamma, \lambda} } \Big|_{\rm frozen-ion} =
r_\lambda \bar{P}^{\rm U}_{\alpha,\beta \gamma}(x) + \bar{P}^{\rm G}_{\alpha \lambda,\beta \gamma}(x).
\mylabel{pmac}
\end{equation}
The two response functions $P^{\rm U}_{\alpha,\beta \gamma}(x)$ and
$P^{\rm G}_{\alpha \lambda,\beta \gamma}(x)$ have the physical meaning of
a \emph{local} piezoelectric and flexoelectric coefficient, respectively.
Note that $P^{\rm U}_{\alpha,\beta \gamma}(x)$ differs from zero only
in the vicinity of the surface, as the bulk material is nonpiezoelectric.
The metric term, on the other hand, reads as
\begin{equation}
\frac{\partial {E}^{\rm met}_\alpha ({\bf r})}{\partial \varepsilon_{\beta \gamma, \lambda} } =  r_\lambda
\left[ \delta_{\beta \gamma} {E}_\alpha (x) - \delta_{\alpha \beta} {E}_\gamma (x)
- \delta_{\gamma \alpha} {E}_\beta (x)
\right].
\mylabel{emet2}
\end{equation}
This quantity, just like ${\bf P}^{\rm U}$, is active only at the surface: the
electric field of the undistorted slab
is nonzero (and directed perpendicular to the surface plane)
only in a small region where the crystal lattice is perturbed by the truncation of the
bonding network.

The details of the derivation differ, from now on, depending
on the specific type of flexovoltage response that one wishes to
calculate.
It can be shown that, given the symmetry of the slab, there are
only three types of strain-gradient deformation that yield a net
open-circuit voltage in a cubic material:
a variation of $\varepsilon_{xx}$ with $x$ (longitudinal),
$\varepsilon_{yy}$ with $x$ (transverse),
or $\varepsilon_{xy}$ with $y$ (shear),
which are responsible for the flexovoltages
$\varphi_{xx,xx}$, $\varphi_{xx,yy}$, and $\varphi_{xy,xy}$ respectively,
in the notation of Eq.~(\ref{varphi1}).

\begin{figure}[!t]
\begin{center}
\includegraphics[width=3in]{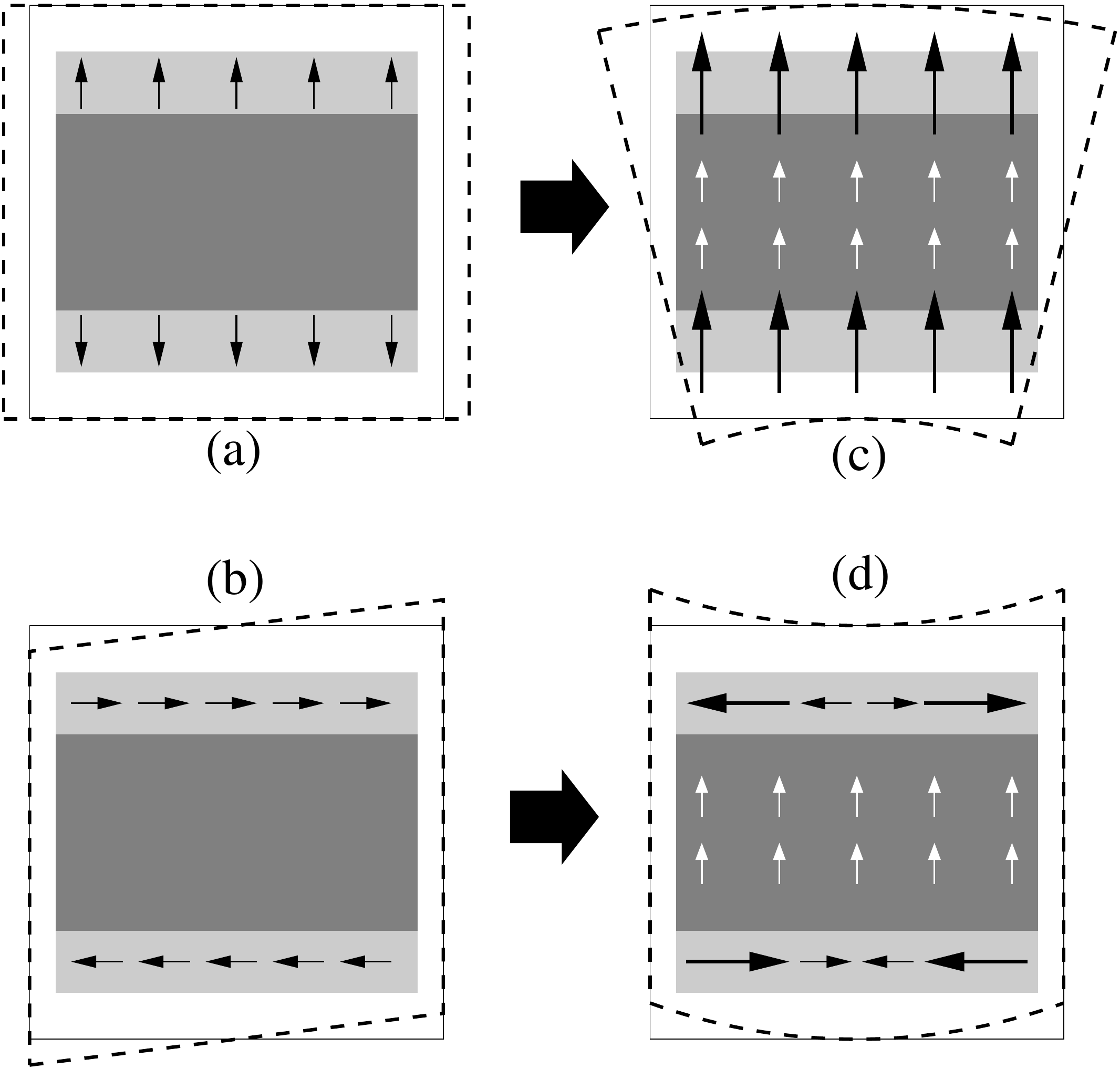}
\end{center}
\caption{ \label{figpolar}
Schematic representation of the polarization fields induced by different
macroscopic deformations of a slab of thickness $t$, drawn in the
undistorted reference frame.  The $x$ coordinate is vertical here.
Left panels (a-b) illustrate uniform strains; right panels (c-d) refer
to uniform strain gradients. Top (a,c) are transverse deformations (the
situation is qualitatively identical in the longitudinal case, not shown),
bottom (b,d) are shear patterns. The surface region and corresponding
polarization field are indicated by light gray shading and black arrows.
White arrows on a dark gray background refer to the bulk region.
The dashed black frames indicate the type of deformation
in each case. (Adapted from Ref.~\citeonline{artgr}.)
}
\end{figure}

{\bf Longitudinal and transverse cases.}
In these two cases
(which we shall indicate as $xx,\alpha \alpha$), the system remains periodic
in-plane, and the problem becomes essentially one-dimensional. We have,
in particular,
\begin{eqnarray}
\frac{\partial \hat{P}_x (x)}{\partial \varepsilon_{\alpha \alpha, x} } \Big|_{\rm FI} &=&
x \bar{P}^{\rm U}_{x,\alpha \alpha}(x) + \bar{P}^{\rm G}_{xx,\alpha \alpha}(x), \\
\frac{\partial {E}^{\rm met}_x (x)}{\partial \varepsilon_{\alpha \alpha, x} } &=&
  x {E}_x (x) (1 - 2 \delta_{\alpha x})
\end{eqnarray}
(where FI is shorthand for `frozen-ion.')
The polarization response functions $P^{\rm U}$ and $x P^{\rm U} + P^{\rm G}$
are schematically illustrated for the transverse case in
Figs.~\ref{figpolar}(a) and (c) respectively.
Given that both vector fields are irrotational and vanish at infinity,
one can safely simplify Eq.~(\ref{ehat2}) by removing the divergence sign
on both sides to get
\begin{equation}
\frac{\partial \hat{E}_x (x)}{\partial \varepsilon_{\alpha \alpha, x} } \Big|_{\rm FI} =
 - \frac{1}{\epsilon_0} \left[ x \bar{P}^{\rm U}_{x,\alpha \alpha}(x) + \bar{P}^{\rm G}_{xx,\alpha \alpha}(x) \right]
 - x {E}_x (x) (1 - 2 \delta_{\alpha x}).
\end{equation}
The frozen-ion flexovoltage coefficient of Eq.~(\ref{varphi1}) can
then be calculated by writing the open-circuit potential
associated with the above field as
\begin{equation}
\Delta V = -\int_{-\infty}^{+\infty} dx \, \left[ x E^{\rm U}_{x,\alpha \alpha}(x) + E^{\rm G}_{xx,\alpha \alpha}(x) \right],
\label{deltav}
\end{equation}
where
\begin{eqnarray}
E^{\rm U}_{x,\alpha \alpha}(x) &=& - \frac{1}{\epsilon_0} \bar{P}^{\rm U}_{x,\alpha \alpha}(x) -
                                       {E}_x (x) (1 - 2 \delta_{\alpha x}), \\
E^{\rm G}_{xx,\alpha \alpha}(x) &=& -\frac{1}{\epsilon_0} \bar{P}^{\rm G}_{xx,\alpha \alpha}(x).
\end{eqnarray}

The above functions enjoy a number of useful properties:
\begin{enumerate}[label=(\roman*)]
\item
$E^{\rm U}_{x,\alpha \alpha}(x)$ vanishes everywhere except for a small
region near the surface at $x \sim \pm t/2$;
\item
$E^{\rm U}_{x,\alpha \alpha}(x-t/2)=-E^{\rm U}_{x,\alpha \alpha}(-x+t/2)$
is antisymmetric, and \emph{independent} of $t$ for a sufficiently
thick slab;
\item
$E^{\rm G}_{xx,\alpha \alpha}(x)$ corresponds to minus the bulk
flexocoupling coefficient in the slab interior,
$$
E^{\rm G}_{xx,\alpha \alpha}(x \sim 0) = - \bar{\varphi}^{\rm bulk}_{xx,\alpha \alpha} =
                                           - \frac{ \mu^{\rm bulk}_{xx,\alpha \alpha} } { \epsilon_0 \bar{\epsilon}_{\rm r} },
$$
and only deviates from this value in a small region near the surface;
\item
$E^{\rm G}_{xx,\alpha \alpha}(x-t/2)=E^{\rm G}_{xx,\alpha \alpha}(-x+t/2)$
is symmetric, and again independent of $t$ for a sufficiently thick slab.
\end{enumerate}
Based on these observations, in the limit of large slab thickness one can
approximate Eq.~(\ref{deltav}) as
\begin{equation}
\Delta V \sim - t \int_{0}^{+\infty} dx \, E^{\rm U}_{x,\alpha \alpha}(x) +
                   t \bar{\varphi}^{\rm bulk}_{xx,\alpha \alpha}.
\end{equation}
(We assume that $x=0$ is the center of the slab
and $x=+\infty$ is deep in the vacuum region.)
As the integral in the last equation is independent of $t$, we can
readily write
\begin{equation}
\bar{\varphi}_{xx,\alpha \alpha} = - \int_{0}^{+\infty} dx \, E^{\rm U}_{x,\alpha \alpha}(x) +
                                          \bar{\varphi}^{\rm bulk}_{xx,\alpha \alpha},
\end{equation}
whence we obtain
\begin{equation}
\bar{\varphi}^{\rm surf}_{xx,\alpha \alpha} = - \int_{0}^{+\infty} dx \, E^{\rm U}_{x,\alpha \alpha}(x).
\mylabel{eqsurf-eu}
\end{equation}
The last equation states that the surface contribution to the flexovoltage
response of a slab corresponds to minus the line integral of the induced
electric field upon application of a uniform strain. The latter is, of course,
the electrostatic potential offset response to uniform strain,
$\partial \phi / \partial \varepsilon_{\alpha \alpha}$, that we already
discussed in Sec.~\ref{sec:surface}. We have, therefore,
rigorously demonstrated that the flexovoltage response of the slab indeed
contains both bulk and surface contributions, and that their nature is correctly
described by Eq.~(\ref{flexotot}).
Note that the derivations presented here, in addition to corroborating the arguments of
Sec.~\ref{sec:surface}, allow us to make one step further and split
$\bar{\varphi}^{\rm surf}_{xx,\alpha \alpha}$ into a polarization current
and a metric term,
\begin{equation}
\frac{\partial \phi}{\partial \varepsilon_{\alpha \alpha}} = \frac{1}{\epsilon_0}
  \int_0^{+\infty} dx \, P^{\rm U}_{x,\alpha \alpha}(x) + (2 \delta_{\alpha x} - 1) \phi_0.
\mylabel{xxaa-surf}
\end{equation}
In the last term on the right, $\phi_0 = -\int_0^{+\infty} dx \,
{E}_x(x)$ is the potential offset before the perturbation.

In summary, in the present case (longitudinal or transverse strain
gradient) the induced electric field in the interior of the film is
a bulk property of the material -- it is given by the flexoelectric
coefficient divided by the macroscopic dielectric constant. The surface
contribution, on the other hand, acts as an induced potential offset
that grows linearly with slab thickness, and therefore scales similarly
to the bulk contribution, as illustrated in Fig.~\ref{slabpotential}.

{\bf Shear case.}
The case of a shear deformation ($xy,xy$) is qualitatively different from the former
two cases.\myfoot{We mention this case for completeness, as it is not
  relevant for an unsupported slab after full atomic relaxation,
  provided that we consider a region that lies far (compared to the
  thickness, $t$) from the edges and/or the mechanical loading points. We
  shall come back to this issue in Sec.~\ref{sec:relax_shear}.}
Here we have
\begin{eqnarray}
\frac{\partial \hat{P}_{\alpha} ({\bf r}) }
      {\partial \varepsilon_{xy, y} } \Big|_{\rm FI} &=&
         \delta_{\alpha y} y  P^{\rm U}_{y, xy} (x) + P^{\rm G}_{\alpha y, xy}(x), \\
\frac{\partial {E}^{\rm met}_{\alpha} ({\bf r})}
      {\partial \varepsilon_{xy,y} } &=& -\delta_{\alpha y} y {E}_x(x).
\end{eqnarray}
Figures~\ref{figpolar}(b) and (d) illustrate
the polarization fields that are linearly induced by shear deformations.
By taking the divergence of the above vector fields, we can write the
curvilinear Poisson's equation, Eq.~(\ref{ehat2}), as a function
of $x$ only,\myfoot{Note that the partial derivatives along $y$ of both
  $P^{\rm U}$ and $P^{\rm G}$ vanish identically, as these nanosmoothed functions
  are periodic in plane, and therefore only depend on $x$.}
\begin{equation}
\epsilon_0 \frac{ \partial \hat{{E}}_{xy,xy} (x) } {\partial x} =
      - \bar{P}^{\rm U}_{y, xy} (x) -
       \frac{\partial \bar{P}^{\rm G}_{xy, xy}(x)}{\partial x} + \epsilon_0 {E}_x(x).
\end{equation}
[We have used the short-hand notation
$\hat{E}_{xy,xy} (x) = \partial \hat{E}_{x} (x) / \partial \varepsilon_{xy,y}$.]
Assuming that the field vanishes inside the slab,\myfoot{This is the
  natural choice for the electrical boundary conditions when considering
  shear strain gradients -- recall that they directly relate to transverse
  acoustic phonons.}
we can then calculate the macroscopic electric field in the vacuum region,
\begin{equation}
\epsilon_0 \frac{\partial \hat{E}_{x} (x=+\infty)} { \partial \varepsilon_{xy,y} } \Big|_{\rm FI,SC} =
\sigma^{\rm surf}_{xy,xy} + \sigma^{\rm bulk}_{xy,xy} + \sigma^{\rm met}_{xy,xy}.
\end{equation}
(SC stands for `short-circuit.')
The three quantities on the right-hand side have the physical dimension of a
surface charge density and are given by
\begin{eqnarray}
\sigma^{\rm bulk}_{xy,xy} &=& \bar{\mu}^{\rm bulk}_{xy,xy}, \\
\sigma^{\rm surf}_{xy,xy} &=& -\int_0^{+\infty} dx \, \bar{P}^{\rm U}_{y,xy}(x), \\
\sigma^{\rm met}_{xy,xy}  &=& \epsilon_0 \int_0^{+\infty} dx \, {E}_x(x) = -\epsilon_0 \phi_0.
\end{eqnarray}
In order to derive the flexocoupling coefficient, we need to switch to open-circuit
boundary conditions by imposing an external electric field that exactly cancels the
above vacuum field.
We obtain
\begin{equation}
\bar{\varphi}_{xy,xy} = \frac{1}{\epsilon_0 \bar{\epsilon}_{\rm r}} \left[ \bar{\mu}^{\rm II}_{xx,\alpha \alpha}
-\int_0^{+\infty} dx \, P^{\rm U}_{y,xy}(x)  -\epsilon_0 \phi_0
\right].
\mylabel{xyxy-tot}
\end{equation}

The total surface contribution coming from both the polarization
currents and from the metric is thus
\begin{equation}
\bar{\varphi}^{\rm surf}_{xy,xy} = \frac{1}{\epsilon_0 \bar{\epsilon}_{\rm r}} \left[
-\int_0^{+\infty} dx \, P^{\rm U}_{y,xy}(x)  -\epsilon_0 \phi_0 \right].
\end{equation}
Note that, in contrast with the transverse and longitudinal cases,
the internal electric field is no longer a bulk property here; the
surface terms contained in $\bar{\varphi}^{\rm surf}_{xy,xy}$ manifest
themselves as surface \emph{charge} densities that tend to a constant
in the limit of large slab thickness $t$, rather than dipole densities
that grow linearly with $t$.
(In either case, the surface contribution to the flexovoltage response of the
slab scales similarly to the bulk term for increasing $t$.)
These, unlike in the previous two cases, need to be divided by the bulk
permittivity, as the bulk material dielectrically screens the additional
electric field produced by surface effects.

\begin{figure}[!t]
\begin{center}
\includegraphics[width=4in]{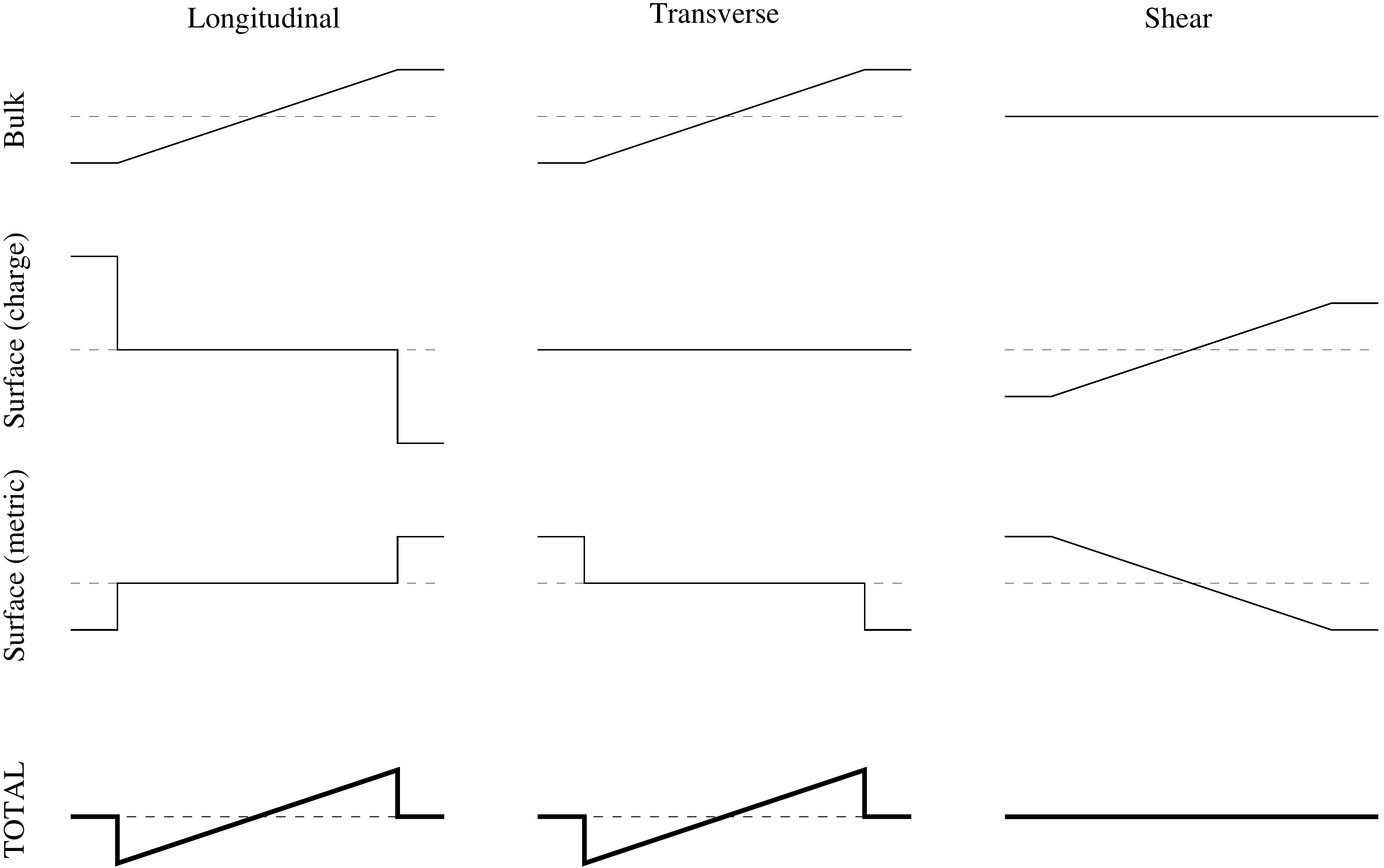}
\end{center}
\caption{ \label{potentials} Flexoelectric response of a slab of noninteracting spheres.
The total induced potential is decomposed into three contributions,
consistent with the formalism developed in the main text. In all panels,
the vertical axis corresponds to the potential (dashed line
indicates the zero), and the horizontal axis
is the spatial coordinate $x$ along the surface normal.
The combined effect of the surface
(including both induced polarization-charge and metric contributions)
and the bulk flexoelectric response yields a vanishing bias potential,
regardless of the type of strain gradient. (From Ref.~\citeonline{artgr}.)}
\end{figure}

{\bf Spherical atom model.}
The formalism that we have developed in this Section allows us to
complete the solution of the toy model that we described at the end of
Section~\ref{sec:surface}, consisting of a finite slab made of a lattice
of rigid (and noninteracting) closed-shell atoms.
The solutions for all the contributions to the flexovoltage response,
now including the shear case and the aforementioned separation of the
surface term into polarization charge and metric terms, are schematically
illustrated in Fig.~\ref{potentials}. (The details of the derivations
can be found in the Supplementary Notes of Ref.~\citeonline{artgr}.)
As expected, in all cases the net voltage vanishes, consistent with the
physical expectations (a rigid displacement of spherical charge distributions
cannot lead to a long-range electrical perturbation).
By the same token, the pseudopotential rigid-core correction
of Eq.~(\ref{rcc}) has no effect on the net flexovoltage.
Note, however, that the bulk, surface-metric and surface-polarization terms
cancel each other in a nontrivial way depending on the strain-gradient
component, indicating that a consistent treatment of all three terms
is crucially important for having a physically meaningful solution.

\subsubsection{Atomic relaxations}
\label{sec:relax_shear}

The contribution of atomic relaxations to the flexovoltage coefficient of
a bent slab has been extensively treated in Sec.~\ref{sec:surf-lattice}.
It is easy to show that, by using the formalism presented in Sec.~\ref{sec:relax}
we recover Eq.~(\ref{rel-ion}), which describes the total response in terms of
bulk- and surface-specific quantities.
What remains to be discussed
is the shear case. In Sec.~\ref{sec:elec_surf}
we postulated that this type of strain-gradient deformation is not relevant
for a fully relaxed unsupported slab. Here we shall substantiate this
statement in light of the results presented
so far.
Recall Eq.~(\ref{atrel}), which describes the microscopic atomic relaxation
pattern induced by a strain gradient in terms of the internal-strain
response tensors $\bm{\Gamma}$ and ${\bf L}$,
and let $X_{l \kappa}$ and $Y_{l \kappa}$ denote the $x$ and $y$
components of ${\bf R}_{l \kappa}$.
In the case of a shear strain gradient of the type $\varepsilon_{xy,y}$
in Fig.~\ref{sketch}(c),
Eq.~(\ref{atrel}) reads as
\begin{equation}
\frac{\partial u^l_{\kappa \alpha}}{\partial \varepsilon_{xy,y}} =
Y_{l \kappa} \, \Gamma^\kappa_{\alpha xy} + L^\kappa_{\alpha y, xy}.
\mylabel{atrel-shear}
\end{equation}
Now, regardless of the microscopic details of the slab, rotational invariance
dictates that
\begin{equation}
\Gamma^\kappa_{\alpha xy} = -X_{0\kappa} \delta_{\alpha y},
\mylabel{gshear}
\end{equation}
i.e., under a uniform shear the slab rigidly rotates to accommodate the
deformation of the supercell, without feeling any restoring force because
the repeated images of the slab are decoupled.
By combining Eqs.~(\ref{atrel-shear}) and (\ref{gshear}) we obtain
\begin{equation}
\frac{\partial u^l_{\kappa \alpha}}{\partial \varepsilon_{xy,y}} =
-Y_{l \kappa} X_{l\kappa} \delta_{\alpha y} + L^\kappa_{\alpha y, xy}.
\mylabel{sheartot}
\end{equation}

\begin{figure}[!t]
\begin{center}
\includegraphics[width=2.5in]{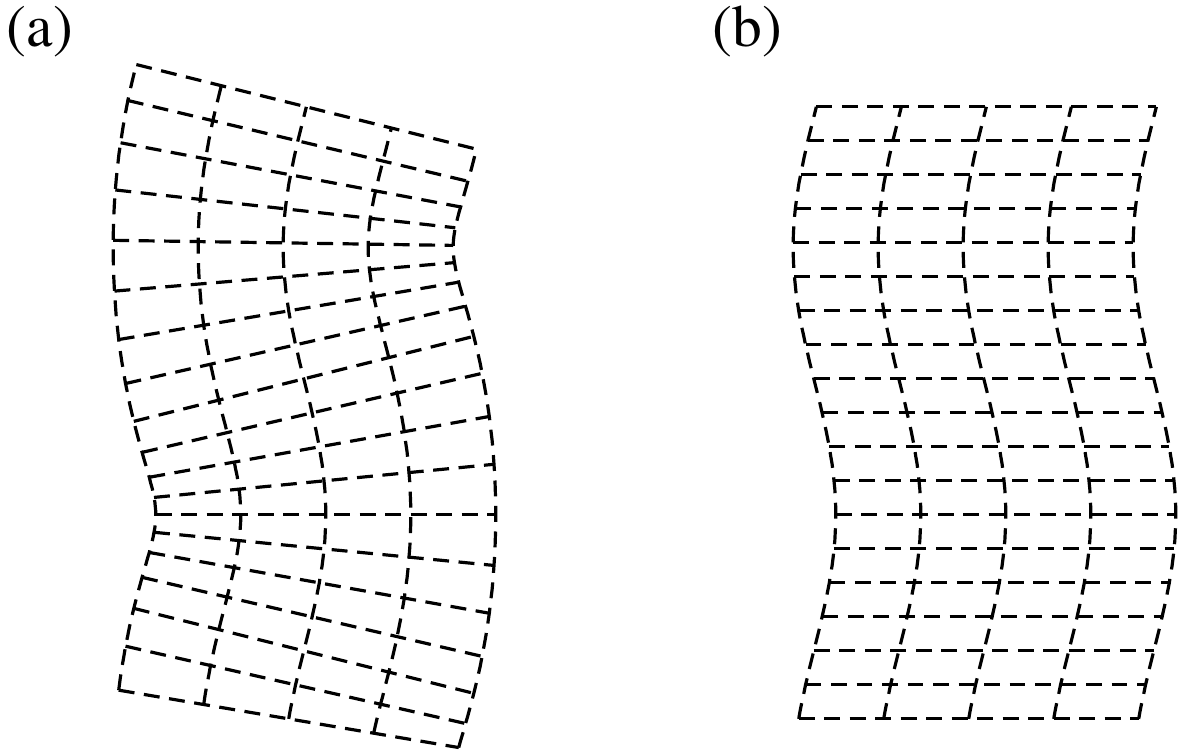}
\end{center}
\caption{ \label{t-s-compar}
Sketch of slab subjected to a periodic transverse strain of the type
shown in Fig.~\ref{sketch}(b), or a negative shear strain of the type
shown in Fig.~\ref{sketch}(c).  After internal atomic relaxations,
the two configurations become equivalent.}
\end{figure}

Now, recall that a macroscopic strain gradient can be written in terms of the
components of the type-I strain-gradient tensor as
\begin{displaymath}
u^l_{\kappa \beta} = \frac{ \eta_{\beta,\gamma \lambda} }{2}
   ({\bf R}_{l \kappa})_\gamma ({\bf R}_{l \kappa})_\lambda.
\end{displaymath}
Eq.~(\ref{sheartot}) states that a shear strain gradient of amplitude
$\eta_{x,yy}=\eta$ is always accompanied, in a fully relaxed unsupported
film, by a second strain gradient component of the type $\eta_{y,xy} =
-2 \eta$.
The overall effect, in type-II notation, is that of a negative
\emph{transverse} strain gradient, $\varepsilon_{yy,x}= -\eta$.
This means that, for a free-standing film, the shear case reduces exactly
to the transverse one.
The basic concept is illustrated in
Fig.~\ref{t-s-compar}, where we compare the configurations
obtained by periodically subjecting a slab to
a transverse strain gradient $\varepsilon_{yy,x}$ as
in Fig.~\ref{sketch}(b), or a (negative) shear strain gradient
$\varepsilon_{xy,y}$ of the kind shown in Fig.~\ref{sketch}(c).
If internal atomic relaxations are allowed while still preserving
the overall undulation along $y$, the two configurations will
clearly relax to the exact same geometry.

Thus, we have rigorously demonstrated the result that we
heuristically presented in Sec.~\ref{sec:surface}: flexoelectric effects in a
free-standing film of sufficiently high symmetry (e.g., cubic or
in-plane hexagonal)
are governed by only one response coefficient $\varphi_{xx,yy}$,
as given by Eq.~(\ref{rel-ion}).
The induced voltage at a given location is then given by $\varphi_{xx,yy}
(t / \xi_y + t / \xi_z)$, where $\xi_y=\varepsilon_{yy,x}^{-1}$ and
$\xi_z=\varepsilon_{zz,x}^{-1}$ are the radii of curvature (along the
Cartesian axes) of the film at that specific point.\myfoot{One could
  equivalently choose different orthogonal axes, e.g., those corresponding
  to the \emph{principal curvatures} of the surface.  Since $1/\xi_y +
  1/\xi_z$ is the trace of the shape operator, the result is independent
  of such a choice.}
This includes the plate-bending and beam-bending limits as special cases.

It is interesting to note that, in contrast with what happens in the bulk,
here we have a notable case where the flexoelectric effects induced
by a sound wave are identical to those associated with a static
deformation. (Equivalently, one can say that the same strain gradient
field can be induced either by dynamic or static means.)
Indeed, any two-dimensional object such as a slab is characterized by a transverse
acoustic phonon branch, usually referred to as ZA, with zero sound velocity,
corresponding to a bending mode.
A long-wavelength ZA phonon coincides, therefore, with the static bending case
described above, and produces the same flexoelectric response.

\subsection{Summary}

In this Section we have presented a fundamental theory of flexoelectricity,
based on a quantum-mechanical description of the electronic and lattice
response to a strain-gradient perturbation.
In particular, we have used a long-wave expansion of acoustic phonons
to derive, in the linear limit, the relevant electromechanical response
functions of a crystalline solid.
Our formalism is fully general, and correctly recovers earlier theories of
piezoelectricity as a special case.

In order to address some conceptual issues (e.g., regarding the role of
the surfaces, or regarding the calculation of some components of the bulk
flexoelectric tensor that are presently difficult to access) we have
gone a step further, and developed a fully \emph{microscopic} theory
of the linear response to an inhomogeneous strain field. In this
context, we have demonstrated that the use of curvilinear coordinate
frames greatly facilitates the representation of the relevant physical
fields and their response to mechanical deformation.
The latter methodological tools are applicable well beyond the specifics
of flexoelectricity, and may find application in related research
areas, such as flexomagnetism.~\cite{Hertel-13}

\section{Application to SrTiO$_3$}
\label{sec:results}

\begin{figure}[!t]
\begin{center}
\includegraphics[width=3.5in,clip]{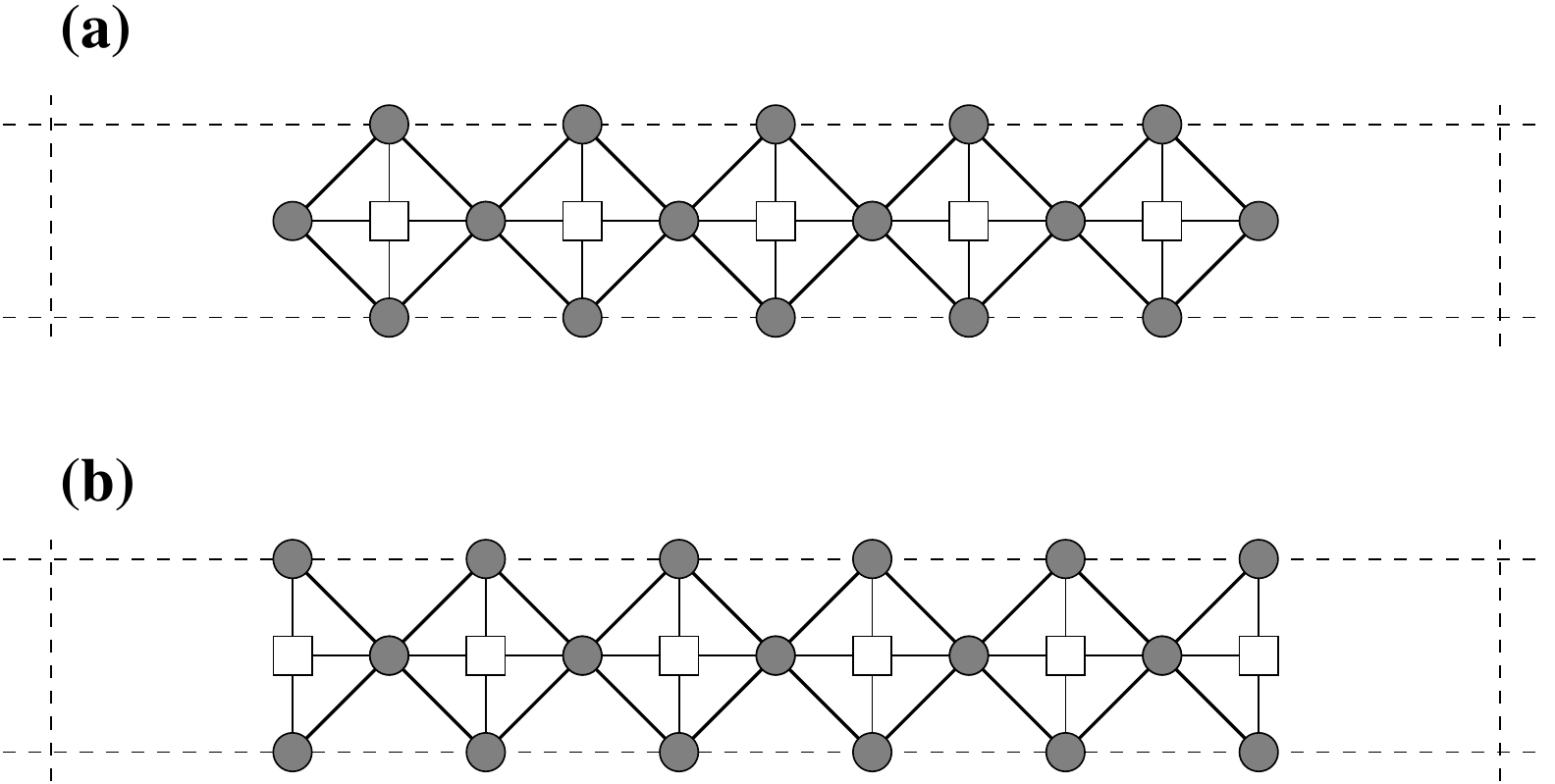} \\
\end{center}
\caption{ \label{figsuper} Supercell models of the SrO- (a) and
TiO$_2$-terminated (b) SrTiO$_3$ slabs. Ti and O atoms are represented
as white squares and gray circles respectively; Sr atoms are not shown.
(Adapted from Ref.~\citeonline{artcalc}.)}
\end{figure}

In this Section we shall demonstrate the theory developed so far by
applying it to SrTiO$_3$, one of the most important materials in the
context of flexoelectricity, and the best known experimentally.
In order to quantify the importance of surface effects, we
shall consider a slab geometry, and two different lattice terminations
(either of the SrO or TiO$_2$ type), as illustrated in Fig.~\ref{figsuper}.

\subsection{General methodology}

Our goal is to calculate the total flexovoltage response of either SrTiO$_3$ slab
to a bending deformation in the limit of large thickness.
We shall do this by taking into account the effect of full atomic relaxation,
under the initial hypothesis that our slab behaves as a
\emph{plate}.\myfoot{This means that along the direction parallel to the
bending axis the system is clamped (i.e., no anticlastic bending is allowed).}
(We shall see in Sec.~\ref{sec:relax-sto} that the beam-bending limit can easily be
recovered by rescaling the plate-bending coefficient by a constant.)
This requires the combination of three different computational frameworks,
as detailed in the following.

\subsubsection{Bulk calculations}
\label{sec:bulkcalc}

Here we perform a number of calculations
on a primitive unit cell of bulk SrTiO$_3$. This is primarily aimed at
calculating the bulk flexoelectric tensor via a long-wave expansion
of acoustic phonons.
Acoustic phonons are treated at the linear-response level
by means of density-functional perturbation theory as
implemented in a modified version of the ABINIT package~\cite{abinit}
in which
the contribution of the macroscopic electric fields has been removed
according to the discussion of Sec.~\ref{sec:e-mac}.
In particular, We choose a small star of wavevectors ${\bf q}$ surrounding
the $\Gamma$ point in the Brillouin zone,
$${\bf q} = \frac{2 \pi \tilde{q}}{a_0} ( \pm 1, 0 ; \, \pm 1, 0; \, 0),$$
and perform a full linear-response calculation for each of these points.
(In practice, we make full use of symmetries to minimize the number of
actual calculations.)
Next, we perform a long-wave expansion of the charge-density response and
interatomic force constants and extract the second-order-in-$\bf q$
coefficients via numerical differentiation with respect
to ${\bf q}$.
We obtain: (i) the flexoelectric force-response tensor
$T^\kappa_{\alpha \lambda, \beta \gamma}$ via Eq.~(\ref{qqUT}),
from which
$C^\kappa_{\alpha \lambda, \beta \gamma}$
and then the internal-strain tensor
$L^\kappa_{\alpha \lambda, \beta \gamma}$ are constructed
via Eqs.~(\ref{massdep}-\ref{CTTT});
and (ii) the charge-density response tensors $Q^{(1,\gamma)}_{\kappa \beta}$
and $Q^{(3,\gamma \lambda \sigma)}_{\kappa \beta}$, corresponding
respectively to the Born effective charge tensor
$Z^*_{\kappa,\beta\gamma}$ and the dynamical octupole tensor,
via Eq.~(\ref{rhoq}).

By combining the internal-strain tensor with the Born effective charges one can readily
obtain the lattice-mediated contributions to the flexoelectric tensor as explained in
Sec.~\ref{sec:longwave}.
The octupole tensor, on the other hand, provides us with only partial information on the
electronic (frozen-ion) flexoelectric tensor.
In particular, only the \emph{longitudinal} component of the electronic flexoelectric tensor,
$\bar{\mu}_{\hat{\bf q}}$, along an arbitrary direction $\hat{\bf q}$ can be inferred
from the two linearly independent entries of $Q^{(3,\gamma \lambda \sigma)}_{\kappa \beta}$.
Following Hong and Vanderbilt~\cite{Hong-13}, we define
$$
\bar{\mu}_{\rm L1} = \bar{\mu}_{(100)}, \qquad \bar{\mu}_{\rm L2} = 2 \bar{\mu}_{(110)} - \bar{\mu}_{(100)}.
$$
These are related to the components of the type-II flexoelectric tensor,
$\mu_{\rm L2}$ by~\cite{artcalc}
\begin{eqnarray}
\bar{\mu}^{\rm II}_{xx,xx} &=& \bar{\mu}_{\rm L1}, \mylabel{mu-L1} \\
\bar{\mu}^{\rm II}_{xx,yy} + 2 \bar{\mu}^{\rm II}_{xy,xy} &=& \bar{\mu}_{\rm L2}. \mylabel{mu-L2}
\end{eqnarray}
Thus, in order to determine the transverse and shear components
$\bar{\mu}^{\rm II}_{xx,yy}$ and $\bar{\mu}^{\rm II}_{xy,xy}$ independently,
an additional calculation is necessary; this will be addressed shortly
in Sec.~\ref{sec:truncslab}.

In addition to the above calculations, which are based on the
methodology described in this Chapter,
we also need a bulk-level calculation of some auxiliary quantities
by means of more established techniques.
Specifically, we extract the high-frequency dielectric constant
$\bar{\epsilon}_{\rm r}$ from a separate linear-response treatment of the
electric-field perturbation.
At the same time we obtain a redundant set of $Z^*_{\kappa,\beta\gamma}$
tensor elements, which are useful for assessing the quality of the numerical
differentiation at first order in $\bf q$
performed in Sec.~\ref{sec:bulkcalc} above.
Similarly, we carry out an independent calculation of the elastic tensor
$C_{\alpha \lambda, \beta \gamma}$ via finite differences with respect to
applied strain; this allows us to check the second-order-in-$\bf q$
calculations of the force-response tensors $C^\kappa_{\alpha \lambda,
\beta \gamma}$, since these quantities are directly related by the sum
rule in Eq.~(\ref{sumrule}).

\subsubsection{Truncated-bulk slab calculations}
\label{sec:truncslab}

Here we
carry out calculations similar to those of Sec.~\ref{sec:bulkcalc}, but now
on a slab supercell. This step is aimed at determining the
transverse and shear components of the \emph{bulk} electronic (frozen-ion) flexoelectric tensor.\myfoot{
  It may seem odd to use a slab supercell to calculate a bulk-specific quantity;
  this is indeed a temporary work-around, which will no longer be necessary
  once a proper theory of the current-density response becomes available.}
In fact, the two independent components of the bulk dynamical octupole tensor
$Q^{(3,\gamma \lambda \sigma)}_{\kappa \beta}$ that
we calculated above are not sufficient to determine
the three independent entries of the bulk
$\bar{\mu}^{\rm II}_{\alpha \lambda, \beta \gamma}$ tensor.
We are able to circumvent this limitation by recourse to a series of
calculations on a slab geometry in which we determine the charge-density
response, both in the bulk and at the surface,
to longitudinal, transverse and shear strain gradients.
A calculation of the flexoelectrically induced open-circuit
electric field in the interior of the film,
which relates [based on Eq.~(\ref{E-transverse})] 
directly to the corresponding component of the bulk
flexoelectric tensor in two cases out of three (longitudinal and transverse),
allows us to obtain the missing component\myfoot{Strictly
  speaking, only the transverse component is really needed, as
  the longitudinal component calculated in this way is redundant with the
  $\mu_{\rm L1}$  value that we already calculated at the bulk level.
  We shall use this as a test to assess the numerical accuracy of our
  calculations.}
of $\bar{\bm{\mu}}^{\rm II}$.
The key point here is that the missing divergence-free component of
the induced polarization
current, which is not currently available from bulk-level calculations,
manifests itself as a surface charge density, whose influence is
readily apparent in the slab supercell geometry.
Note that the specifics of the surface structure should
not matter in these calculations.
Thus, we choose the geometry that ensures the best convergence
of the inner open-circuit field as a function of slab thickness, i.e., a truncated-bulk
structure. (We perform such an analysis on both SrO- and
TiO$_2$-terminated slabs, in order to verify that the results are indeed
surface-independent as we expect.)

In practice, we use the same star of ${\bf q}$-points surrounding $\Gamma$ as in the
bulk calculations described above. This time, however, we neglect the information on
the force constants and only focus on the charge-density response of the system.
We need to analyze such a response at the microscopic level, by using the
curvilinear-coordinate formalism of Sec.~\ref{sec:microscopic} and
Sec.~\ref{sec:surf_micro}.
Of the two relevant response functions, $\rho^{\rm U}(x)$ and $\rho^{\rm G}(x)$, only the
latter is really an issue, as $\rho^{\rm U}(x)$ can be straightforwardly calculated as the
response to a uniform strain.\myfoot{We calculated $\rho^{\rm U}(x)$ separately by using
  standard ground-state calculations where we took finite differences in the strain.
  We found that this latter procedure yields slightly better accuracy than the long-wave
  method described above.}
The result of the second-order Taylor expansion in ${\bf q}$ yields $\rho^{\rm G}(x)$,
and this (together with $\rho^{\rm U}$) is then used to calculate the electric-field
response functions $E^{\rm U,G}(x)$.

Note, however, that due to the removal
of the macroscopic electric fields in
the phonon calculations~\cite{artlin,artgr,artcalc} (as required to perform the
aforementioned Taylor expansions in ${\bf q}$, see Sec.~\ref{sec:e-mac}), short-circuit electrical boundary
conditions are enforced by construction on the calculated $\rho^{\rm G}$ and $E^{\rm G}$.
This means that there are nonvanishing macroscopic electric fields in both the vacuum
and the slab interior, and these fields show an undesirable dependence on the supercell
geometry (vacuum and slab thicknesses).
To have a physically well-defined (and geometry-independent) value of the internal
field we need to enforce open-circuit electrical boundary conditions.
We do this by applying an \emph{external} field to the system that is
exactly opposite to the calculated  vacuum field.
To determine the charge redistribution induced in the system upon
application of an external field, we perform a separate linear-response
calculation of the \emph{local} electric field response to a macroscopic
electric displacement field $D$.
This is nothing but the local inverse dielectric permittivity of the slab
supercell,
$$
\frac{\partial E_x(x)}{\partial D_x} = \frac{1}{\epsilon_0}
\bar{\epsilon}_{\rm r}^{\,-1}(x).
$$
We then use
$$
E^{\rm G,OC}_x(x) = E^{\rm G,SC}_x(x) - E^{\rm G,SC}_x(+\infty)
\bar{\epsilon}_{\rm r}^{\,-1}(x),
$$
where $x=+\infty$ corresponds, as usual, to the vacuum region. When referring
to $E^{\rm G}(x)$ in the following, we shall implicitly assume that we are
speaking of the open-circuit version $E^{\rm G,OC}_x(x)$.

\subsubsection{Relaxed-ion slab calculations}

Now that we have all the necessary bulk-specific information in
hand, we still need to determine the surface-specific contributions
to the flexovoltage coefficient $\varphi^{\rm surf}_{xx,{\rm
eff}}$.\myfoot{Recall that we need to consider,
  for a bent slab at mechanical equilibrium, an \emph{effective}
  combination of transverse and longitudinal strain-gradient deformations,
  $\varepsilon_{yy,x} = \varepsilon_{{\rm eff},x}; \, \, \varepsilon_{xx,x} = -\nu \varepsilon_{{\rm eff},x}$,
  where $\nu=\mathcal{C}_{yy,xx} / \mathcal{C}_{xx,xx}$.}
We shall compute $\varphi^{\rm surf}$ as the induced electrostatic
potential offset upon application of a \emph{uniform} effective strain
($\varepsilon_{yy} = \varepsilon_{{\rm eff}};
\,\,\varepsilon_{xx} = - \nu \varepsilon_{{\rm eff}}$)
to a free-standing slab with (001) surface orientation.
This quantity can be conveniently accessed by means of a standard plane-wave code;
no linear-response features are needed. In particular, we take a slab supercell
corresponding to a periodic lattice of alternating SrTiO$_3$ and vacuum layers,
and first calculate the electronic and structural ground state by setting the
in-plane lattice parameter to the equilibrium bulk value.
We then apply a small positive or negative strain of the type
$$
\bm{\varepsilon} = \frac{ \varepsilon_{\rm eff} }{2} \left( \begin{array}{r @{\quad} r @{\quad} r}
-2\nu & 0 & 0 \\
    0 & 1 & 0 \\
    0 & 0 & 1
  \end{array} \right),
$$
where $\varepsilon_{\rm eff}$ is a small dimensionless number, typically
$\varepsilon_{\rm eff} (\pm) = \pm 0.001$.
(We find it computationally advantageous to preserve the fourfold axis of the
SrTiO$_3$ surface by applying an isotropic in-plane strain.)

In each perturbed configuration, we first calculate the electronic
ground state with the reduced coordinates of the atoms kept fixed
to their unperturbed values; the resulting electrostatic potential
profile is then processed by means of macroscopic
averaging~\cite{Baldereschi-88,Junquera-07} to extract the perturbed
frozen-ion (FI) surface potential offsets.
Next, we let the atoms relax to their new equilibrium positions in
the strained lattices, and repeat the macroscopic averaging procedure
to obtain the relaxed-ion (RI) offsets.
Finally, we numerically differentiate the perturbed offsets (both
FI and RI) to obtain their corresponding first-order variation,
$$
\varphi^{\rm surf} = \frac{\phi(+) - \phi(-)}{2 |\varepsilon_{\rm eff}| },
$$
where $\phi(\pm)$ refers to the surface
potential offset at positive or negative strain.\myfoot{
  As a technical note, many first-principles codes use the Ewald procedure to calculate
  the self-consistent electrostatic potential. This involves adding to the electronic
  density a lattice of spherical Gaussian compensating charges, whose spurious contribution
  must be removed from the calculated value of $\varphi^{\rm surf}$. See the Supplementary
  Notes of Ref.~\citeonline{artcalc} for details.}
This procedure readily yields the RI and FI values of $\varphi^{\rm surf}$.
The lattice-mediated (LM) values are simply calculated as the difference
of the RI and FI ones.
%
Of course, the slab needs to be sufficiently thick in order for the inner layers
to be truly bulk-like, i.e., unaffected by the atomic distortions that originate
from the surface truncation of the bonding network.


\subsection{Computational parameters}

We use the local-density approximation~\cite{Perdew/Wang:1992}
to density-functional theory. The interactions between valence
electrons and ionic cores are described by separable norm-conserving
pseudopotentials in the Troullier-Martins~\cite{troullier} form,
generated with the fhi98PP code.~\cite{fhi98pp}
The $4s^24p^6$ and $3s^23p^6$ shells of Sr and Ti,
respectively, are explicitly treated as 
valence electrons.
The reference states
(numbers in parentheses indicate the core radius in bohr) of the
isolated neutral atom used in the pseudopotential generation
are $2s$(1.4), $2p$(1.4) and $3d$(1.4) for O, $4s$(1.5), $4p$(1.5)
and $4d$(2.0) for Sr and $3s$(1.3), $3p$(1.3) and $3d$(1.3) for Ti.  The
local angular-momentum channel is $l = 2$ for Sr and O and $l = 0$ for Ti.
The rigid-core corrections of Eq.~(\ref{rcc}) are not included
in the presented results.
The cutoff for the wavefunction plane-wave basis is set to 150 Ry in the
slab calculations. (A test calculation with a 300 Ry cutoff did not show
appreciable changes in the calculated electronic response functions;
the 300 Ry cutoff was, nonetheless, necessary to ensure satisfactory
accuracy in the force-response tensor at the bulk level.) The surface
Brillouin zone of the slab supercell is sampled by means of a $8 \times 8$
Monkhorst-Pack grid~\cite{Monkhorst/Pack:1976}; for the bulk primitive
cell we use a sampling of up to $12 \times 12 \times 12$ $k$-points. The
finite-difference parameter in the long-wave expansion, $\tilde{q}$, is
set to 0.01 (tests with $\tilde{q}\!=\!0.02$ or $\tilde{q}\!=\!0.03$ indicated
a convergence better than 1\% in the calculated electronic response
functions; smaller values of $\tilde{q}$ were found to yield less accurate
results because of the excessive numerical noise).
The lattice parameter of the cubic cell is set to $a_0$=7.268 bohr,
which corresponds to the calculated equilibrium value.

The supercell models are based on the schematic illustrations of
Fig.~\ref{figsuper}(a-b).
For the truncated-bulk linear-response calculations we use
5.5-unit-cell (uc) thick SrTiO$_3$ slabs alternating with vacuum layers
whose thickness is set to 2.5\,uc. Of course, both (slab and vacuum)
thicknesses are intended as convergence parameters in our calculations,
whose scope is to describe the thermodynamic limit of a macroscopic
slab. Tests with thinner slabs and thicker vacuum layers (up to 3.5\,uc)
showed optimal convergence for the aforementioned values of these
parameters (again, better than 1\%). For the relaxed-ion slabs, we use
7.5\,uc-thick slabs with 3.5\,uc-thick vacuum layers.


\subsection{Results}

\subsubsection{Bulk calculations}

\begin{table}
\tbl{Force-response tensor $C^{\kappa}_{\alpha \lambda, \beta \gamma}$
     of bulk SrTiO$_3$ in short-circuit boundary conditions. O1, O2 and O3
     refer to oxygen atoms forming $x$-, $y$- or $z$-oriented Ti-O-Ti
     bonds, respectively. All values are in eV.}
{\begin{tabular}{ c  r @{.} l  r @{.} l r @{.} l } \toprule
 Atom      &  \multicolumn{2}{c}{$(xx,xx)$} & \multicolumn{2}{c}{$(xy,xy)$} & \multicolumn{2}{c}{$(xx,yy)$} \\
\colrule 
   Sr &   $-$24&9  &    7&9 &  $-$28&7 \\
   Ti &   $-$67&9  &    3&8 & $-$102&3 \\
   O1 &     159&3  &   15&3 &     97&4 \\
   O2 &      35&2  &   17&3 &     42&3  \\
   O3 &      35&2  & $-$0&9 &     30&9 \\
   \botrule
\end{tabular}}
\label{tab:force-resp}
\end{table}

\begin{table}
\tbl{Summary of the linear-response data obtained from the
long-wave (LW) approach at the bulk level, compared to the
results of Hong and Vanderbilt~\cite{Hong-13} (HV) for the same quantities.
Open-circuit electrical boundary conditions are enforced on the longitudinal response functions (L1 and L2).
The force response to a shear strain gradient (S) is quoted in short circuit.
The oxygen modes $\xi_3= x_{\rm O1}$ and $\xi_4 = (x_{\rm O2} +x_{\rm O3}) / \sqrt{2}$ are defined
following Ref.~\citeonline{Hong-13}.
$\bar{\varphi}^{\rm bulk}$ is in V; other values are reported in eV.}
{\begin{tabular}{ c  r @{.} l  r @{.} l  r @{.} l  r @{.} l  r @{.} l  r @{.} l } \toprule
     &  \multicolumn{2}{c}{L1(LW)} & \multicolumn{2}{c}{L1(HV)}
                  &  \multicolumn{2}{c}{L2(LW)} & \multicolumn{2}{c}{L2(HV)} &  \multicolumn{2}{c}{S(LW)} & \multicolumn{2}{c}{S(HV)}\\
\colrule 
 $\bar{\varphi}^{\rm bulk}$ &  $-$16&15 & $-$16&25 & $-$18&07 & $-$18&17  &\multicolumn{2}{c}{-} & \multicolumn{2}{c}{-} \\
\colrule 
   Sr         &   16&3 & 17&0 &    33&2 & 35&7  & 7&9 &  8&4 \\
   Ti         &  49&1 & 52&3 &    36&3 & 38&9   & 3&8 &  3&0 \\
   $\xi_3$    &  67&2 & 68&7 &    24&9 & 13&1  & 15&3 & 15&7 \\
   $\xi_4$    &   3&0 &  3&6 &    22&7 & 18&2  & 11&6 & 12&0 \\ \botrule
\end{tabular}}
\label{tab:lr}
\end{table}

\begin{table}
\tbl{Calculated Born effective charges and dielectric properties
of bulk SrTiO$_3$.}
{\begin{tabular}{c  r @{.} l  r @{.} l  r @{.} l  r @{.} l r @{.} l  r @{.} l  c} \toprule
    &   \multicolumn{2}{c}{$Z^*_{\rm Sr}$} & \multicolumn{2}{c}{$Z^*_{\rm Ti}$} & \multicolumn{2}{c}{$Z^*_{\rm O1}$}
    &   \multicolumn{2}{c}{$Z^*_{\rm O2}$} & \multicolumn{2}{c}{$Z^*_{\rm O3}$} & \multicolumn{2}{c}{$\bar{\epsilon}_{\rm r}$} & $\epsilon_{\rm r}$ (static)\\
\colrule 
  &  2&5548 & 7&2455 & $-$5&7027 & $-$2&0488  & $-$2&0488 & 6&1785 & 1657 \\ \botrule
\end{tabular}}
\label{tab:born}
\end{table}

In Table~\ref{tab:force-resp} we report the relevant values of the
force-response tensor of bulk SrTiO$_3$, calculated by using the
long-wave method described in Sec.~\ref{sec:longwave}.
In Table~\ref{tab:lr} we compare the above physical quantities to the analogous
ones that were calculated in Ref.~\citeonline{Hong-13}. To perform the comparison we first
recast the force-response components into a tensorial representation that
follows the same prescriptions as Eq.~(\ref{mu-L1}) and (\ref{mu-L2}),
\begin{eqnarray}
C^\kappa_{\rm L1} &=& C^\kappa_{xx,xx}, \\
C^\kappa_{\rm L2} &=& C^\kappa_{xx,yy} + 2 C^\kappa_{xy,xy}.
\end{eqnarray}
Then, we convert the longitudinal quantities L1 and L2 from
fixed-$\bf E$ or short-circuit (SC) to
fixed-$\bf D$ or open-circuit (OC) boundary conditions by using
[see Eq.~(106) of Ref.~\citeonline{Hong-13}]
\begin{equation}
C^\kappa_{\rm L}({\rm OC}) = C^\kappa_{\rm L}({\rm SC}) -
\bar{\varphi}^{\rm bulk}_{\rm L} Z^*_\kappa,
\mylabel{ocsc}
\end{equation}
where $Z^*_\kappa$ is the Born effective charge (calculated values are reported in
Table~\ref{tab:born}), $\bar{\varphi}^{\rm bulk}_{\rm L}$ is
the purely electronic flexovoltage coefficient, and L stands for either L1 or L2.
The calculated values of
$\bar{\varphi}^{\rm bulk}_{\rm L1,L2}$ are also reported in Table~\ref{tab:lr} for
direct comparison to those reported by Hong and Vanderbilt~\cite{Hong-13}.
The agreement is overall very good, especially considering the different
computational strategy, first-principles code and pseudopotentials that were
used in Ref.~\citeonline{Hong-13}.

\begin{table}
\tbl{Calculated elastic tensor of bulk SrTiO$_3$. The two rows
refer to the bulk force-response calculation (``Force'')
and to a direct bulk calculation where we took finite
differences of the calculated stress tensor while varying the strain
around the equilibrium cubic configuration (``Strain''). Values are in GPa.}
{\begin{tabular}{ l  r @{.} l  r @{.} l r @{.} l } \toprule
Method &  \multicolumn{2}{c}{$(xx,xx)$} & \multicolumn{2}{c}{$(xy,xy)$} & \multicolumn{2}{c}{$(xx,yy)$} \\
\colrule 
 Force &   385&3  &   122&2  &   111&7 \\
 Strain &   386&2  &   122&4  &   112&6 \\ \botrule
\end{tabular}}
\label{tab:elas}
\end{table}

As a numerical test of the calculated force-response tensor
(Table~\ref{tab:force-resp}),
in Table~\ref{tab:elas} we report the elastic constants of bulk SrTiO$_3$ that we
computed in two different ways: either as a first derivative of the stress with
respect to the applied strain (``strain'') or by using the sum rule of
Eq.~(\ref{sumrule}) (``force'').
The agreement is excellent (better than 1\%), confirming the high numerical
quality of the calculation.
Note that the choice of the electrical boundary conditions is irrelevant for
this test, as the sublattice sum of the atomic forces induced by a hypothetical
electric field vanishes due to the acoustic sum rule.


\subsubsection{Truncated-bulk slab calculations}

\begin{figure}[!t]
\begin{center}
\includegraphics[width=4in,clip]{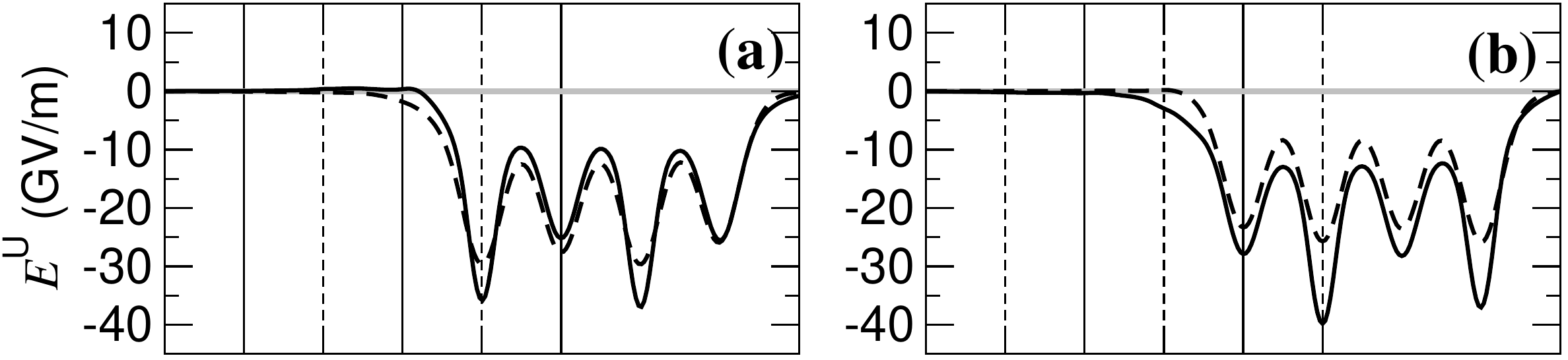} \\
\includegraphics[width=4in,clip]{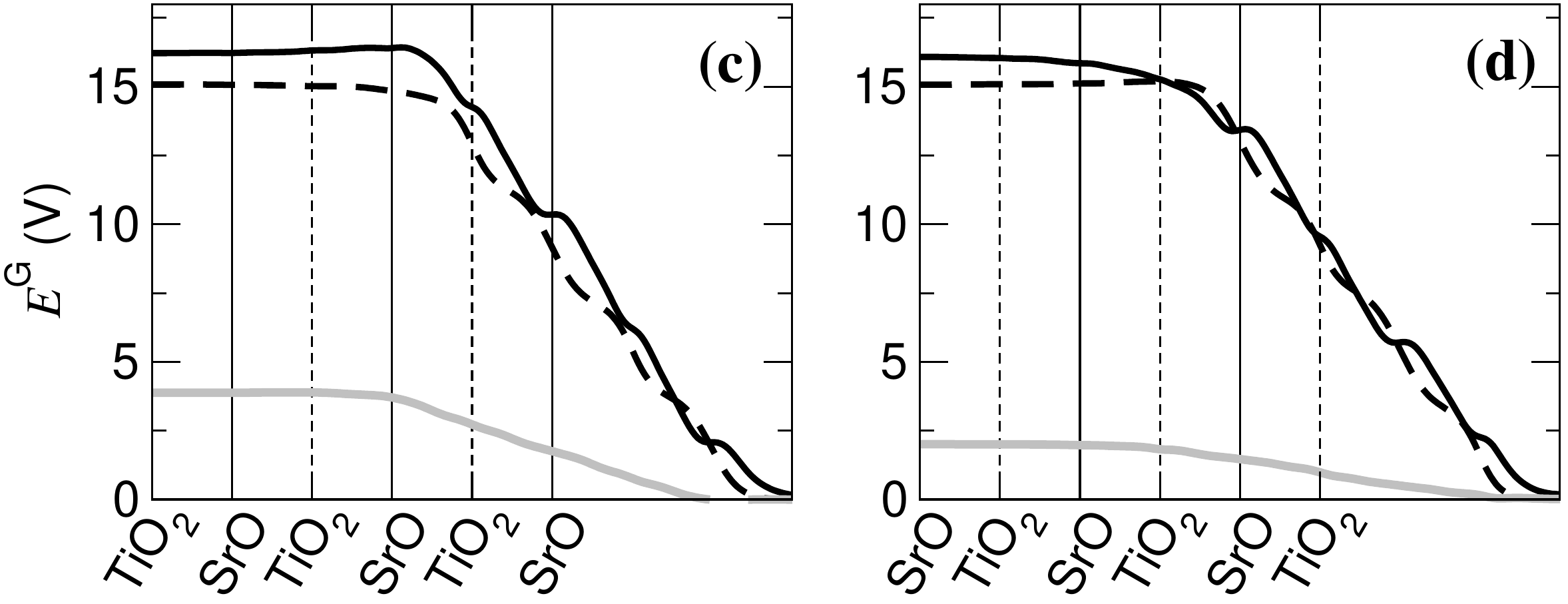}
\end{center}
\caption{ \label{efields} Electric field response to mechanical
deformations.  The $E^{\rm U}_x$ (a-b) and $E^{\rm G}_x$ (c-d)
response functions are shown for a SrO- (a,c) and TiO$_2$-terminated
(b,d) slab. Solid black, dashed black and solid gray curves
refer to longitudinal, transverse and shear deformations, respectively.
The location of the SrO (dashed) and TiO$_2$ (solid) atomic layers
is indicated by vertical lines (only half of the symmetric slab is
shown). (Adapted from Ref.~\citeonline{artcalc}.)
}
\end{figure}

\begin{table}
\tbl{Frozen-ion flexovoltage coefficients of a truncated-bulk SrTiO$_3$ slab.
To compute $\varphi^{\rm bulk}$ we used $\varphi^{\rm bulk}_{\rm L1,L2}$
as reported in Table~\ref{tab:lr}
and $E^{\rm slab}_{xx,yy} = 15.08$ V [extracted from Fig.~\ref{efields}(c-d)].
(L), (T) and (S) stands for longitudinal, transverse and shear, respectively.
Units of Volts are used throughout.}
{\begin{tabular}{crrrrr} \toprule
&  \multicolumn{1}{c}{$\varphi^{\rm bulk}$} & \multicolumn{2}{c}{$\varphi^{\rm surf}$} &  \multicolumn{2}{c}{$\varphi$ (total)} \\
      &      & \multicolumn{1}{c}{SrO} &  \multicolumn{1}{c}{TiO$_2$}
                                  & \multicolumn{1}{c}{SrO} &  \multicolumn{1}{c}{TiO$_2$} \\
\colrule 
$xx,xx$ (L)  & $-$16.15  &    14.36    &     16.95  &  $-1$.80   &       0.80 \\
$xx,yy$ (T)  & $-$15.08  &    15.68    &     12.45  &     0.61   &    $-2$.63 \\
$xy,xy$ (S)  &  $-$1.50  &  $-2$.38    &   $-$0.51  &  $-3$.88   &    $-2$.01 \\ \botrule
\end{tabular}}

\caption{ }

\label{tab4}
\end{table}

In Fig.~\ref{efields}(a-d) we plot the calculated $E_x^{\rm U,G}(x)$,
corresponding to either a SrO- or a TiO$_2$-terminated slab and
to each of the three types of imposed strain gradients shown
in Fig.~\ref{sketch} (with no internal relaxations allowed).
As anticipated in Sec.~\ref{sec:elec_surf}, there is an important qualitative
difference between the case of the longitudinal or transverse response, where the
strain gradient is oriented along the surface normal, and that of the shear response,
where it is directed in plane.

In the former case, $E_{x,\beta \beta}^{\rm U}(x)$ (describing the ${\bf E}$-field response
to a \emph{uniform} strain) yields the surface contribution to the flexovoltage coefficient,
$\bar{\varphi}^{\rm surf}_{xx,\beta \beta}$, via Eq.~(\ref{eqsurf-eu}),\myfoot{
  Note, however, that here we are dealing with a truncated-bulk slab, whose surface
  atomic coordinates were artificially frozen to ideal bulk positions. The surface
  contributions that one extracts from such a geometry do not necessarily reflect,
  therefore, the response of a realistic system; they are quoted here mostly for
  illustrative purposes.}
while
%
%
%
the functions $E_{x x,\beta \beta}^{\rm G}(x)$ provide us with the sought-after
information on the \emph{bulk} flexovoltage coefficient of SrTiO$_3$,
\begin{displaymath}
\varphi^{\rm bulk}_{xx,\beta \beta} = -E^{\rm G}_{xx,\beta \beta}(x=0).
\end{displaymath}
Note that the $E_{x x,\beta \beta}^{\rm G}(x)$ functions are roughly uniform inside the film,
which indicates that the slab is thick enough to display bulk properties therein, and zero outside,
consistent with the open-circuit electrical boundary conditions that were enforced.
Moreover, the uniform internal field appears to be nicely \emph{independent} of
the surface termination
for the longitudinal and transverse deformations,
which is a further important consistency test for
our computational approach.

In the shear case, however, the flexoelectric field depends on both
bulk and surface-specific properties,~\cite{artgr} and such a
termination dependence is clear from a comparison of the
gray curves in Fig.~\ref{efields}(c) and (d). From the electric-field response
functions of Fig.~\ref{efields}(c-d) we can thus only extract the \emph{total}
flexovoltage coefficient of the slab, $\varphi_{x y,xy} = -E^{\rm G}_{xy,xy}(x=0)$.
To separate $\varphi_{x y,xy}$ into bulk and surface terms it suffices, however,
to complement the above data with the $\varphi^{\rm bulk}_{\rm L1, L2}$ values that
we calculated at the bulk level. Indeed, by replacing the flexoelectric tensor
components in Eq.~(\ref{mu-L1}) and Eq.~(\ref{mu-L2}) with the corresponding
flexovoltage coefficients, we have
\begin{eqnarray}
\varphi^{\rm bulk}_{\rm L1} &=& \varphi^{\rm bulk}_{xx,xx}, \label{phil1} \\
\varphi^{\rm bulk}_{\rm L2} &=& \varphi^{\rm bulk}_{xx,yy} + 2 \varphi^{\rm bulk}_{xy,xy}.
\label{phil2}
\end{eqnarray}
Eq.~(\ref{phil1}) constitutes a useful consistency check of the methodology,
as $\varphi^{\rm bulk}_{\rm L1}$ is redundant with the already calculated
value of $\varphi^{\rm bulk}_{xx,xx}$.
Eq.~(\ref{phil2}), on the other hand, yields the desired value
of $\varphi^{\rm bulk}_{xy,xy}$ since we already know $\varphi^{\rm bulk}_{xx,yy}$
from the slab calculations.
Finally, we use $\varphi_{xy,xy} = -E^{\rm slab}_{xy,xy}$ to infer
$\varphi^{\rm surf}_{x y, x y}=\varphi_{xy,xy} - \varphi^{\rm bulk}_{xy,xy}$.

Our results for the bulk, surface, and total flexovoltage
coefficients of the truncated-bulk, frozen-ion deformation of a SrTiO$_3$
slab are summarized in Table~\ref{tab4}.
At the bulk level, it is interesting to note the relatively small magnitude
of the shear coefficients $\varphi^{\rm bulk}_{xy,xy}$ and $\varphi^{\rm surf}_{xy,xy}$
compared to both the longitudinal and the transverse ones.
Meanwhile, in the latter two cases there is a substantial cancellation
between bulk and surface terms;
as a result, the values of the total flexovoltage coefficients $\varphi$
are all comparable in magnitude.
This fact can be rationalized by observing that the linear
response to atomic displacements in a ionic (or partially ionic) solid
is largely
dominated by the rigid displacement of an approximately spherical charge
density distribution surrounding each atom.
The spherical contribution, which is typically large and negative,~\cite{hong-11}
shows up in  $\varphi^{\rm bulk}_{xx,\beta \beta}$, and with opposite sign
in $\varphi^{\rm surf}_{xx,\beta \beta}$; in the shear case neither the bulk
nor the surface term are affected (see Sec.~\ref{sec:spherical}).
Remarkably, the resulting values of $\varphi$ depend
strongly on the details of the surface, and in some cases even have
opposite signs in the SrO- and TiO$_2$-terminated slabs.
Such a conclusion, in fact, persists after we take into account the
full relaxation of the atomic structure; we shall demonstrate this point in
the following paragraphs.


\subsubsection{Relaxed-ion slab calculations}

%

\label{sec:relax-sto}

\begin{table}

\tbl{Flexovoltage coefficients of a relaxed SrTiO$_3$ slab. The frozen-ion (FI),
lattice-mediated (LM) and total relaxed-ion (RI=FI+LM) values of the bulk, surface and
total slab response are reported. Units of Volts are used throughout.}
{\begin{tabular}{crrrrr} \toprule
&  \multicolumn{1}{c}{$\varphi^{\rm bulk}$} & \multicolumn{2}{c}{$\varphi^{\rm surf}$} &  \multicolumn{2}{c}{$\varphi$ (total)} \\
&      & \multicolumn{1}{c}{SrO} &  \multicolumn{1}{c}{TiO$_2$}
                                  & \multicolumn{1}{c}{SrO} &  \multicolumn{1}{c}{TiO$_2$} \\
\colrule 
FI &  $-$10.37  &      13.47    &      6.84   &        3.10  &   $-$3.53 \\
LM &   $-$0.44  &    $-$4.93   &       5.34   &     $-$5.38  &      4.90 \\
\colrule 
RI &  ${-10.81}$  &     8.53     &    12.18    &     ${\bf -2.28}$  &    ${\bf 1.37}$ \\ \botrule
\end{tabular}}


\label{tab5}
\end{table}

The results of the relaxed-ion slab calculations allow us to complete the
picture of the fully relaxed flexovoltage response of a SrTiO$_3$
slab in the plate-bending limit.
[The beam-bending case
is easily recovered by multiplying the reported values by
$\tau =\mathcal{C}_{xx,xx} /  (\mathcal{C}_{xx,xx} + \mathcal{C}_{xx,yy})$.
By using the calculated elastic constants of bulk SrTiO$_3$, reported in
Table~\ref{tab:elas}, we find $\tau=0.77$.]
A summary of the results is reported in Table~\ref{tab5}.
The respective contributions of the bulk and surface are, overall, in line with the
available order-of-magnitude estimates.~\cite{Yudin-13}
The values shown in bold font, i.e., the total flexovoltage coefficients of the
two types of slab, comprise the main result of this work.
Note that they depart substantially from the
corresponding bulk coefficient, confirming the dramatic impact of the
surface structural and electronic properties on the electromechanical
response of the system.
In fact, the aforementioned response coefficients are even opposite in sign
depending on whether a SrO- and TiO$_2$-terminated slab is considered.
This is a remarkable result, as it means that an atomically thin surface
termination layer can modify, and even reverse, the flexovoltage
response of a macroscopically thick sample.
This constitutes a rather drastic departure from the characteristics
of other electromechanical phenomena (e.g., piezoelectricity), where the
details of the surfaces typically become irrelevant in the thermodynamic limit.

\begin{figure}[!h]
\begin{center}
\begin{tabular}{r}
\includegraphics[width=3.5in]{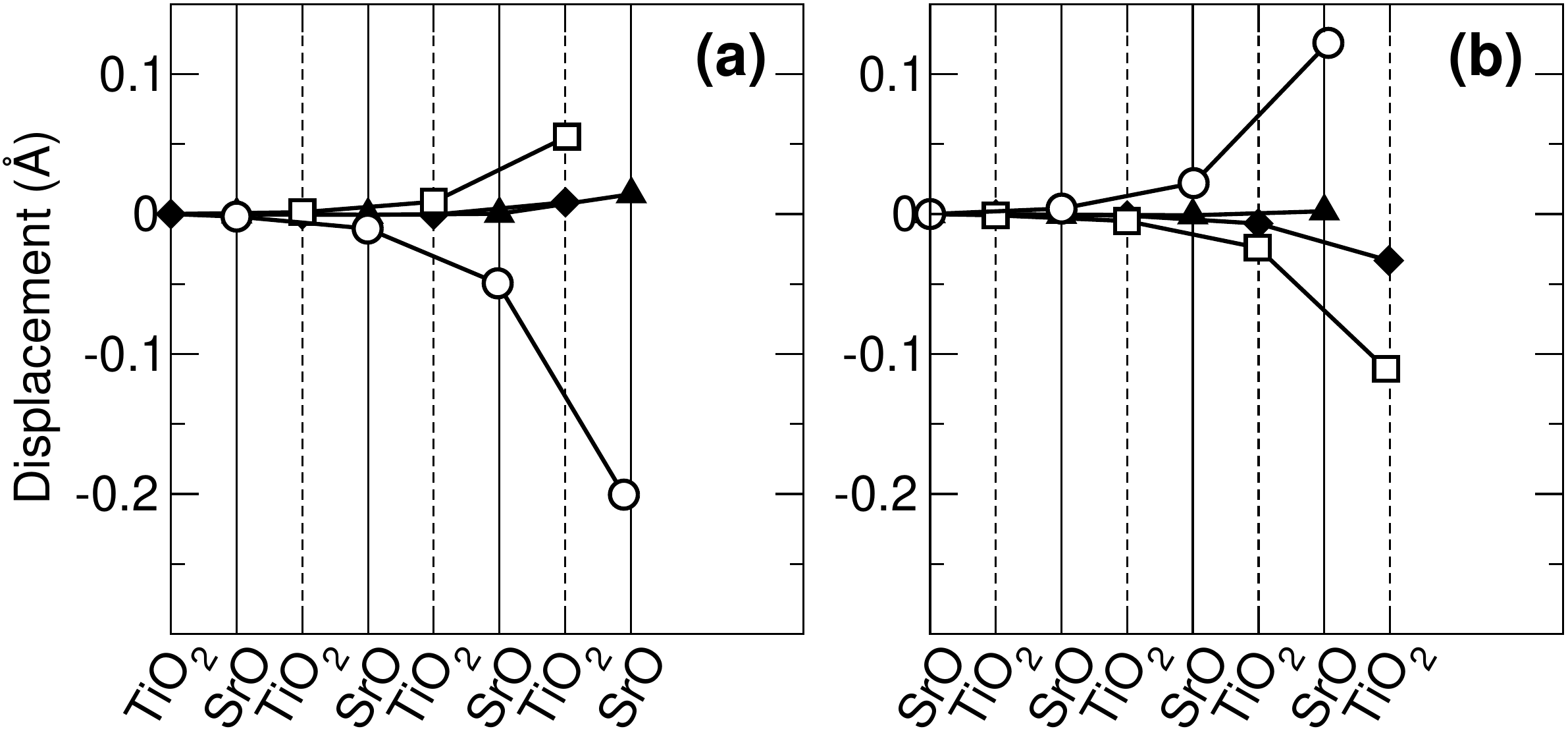} \\
\vspace{10pt}
\includegraphics[width=3.5in]{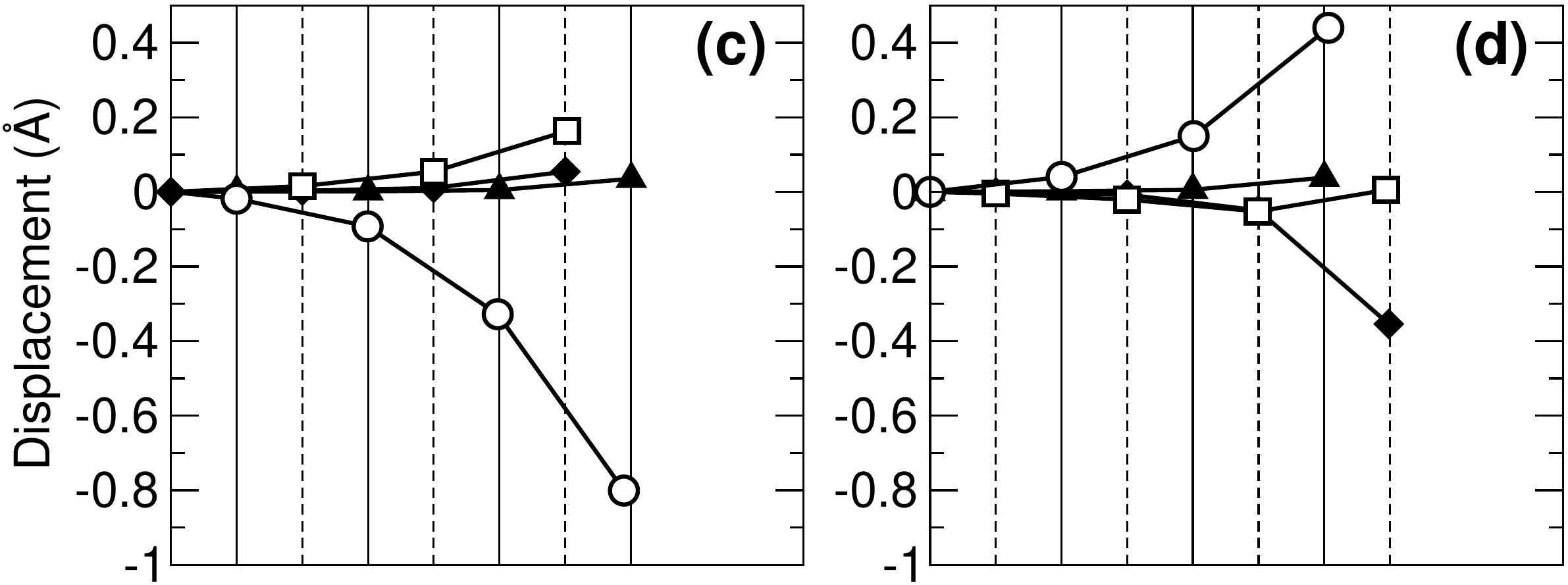} \\
\includegraphics[width=3.4in]{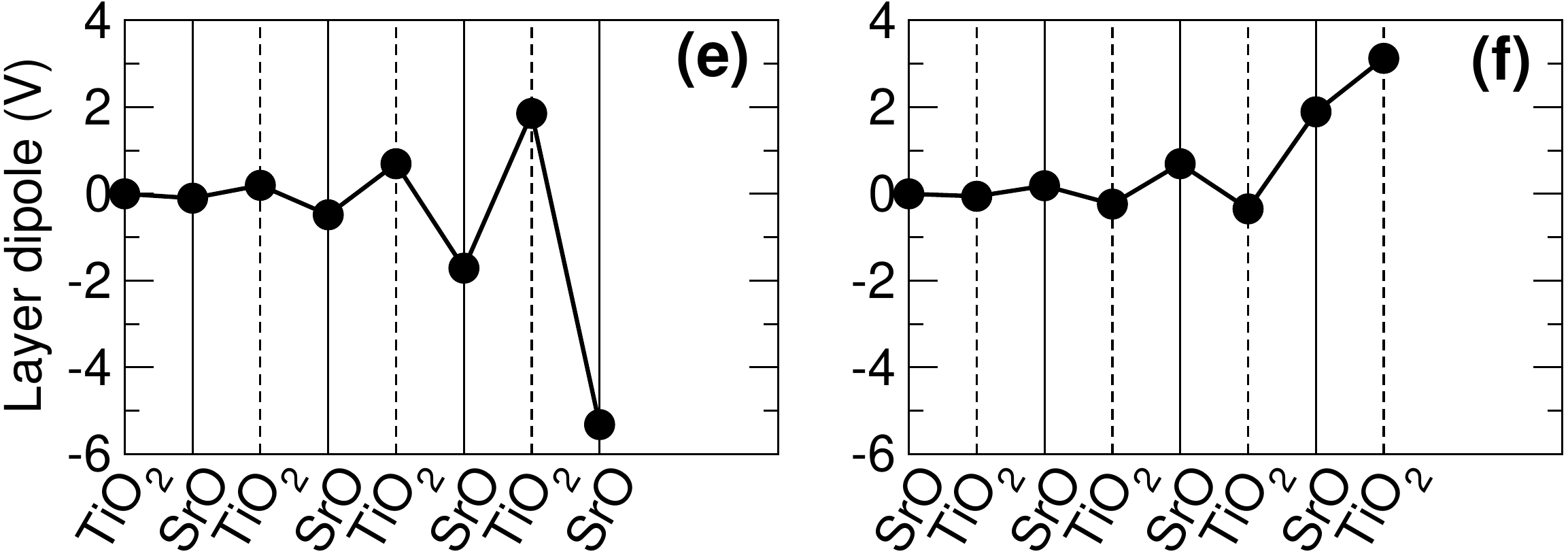}
\end{tabular}
\end{center}
\caption{Static and induced ionic relaxations at the SrTiO$_3$ surface.
(a-b): Ionic relaxations in the unperturbed slabs (displacements from ideal bulk-like sites).
Circles, squares, diamonds and triangles correspond, respectively, to Sr, Ti, O(Ti) and O(Sr)
atoms. (Cations are indicated by empty symbols, oxygen atoms by filled ones.)
Negative values indicate inward displacements (i.e., towards the slab center).
(c-d): Displacements induced by a uniform strain of the type
$\varepsilon_{yy} - \nu \varepsilon_{xx}$; for the two oxygen atoms in the TiO$_2$ layers,
only one value (their average displacement) is shown. (e-f) Layer-by-layer decomposition of the
lattice-mediated contribution to the induced surface potential offset. Vertical lines
indicate the position of the SrO (solid) and TiO$_2$ (dashed) atomic planes.
(Adapted from Ref.~\citeonline{artcalc}.)
\label{figrel} }
\end{figure}

It is interesting to note that the surface shows an even larger termination
dependence at the frozen-ion level, but with \emph{opposite} sign.
The LM contribution to $\varphi^{\rm surf}$ is indeed large,
and depends so strongly on the termination that its inclusion results
in a voltage reversal, both in the TiO$_2$- and SrO-type slabs.
(By contrast, the LM contribution to the bulk flexovoltage coefficient
is relatively minor, about one order of magnitude smaller than any other value
reported in the table, and has little impact on the final results.)
To illustrate the reason for such a strong dependence, a microscopic
analysis of the surface relaxations is provided in Fig.~\ref{figrel}.
In the SrO case, the layer-by-layer decomposition of the induced dipole
shown in Fig.~\ref{figrel}(e) has an oscillatory behavior whose amplitude
decays exponentially as a function of the distance from the surface;
as a consequence, the surface layer clearly dominates the overall
response.\myfoot{
  Interestingly, the structural relaxation pattern in the unperturbed state,
  Fig.~\ref{figrel}(a), appears very similar to the \emph{induced} relaxation
  pattern under an applied tensile strain. This suggests that the former might be,
  in fact, rationalized as a response of the system to a large surface stress.}
Instead, for the TiO$_2$-terminated slab shown in Fig.~\ref{figrel}(f),
the surface layer responds with a positive dipole instead of a negative
one, in sharp contrast to the ``underdamped'' oscillatory behavior
in Fig.~\ref{figrel}(e).
This behavior is probably due to the alteration of the bonding
network, which we speculate to be much more profound at the TiO$_2$-type
surface than at the SrO-type one, whereby the boundary atoms no longer
behave as bulk-like but rather as a distinct chemical entity.

Apart from the obvious relevance of the above observations to
the physics of SrTiO$_3$ surfaces, the analysis of Fig.~\ref{figrel}(e-f)
carries a general message that we have already anticipated in
the above paragraphs. Any single atomic layer near the surface
has a remarkably large contribution to $\varphi^{\rm surf}$, sometimes
of the same order as (or even larger than) the overall flexo\-voltage
response of the slab.
In fact, the total open-circuit voltage results from the subtle cancellation
of many contributions of dissimilar physical nature. This implies that
exceptional care is needed when dealing with flexoelectric phenomena,
either when performing the calculations or when interpreting the experiments.

\section{Conclusions and outlook}

In this Chapter we have described the main advances in the
first-principles theory of flexoelectricity that have taken place
during the past five years.
The progress that emerges from these pages is undoubtedly impressive --
we are at the stage where the full flexoelectric response
of real materials, including bulk and surface effects, can be
calculated \emph{ab initio} with great accuracy.
Still, much remains to be done before the field can be
regarded as mature. 
We discuss here several research avenues that we identify as being of
pivotal importance for future progress.

\begin{itemize}[leftmargin=0.8cm]

\item{\bf Theory of the current-density response.}
The most fundamental and complete framework for the theory of
flexoelectricity is the current-response formalism introduced
in Sec.~\ref{sec:longwave}.  Unlike the charge-response
formalism summarized in Sec~\ref{sec:electronic}, the
current-response approach is capable in principle of resolving
all independent components of the flexoelectric tensor.
However, two issues remain to be settled in relation to this
approach.  First, direct methods for obtaining the
current response functions $\overline{P}_{\kappa \beta}^{\bf q}$
of Eq.~(\ref{Pbar}) by computing
the linear response to a phonon of small but finite wavevector
$\bf q$ have not yet been developed and tested.  Once implemented,
this would allow for a finite-difference calculation of the
$\overline{P}_{\alpha, \kappa \beta} ^{(2,\gamma \lambda)}$ of
Eq.~(\ref{pq}), and thence, the electronic contribution in
Eq.~(\ref{mui-el}).  Second, some aspects of the connection
between the current-response theory and the theory of charge
responses (including surface charges) remain to be clarified, as
discussed in the context of Eq.~(\ref{E-transverse}) and following
Eq.~(\ref{delhp}). A solution of these two issues would help
put the theory of flexoelectricty on a truly sound footing.

\smallskip
\item{\bf Analytic derivation of the ${\bf q}$-expansions.}
The conceptual foundation of most of the material treated in this Chapter
is a long-wave expansion of certain physical observables as a function of
the wavevector ${\bf q}$ of an acoustic phonon.
The calculations described in Sec.~\ref{sec:results} were performed by
taking such a ${\bf q}$-expansion numerically via finite differences,
which is computationally cumbersome. Ideally, it would be
best to perform the expansion analytically, i.e., to derive the DFPT equations
that directly yield the wavefunction response to a strain gradient
perturbation.
This would also be desirable in the context of the direct
current-density implementation sketched just above.
When implemented in an existing DFPT code,
such methods would allow for a
more straightforward calculation of flexoelectric properties of materials,
and thus foster a more widespread application of these techniques
within the research community.

\smallskip
\item{\bf Application to complex materials.}
Our focus in this chapter has been on materials with cubic
symmetry.  Clearly a proper theory that also covers crystals of
lower symmetry is strongly required.  The extension of the theory
to such materials will require attention not just to
the proliferation of independent parameters in the flexoelectric
tensor, but also to subtle physical issues having to do, for
example, with the anisotropic electronic screening that occurs
when the symmetry is reduced.  In the case of crystals that are
piezoelectric (and possibly also polar), care will be needed to
separate the higher-order flexoelectric from dominant piezoelectric
(and possibly spontaneous) polarization response.  The application
to insulating ferromagnets or antiferromagnets should introduce no
special difficulties in most cases, but may involve subtleties
for magnetoelectric crystals or when spin-orbit coupling is strong.
A first-principles theory of {\it flexomagnetism} has yet to be developed.

\smallskip
\item{\bf Compositional gradients.}
An electric polarization can also arise in the presence of a
{\it compositional} gradient, e.g., in Ba$_{1-x}$Ti$_x$O$_3$
films.\cite{zhang-prb14}
To our knowledge, a proper theory of such an effect is lacking.
Since a compositional gradient generally also entails a strain gradient,
some care will be called for in separating these effects and computing
them independently before combining the contributions to make physically
meaningful predictions.

\smallskip
\item{\bf Connection to higher-level models.}
With the techniques described here, one can in principle calculate
the fundamental flexoelectric properties of an arbitrary material.
To use this information in real physical problems, however, one typically
has to deal with many additional issues that are intractable by means
of direct first-principles simulation: large samples with complex shapes,
temperature effects, etc.
It would be very desirable in this context to be able to extract the
relevant physical parameters from the \emph{ab initio} calculations,
and incorporate them in some higher-level theory (e.g. atomistic,
effective Hamiltonian, or continuum) where length- and time-scale
limitations are much less stringent.
A successful attempt in this sense has already been reported;~\cite{ponomareva}
still, consistently incorporating the latest first-principles developments into
macroscopic theories remains an open challenge.
For example, it would be of crucial importance, for a realistic
description of the flexoelectric effect, to extract the relevant
surface-specific properties from the density-functional calculations,
and incorporate them into the higher-level model.
Making progress in this direction will also promote a closer interaction
between different communities working on flexoelectricity (continuum numerical
modeling, Landau theory, etc.), which we believe would have a strong positive
impact on the field.

\end{itemize}


In summary, there has been dramatic progress in the development of a full
first-principles theory of flexoelectricity. Several important
challenges remain, as discussed above, but at least these have
been identified, and solutions appear to be within reach.
In any case,
the development of the theory of flexoelectricity has already revealed 
many fascinating links to other, at first sight unrelated, research areas
(e.g., the relationship to transformation optics, where the use of curvilinear
coordinates facilitates the solution of complex electrical engineering problems).
We believe that more surprises are in store, and will progressively emerge
while further progress is made along the above lines.
As the study of flexoelectricity touches so many subfields of condensed matter
physics, we expect cross-cutting progress that will 
likely benefit the first-principles materials theory community at large.
All in all, we look forward to the day when predictive calculations of
flexoelectric responses can become a routine part of the
tool-kit of first-principles computational materials theory.

\acknowledgments

We thank Jiawang Hong for useful discussions.  We acknowledge support from
ONR Grant N00014-12-1-1035 (D.V.), a grant from the Simons Foundation
(\#305025 to D.V.), MINECO-Spain Grant FIS2013-48668-C2-2-P
(M.S.) and Generalitat de Catalunya Grant 2014 SGR 301 (M.S.).

\bibliographystyle{ws-rv-van}
\bibliography{merged}

\begin{thebibliography}{36}
\providecommand{\natexlab}[1]{#1}
\providecommand{\url}[1]{\texttt{#1}}
\expandafter\ifx\csname urlstyle\endcsname\relax
  \providecommand{\doi}[1]{doi: #1}\else
  \providecommand{\doi}{doi: \begingroup \urlstyle{rm}\Url}\fi

\bibitem{jones-rmp89}
R.~O. Jones and O.~Gunnarsson, The density functional formalism, its
  applications and prospects, \emph{Reviews of Modern Physics}. {\bf 61},
  \penalty0 689  (1989).

\bibitem{Baroni/deGironcoli/DalCorso:2001}
S.~Baroni, S.~de~Gironcoli, and A.~D. Corso, Phonons and related crystal
  properties from density-functional perturbation theory., \emph{Rev. Mod.
  Phys.} {\bf 73}, \penalty0 515  (2001).

\bibitem{Martin}
R.~M. Martin, Piezoelectricity, \emph{Phys. Rev. B}. {\bf 5}, \penalty0
  1607--1613  (1972).

\bibitem{resta:92}
R.~Resta, Theory of the electric polarization in crystals,
  \emph{Ferroelectrics}. {\bf 136}, \penalty0 51--55  (1992).

\bibitem{King-Smith/Vanderbilt:1993}
R.~D. King-Smith and D.~Vanderbilt, Theory of polarization of crystalline
  solids, \emph{Phys. Rev. B}. {\bf 47}, \penalty0 R1651--R1654  (1993).

\bibitem{ferro:2007}
R.~Resta and D.~Vanderbilt.
\newblock Theory of polarization: A modern approach.
\newblock In eds. K.~M. Rabe, C.~H. Ahn, and J.-M. Triscone, \emph{Physics of
  Ferroelectrics: A Modern Perspective}. Springer-Verlag, Berlin Heidelberg
  (2007).

\bibitem{resta-jpcm2010}
R.~Resta, Electrical polarization and orbital magnetization: the modern
  theories, \emph{Journal of Physics: Condensed Matter}. {\bf 22}, \penalty0
  123201  (2010).

\bibitem{Tagantsev}
A.~K. Tagantsev, Piezoelectricity and flexoelectricity in crystalline
  dielectrics, \emph{Phys. Rev. B}. {\bf 34}, \penalty0 5883  (1986).

\bibitem{tagantsev-pt91}
A.~Tagantsev, Electric polarization in crystals and its response to thermal and
  elastic perturbations, \emph{Phase Transit.} {\bf 35}, \penalty0 119--203
  (1991).

\bibitem{hong-jpcm10}
J.~Hong, G.~Catalan, J.~F. Scott, and E.~Artacho, The flexoelectricity of
  barium and strontium titanates from first principles, \emph{J.~Phys.:
  Condens. Matter}. {\bf 22}, \penalty0 478--492  (2010).

\bibitem{Resta-10}
R.~Resta, Towards a bulk theory of flexoelectricity, \emph{Phys. Rev. Lett.}
  {\bf 105}, \penalty0 127601  (2010).

\bibitem{hong-11}
J.~Hong and D.~Vanderbilt, First-principles theory of frozen-ion
  flexoelectricity, \emph{Phys. Rev. B}. {\bf 84}, \penalty0 180101(R)  (2011).

\bibitem{artlin}
M.~Stengel, Flexoelectricity from density-functional perturbation theory,
  \emph{Phys. Rev. B}. {\bf 88}, \penalty0 174106  (2013).

\bibitem{Hong-13}
J.~Hong and D.~Vanderbilt, First-principles theory and calculation of
  flexoelectricity, \emph{Phys. Rev. B}. {\bf 88}, \penalty0 174107  (2013).

\bibitem{artgr}
M.~Stengel, Microscopic response to inhomogeneous deformations in curvilinear
  coordinates, \emph{Nature Communications}. {\bf 4}, \penalty0 2693  (2013).

\bibitem{Pavlo}
P.~Zubko, G.~Catalan, A.~Buckley, P.~R.~L. Welche, and J.~F. Scott,
  Strain-gradient-induced polarization in {SrTiO}$_{3}$ single crystals,
  \emph{Phys. Rev. Lett.} {\bf 99}, \penalty0 167601  (2007).

\bibitem{Baldereschi-88}
A.~Baldereschi, S.~Baroni, and R.~Resta, Band offsets in lattice-matched
  heterojunctions: a model and first-principles calculations for {GaAs/AlAs},
  \emph{Phys. Rev. Lett.} {\bf 61}, \penalty0 734--737  (1988).

\bibitem{resta-prb91}
R.~Resta, Deformation-potential theorem in metals and in dielectrics,
  \emph{Phys. Rev. B}. {\bf 44}, \penalty0 11035--11041  (1991).

\bibitem{Resta-DP}
R.~Resta, L.~Colombo, and S.~Baroni, Absolute deformation potentials in
  semiconductors, \emph{Phys. Rev. B}. {\bf 41}, \penalty0 12358--12361
  (1990).

\bibitem{Gygi-93}
F.~Gygi, Electronic-structure calculations in adaptive coordinates, \emph{Phys.
  Rev. B}. {\bf 48}, \penalty0 11692--11700  (1993).

\bibitem{Hamann-metric}
D.~R. Hamann, X.~Wu, K.~M. Rabe, and D.~Vanderbilt, Metric tensor formulation
  of strain in density-functional perturbation theory, \emph{Phys. Rev. B}.
  {\bf 71}, \penalty0 035117  (2005).

\bibitem{cloaks}
W.~Yan, M.~Yan, Z.~Ruan, and M.~Qiu, Coordinate transformations make perfect
  invisibility cloaks with arbitrary shape, \emph{New Journal of Physics}. {\bf
  10}, \penalty0 043040  (2008).

\bibitem{genrel}
U.~Leonhardt and T.~G. Philbin, General relativity in electrical engineering,
  \emph{New Journal of Physics}. {\bf 8}, \penalty0 247  (2006).

\bibitem{fixedd}
M.~Stengel, N.~A. Spaldin, and D.~Vanderbilt, Electric displacement as the
  fundamental variable in electronic-structure calculations, \emph{Nature
  Physics}. {\bf 5}, \penalty0 304--308  (2009).

\bibitem{puma}
P.~Umari, A.~D. Corso, and R.~Resta, Inside dielectrics: Microscopic and
  macroscopic polarization, \emph{AIP Conference Proceedings}. {\bf 582},
  \penalty0 107--117  (2001).

\bibitem{Junquera-07}
J.~Junquera, M.~H. Cohen, and K.~M. Rabe, Nanoscale smoothing and the analyis
  of interfacial charge and dipolar densities, \emph{J. Phys.: Condens.
  Matter}. {\bf 19}, \penalty0 213203  (2007).

\bibitem{artcalc}
M.~Stengel, Surface control of flexoelectricity, \emph{Phys. Rev. B}. {\bf 90},
  \penalty0 201112(R)  (2014).

\bibitem{Hertel-13}
R.~Hertel, Flexomagnetism and curvature-induced magnetochirality, \emph{Spin}.
  {\bf 3}, \penalty0 1340009  (2013).

\bibitem{abinit}
X.~Gonze, B.~Amadon, P.-M. Anglade, J.-M. Beuken, F.~Bottin, P.~Boulanger,
  F.~Bruneval, D.~Caliste, R.~Caracas, M.~{C\^ot\'e}, T.~Deutsch, L.~Genovese,
  P.~Ghosez, M.~Giantomassi, S.~Goedecker, D.~Hamann, P.~Hermet, F.~Jollet,
  G.~Jomard, S.~Leroux, M.~Mancini, S.~Mazevet, M.~Oliveira, G.~Onida,
  Y.~Pouillon, T.~Rangel, G.-M. Rignanese, D.~Sangalli, R.~Shaltaf, M.~Torrent,
  M.~Verstraete, G.~Zerah, and J.~Zwanziger, {ABINIT}: {F}irst-principles
  approach to material and nanosystem properties, \emph{Computer Phys. Commun.}
  {\bf 180}, \penalty0 2582 -- 2615  (2009).

\bibitem{Perdew/Wang:1992}
J.~P. Perdew and Y.~Wang, Accurate and simple analytic representation of the
  electron-gas correlation energy, \emph{Phys. Rev. B}. {\bf 45}, \penalty0
  13244  (1992).

\bibitem{troullier}
N.~Troullier and J.~L. Martins, Efficient pseudopotentials for plane-wave
  calculations, \emph{Phys. Rev. B}. {\bf 43}, \penalty0 1993--2006  (1991).

\bibitem{fhi98pp}
M.~Fuchs and M.~Scheffler, Ab initio pseudopotentials for electronic structure
  calculations of polyatomic systems using density-functional theory,
  \emph{Computer Phys. Commun.} {\bf 119}, \penalty0 67--98  (1999).

\bibitem{Monkhorst/Pack:1976}
H.~J. Monkhorst and J.~D. Pack, Special points for brillouin-zone integrations,
  \emph{Phys. Rev. B}. {\bf 13}, \penalty0 5188--5192  (1976).

\bibitem{Yudin-13}
P.~V. Yudin and A.~K. Tagantsev, Fundamentals of flexoelectricity in solids,
  \emph{Nanotechnology}. {\bf 24}, \penalty0 432001  (2013).

\bibitem{zhang-prb14}
J.~Zhang, R.~Xu, A.~R. Damodaran, Z.-H. Chen, and L.~W. Martin, Understanding
  order in compositionally graded ferroelectrics: Flexoelectricity, gradient,
  and depolarization field effects, \emph{Phys. Rev. B}. {\bf 89}, \penalty0
  224101  (2014).

\bibitem{ponomareva}
I.~Ponomareva, A.~K. Tagantsev, and L.~Bellaiche, Finite-temperature
  flexoelectricity in ferroelectric thin films from first principles,
  \emph{Phys. Rev. B}. {\bf 85}, \penalty0 104101  (2012).

\end{thebibliography}

\end{document}